\documentclass[11pt,a4paper]{article}
\pdfoutput=1

\usepackage[LGRx,T1]{fontenc}
\usepackage[utf8]{inputenc}
\usepackage[greek,english]{babel}
\usepackage{jcappub}
\usepackage{bm}
\usepackage{lmodern}
\usepackage{booktabs}
\usepackage{array}
\usepackage[table]{xcolor}
\usepackage{relsize}
\usepackage{mathrsfs}
\usepackage{bm}
\usepackage{feynmp}

\RequirePackage{ifpdf}
\ifpdf
\else 
\fi

\newcommand{\Hint}{H_{\text{int}}}
\newcommand{\Lint}{L_{\text{int}}}

\newcommand{\tinit}{t_{\text{init}}}
\newcommand{\tauinit}{\tau_{\text{init}}}
\newcommand{\Hinit}{H_{\text{init}}}

\newcommand{\Veff}{V_{\text{eff}}}

\renewcommand{\d}{\mathrm{d}}
\newcommand{\im}{\mathrm{i}}
\newcommand{\e}[1]{\mathrm{e}^{{#1}}}

\newcommand{\Npre}{N_{\text{pre}}}

\newcommand{\ext}[1]{{\mathsf{{#1}}}}
\newcommand{\extgreek}[1]{{\text{\textsf{\textgreek{{#1}}}}}}
\newcommand{\flip}[1]{\bar{{#1}}}
\newcommand{\tensoridx}[1]{{\mathfrak{{#1}}}}

\newcommand{\Mp}{M_{\mathrm{P}}}

\newcommand{\vect}[1]{\bm{\mathrm{{#1}}}}

\newcommand{\Aslow}{A_{\text{slow}}}
\newcommand{\massmatrix}{m}

\newcommand{\fNL}{f_{\mathrm{NL}}}
\newcommand{\fNLlocal}{f_{\mathrm{NL}}^{\text{local}}}

\newcommand{\dimlessP}{\mathcal{P}}
\newcommand{\dimlessB}{\mathcal{B}}

\newcommand{\packagefont}{\sffamily\fontseries{sbc}\selectfont}

\newcommand{\MadGraph}{{\packagefont MadGraph}}

\newcommand{\FormCalc}{{\packagefont FormCalc}}
\newcommand{\GiNaC}{{\packagefont GiNaC}}
\newcommand{\xloopsginac}{{\packagefont XLOOPS-GiNaC}}
\newcommand{\LanHEP}{{\packagefont LanHEP}}
\newcommand{\CompHEP}{{\packagefont CompHEP}}
\newcommand{\CalcHEP}{{\packagefont CalcHEP}}
\newcommand{\FeynRules}{{\packagefont FeynRules}}

\newcommand{\ModeCode}{{\packagefont ModeCode}}
\newcommand{\MultiModeCode}{{\packagefont MultiModeCode}}
\newcommand{\FieldInf}{{\packagefont FieldInf}}
\newcommand{\PyFlation}{{\packagefont PyFlation}}

\newcommand{\BINGO}{{\packagefont BINGO}}

\newcommand{\PyTransport}{{\packagefont PyTransport}}
\newcommand{\CppTransport}{{\packagefont CppTransport}}

\newcommand{\SymPy}{{\packagefont SymPy}}
\newcommand{\Matplotlib}{{\packagefont Matplotlib}}
\newcommand{\Mayavi}{{\packagefont Mayavi}}

\newcommand{\MpiForPy}{{\packagefont Mpi4Py}}
\newcommand{\MPI}{{\packagefont MPI}}

\newcommand{\SQLite}{{\packagefont SQLite}}

\DeclareMathOperator{\TimeOrder}{\mathsf{T}}
\newcommand{\AntiTimeOrder}{\bar{\TimeOrder}}

\DeclareMathOperator{\Or}{\mathrm{O}}
\newcommand{\Operator}{\mathscr{O}}

\newcommand{\semibold}[1]{{\fontseries{b}\selectfont{#1}}}
\newcommand{\para}[1]{\par\vspace{2mm}\noindent\semibold{{#1.}---}\ignorespaces}

\DeclareMathOperator{\newRe}{Re}
\DeclareMathOperator{\newIm}{Im}
\renewcommand{\Re}{\newRe}
\renewcommand{\Im}{\newIm}

\renewcommand{\leq}{\leqslant}

\DeclareMathOperator{\tr}{tr}

\newcommand{\GitRevision}[2]{{\small\sffamily \packagefont\href{#2}{{#1}}}}

\newcommand\CC{C\nolinebreak\hspace{-.05em}\raisebox{.4ex}{\relsize{-3}{\textbf{+}}}\nolinebreak\hspace{-.10em}\raisebox{.4ex}{\relsize{-3}{\textbf{+}}}}

\usepackage{color}

\graphicspath{{./}}

\begin{document}

\begin{center}
\rightline{\small DESY-16-171}
\vskip -1.3cm
\end{center}

\title{\fontseries{s}\selectfont Numerical evaluation of the bispectrum in multiple field inflation \\
{\Large \it -- the transport approach with code}
\vspace{5mm}\hrule}
\author{\fontseries{s}\selectfont\large Mafalda Dias$^{1,2}$, Jonathan Frazer$^{1,3,4}$, David J. Mulryne$^5$ and David Seery$^2$}

\affiliation{\vspace{2mm}\small
$^1$ Theory Group, Deutsches Elektronen-Synchrotron, DESY, D-22603, Hamburg, Germany \\
$^2$ Astronomy Centre, University of Sussex, Falmer, Brighton BN1 9QH, UK \\
$^3$ Department of Theoretical Physics, University of the Basque Country, \\ UPV/EHU
48040 Bilbao, Spain \\
$^4$ \textsc{ikerbasque}, Basque Foundation for Science, 48011 Bilbao, Spain \\
$^5$ School of Physics and Astronomy, Queen Mary
University of London, Mile End Road, \\ London E1 4NS, UK
}

\emailAdd{mafalda.dias@desy.de}
\emailAdd{jonathan.frazer@desy.de}
\emailAdd{d.mulryne@qmul.ac.uk}
\emailAdd{D.Seery@sussex.ac.uk}

\abstract{We present a complete framework for numerical calculation of
the power spectrum and bispectrum in canonical inflation with
an arbitrary number of light or heavy fields.
Our method includes all relevant effects at tree-level in the loop expansion,
including (i) interference between growing and decaying modes near horizon exit;
(ii) correlation and coupling between species near horizon exit and on
superhorizon scales; (iii) contributions from mass terms;
and (iv) all contributions from
coupling to gravity.
We track the evolution of each correlation function
from the vacuum state
through horizon exit and the
superhorizon regime, with no need to match quantum and classical parts of the
calculation;
when integrated, our approach corresponds exactly with the tree-level
Schwinger or `in--in' formulation of quantum field theory.
In this paper we give the equations necessary to evolve
all two- and three-point correlation functions
together with suitable initial conditions.
The final formalism is suitable to compute the amplitude, shape, and scale
dependence of the bispectrum in models with $|\fNL|$ of order unity or less,
which are a target for future galaxy surveys
such as Euclid, DESI and LSST.
As an illustration we apply our framework to a number of examples,
obtaining quantitatively accurate
predictions for their bispectra for the first time.
Two accompanying reports describe
publicly-available software packages that implement the method.
\par\vspace{6mm}\mbox{}\hfill
\includegraphics[scale=0.1]{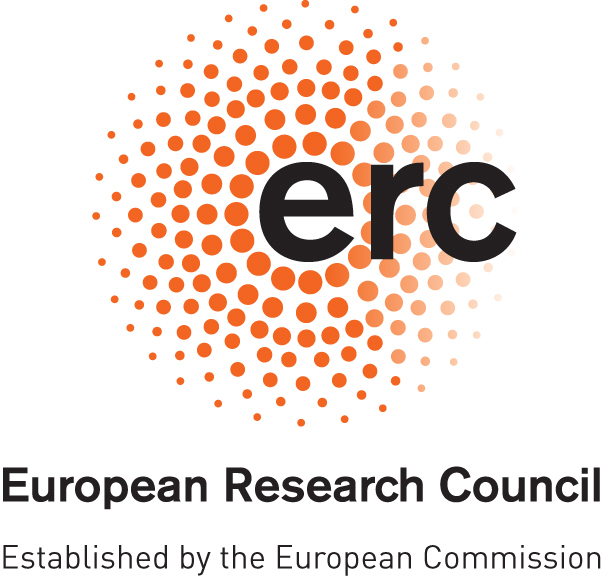}
\quad
\includegraphics[scale=0.5]{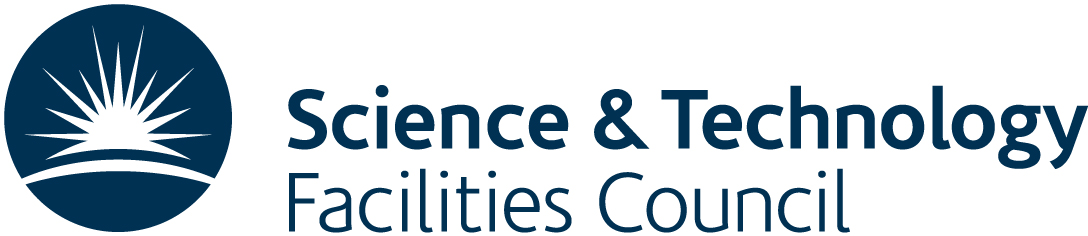}
\quad
\includegraphics[scale=0.3]{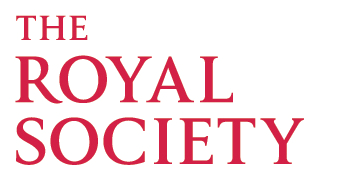}}

\maketitle
\newpage

\begin{fmffile}{feyn_diags}

\section{Introduction}
\label{sec:introduction}
In the inflationary scenario,
quantum-mechanical processes seeded an early distribution of gravitational
potential wells.
As matter sank into these wells it formed the largest structures we observe,
with the result that any
observable tracing this structure can be used to infer details
of the seeding process.
To do so we require
measurements on cosmological scales,
which continue to improve at a remarkable
rate---and, soon, we can expect
the cosmic microwave background temperature
and polarization anisotropies to be joined as a precision
probe
by the galaxy density field, intrinsic alignments, weak lensing shear maps
and perhaps others.

The raw materials for these analyses are the correlation functions
describing the primordial distribution of potential wells,
and
to calculate them we need the methods of quantum field theory in curved
spacetime.
We will discuss these methods in detail below, and explain
why such calculations are challenging and why analytic methods are limited.
But two of the reasons are easy to state and
especially difficult to overcome:
first, the algebraic complexity
arising when even the simplest models are coupled to gravity;
and second,
the occurrence of large hierarchies
that compensate for the smallness
of any natural expansion parameters
and render na\"{\i}ve perturbation theory useless.
The most straightforward solution to both these
issues is to switch to a numerical method.
Exactly this approach has been adopted in other areas of physics---%
including collider phenomenology, the paradigmatic example
of extracting observational predictions from quantum
field theory---%
when the same difficulties are encountered.

The tools available to assist cosmologists
in these calculations
are substantially less sophisticated than in collider phenomenology,
where not only the numerical computations are implemented by software---%
it is possible to combine
powerful computer packages
that partially \emph{automate} the calculation of LHC observables
directly from a Lagrangian~\cite{Degrande:2011ua}.
Examples include
the Feynman diagram generators
\href{http://theory.sinp.msu.ru/~semenov/lanhep.html}{\LanHEP}~\cite{Semenov:1998eb,Semenov:2008jy}
or
\href{https://feynrules.irmp.ucl.ac.be}{\FeynRules}~\cite{Christensen:2008py,Christensen:2009jx}
combined with
\href{http://theory.sinp.msu.ru/~pukhov/calchep.html}{{\CompHEP}/{\CalcHEP}}~\cite{Boos:2004kh,Pukhov:1999gg,Pukhov:2004ca},
\href{http://madgraph.hep.uiuc.edu}{\MadGraph}~\cite{Alwall:2014hca}
or
\href{http://www.feynarts.de/formcalc/}{\FormCalc}~\cite{Hahn:1998yk,Hahn:2006qw}.
Numerical tools for inflationary calculations have been developed,
but typically they work on a case-by-case basis
where derivatives of the potential must be obtained by hand and
supplied as subroutines.
Examples include the Fortran codes
\href{http://theory.physics.unige.ch/~ringeval/fieldinf.html}{\FieldInf}
\cite{Ringeval:2005yn,Martin:2006rs,Ringeval:2007am},
\href{http://modecode.org}{\ModeCode} and
\href{http://modecode.org}{\MultiModeCode}
\cite{Mortonson:2010er,Easther:2011yq,Norena:2012rs,Price:2014xpa},
and the Python package
\href{http://pyflation.ianhuston.net}{\PyFlation}~\cite{Huston:2009ac,Huston:2011vt,Huston:2011fr},
which are all solvers
for the two-point function in
models of varying generality.
For the three-point function,
the only public code of which we are aware is
the Fortran90 solver
\href{https://sites.google.com/site/codecosmo/bingo}{\BINGO}~\cite{Hazra:2012yn,Sreenath:2014nca}
which is restricted to single-field canonical models.

Providing derivatives of the potential by hand
becomes burdensome when the potential is complex
or there are many fields---unfortunately,
precisely the cases where numerical
methods have most value.
The situation is worse where a nontrivial field-space metric or kinetic structure means that
further derivatives are required, such as the field-space Riemann tensor.
It would be preferable,
as in collider phenomenology,
to automate these calculations.
But
automated tools have
advantages beyond mere convenience,
providing a fair basis for comparison between models by dropping simplifying assumptions
and enabling researchers whose primary interest may be model building
(rather than the calculation of correlation functions
as an end in themselves)
to obtain observable predictions close to
the state-of-the-art in technical sophistication.
More ambitiously, once $n$-point functions are available there is no need to
stop: we can extend the analysis to
include automated calculation of
late-universe observables such as the
CMB angular correlation functions, the dark matter or galaxy correlation functions,
estimates of scale-dependent bias, and so on.

In Ref.~\cite{Dias:2015rca}, three of us presented a Mathematica
\href{http://transportmethod.com}{`transport' code}
that can automate (in the sense just described)
calculation of the inflationary two-point function
in a multiple-field model with nontrivial field-space metric,
given only symbolic expressions for the metric and potential.
Like those described above this code is a two-point function solver.
But present-day datasets already have some sensitivity to
inflationary \emph{three}-point functions,
and this sensitivity is expected to improve
as large-scale galaxy surveys such as Euclid, DESI and LSST
become available.
To enable model-building theorists to compare their
scenarios with these datasets it would be very convenient to
automate calculation of the
three-point functions.
In this paper we describe a numerical method for doing so---%
currently,
applied to an arbitrary
multifield model with canonical kinetic terms---%
and collect a number of results showcasing its utility.
The method generalizes to nontrivial kinetic terms but the
calculations necessary to implement it have not yet been performed.
We intend to return to this in the future.

For those wishing to replicate our analyses,
or apply our methods to their own models,
we have made our computer codes available.
The transport code described in Ref.~\cite{Dias:2015rca}
was implemented in Mathematica, which trades speed and flexibility for
a certain kind of convenience
inherited from access to Mathematica's symbolic engine and visualization
capabilities.
However, because numerical calculation of three-point functions is
substantially slower
than for two-point functions---and the complexity of implementation and maintenance
substantially higher---%
it is not clear that Mathematica will continue to provide
a suitable platform.
To accommodate this our
bispectrum codes are mostly written in compiled languages
and optionally can be parallelized.
They have been tested against each other but do not share code,
and have different specializations and use cases:
\begin{itemize}
	\item {\PyTransport} is developed by a team at Queen Mary, University of London.
	It has a {\CC} core but is intended to be used through a Python interface,
	and uses the
	\href{http://www.sympy.org/en/index.html}{\SymPy} package to provide its
	symbolic algebra support.
	It has minimal prerequisites, supports rapid development,
	and
	its Python interface means that it is easily scriptable.
	It can be used with
	libraries such as
	\href{http://matplotlib.org}{\Matplotlib} or
	\href{http://code.enthought.com/projects/mayavi/}{\Mayavi}
	for visualization
	and can be parallelized using
	\href{http://pythonhosted.org/mpi4py/}{\MpiForPy} or similar packages.

	\item {\CppTransport} is developed by a team at the University of Sussex.
	It is a pure {\CC} platform
	using the
	\href{http://www.ginac.de}{\GiNaC} library
	(originally developed as part of the
	one-loop particle physics project
	\href{http://wwwthep.physik.uni-mainz.de/~xloops/}{\xloopsginac}~\cite{Bauer:2001ig})
	for symbolic algebra.
	It uses
	\href{http://www.mpi-forum.org}{\MPI}
	multi-process communication
	to parallelize calculations and scales from laptop-class
	hardware up to many cores on a HPC cluster, without
	requiring a shared-memory architecture.
	Results are stored as
	\href{https://www.sqlite.org}{\SQLite} databases,
	enabling sophisticated postprocessing using SQL database queries.
	For simple analyses a suite of built-in visualization and analysis tools are
	provided.
\end{itemize}
Both codes automate
the calculation of two- and three-point field-space
correlation functions
(and hence two- and three-point $\zeta$ correlation functions)
directly from a potential.
They implement the same computational scheme, but in slightly
different ways.
In this paper our intention is to
describe this scheme,
illustrate its utility,
and explain its relationship to earlier work.
We refer to the codes when giving examples,
but they will be described more fully
elsewhere~\cite{CppTransportUserGuide, PyTransportUserGuide}.
They can be downloaded by following
the links given in~\S\ref{sec:automated-codes}.

\para{Synopsis}
This paper is divided into three parts.
First, in~\S\ref{sec:why-automated} we review the current state-of-the-art in
computing inflationary three-point functions.
This depends
on accurate calculation
of vacuum fluctuations over
a time-dependent cosmological background
for a variety of masses and parameter regimes.
In some regimes it is possible to find successful analytic approximations,
but in others the success of such approximations has been meagre.
We argue that dramatic short-term
improvements in our ability to make analytic
estimates are unlikely, making a numerical method essential.

Second, in~\S\S\ref{sec:numerical-computation}--\ref{sec:gauge-transform}
we describe our numerical approach.
In~\S\S\ref{sec:numerical-computation}--\ref{sec:transport-equations}
we show that it can be regarded
as a reformulation of the Schwinger or `in--in' method to compute
expectation values---as distinct from transition amplitudes,
which are the output of the Feynman calculus described in
particle physics textbooks.
As with all formulations of field theory, ours has advantages
and disadvantages.
A major advantage is that it is very well suited to numerical evaluation.
For example,
challenging steps such as Wick rotation
(that are best handled analytically)
appear only in the calculation of initial conditions
but \emph{not} in the subsequent evolution equations.
Suitable initial conditions need be calculated only once
and are universal for all models,
no matter what parameter regime they inhabit.
We describe this calculation in~\S\ref{sec:initial-conditions},
and the transition to $\zeta$ correlation functions in~\S\ref{sec:gauge-transform}.
The major disadvantage
of our method is that it obscures the clear physical
interpretation of
Feynman diagrams as processes occurring
in spacetime~\cite{Seery:2009hs,Arkani-Hamed:2015bza}.
But since we intend to apply our method for numerical evaluation this is not
so important.

Third, in~\S\S\ref{sec:automated-codes}--\ref{sec:performance}
we briefly describe our implementations
{\PyTransport} and {\CppTransport} before using them to showcase the
utility of the
method.
In~\S\ref{sec:examples}
we verify numerically that our formalism
successfully tracks the evolution of the two-
and three-point functions from sub- to super-horizon scales.
We demonstrate
that it can be used to extract very delicate variation with shape and scale,
and produce examples that exemplify some of the physical processes
discussed in~\S\ref{sec:why-automated}.
In each case, our analysis
yields results that can be obtained only using numerical techniques.
In~\S\ref{sec:performance} we discuss the numerical characteristics
of our method and its performance as implemented in
{\PyTransport} and {\CppTransport}.
In particular, we explain
how the
integration time and convergence properties scale with key adjustable
parameters.
Finally, we conclude in~\S\ref{sec:conclusions}.

\para{Notation}
We work in units where $c = \hbar = 1$
and use the reduced Planck mass
$\Mp^2 = (8\pi G)^{-1}$,
where $G$ is Newton's gravitational constant.
Our metric signature is
$(-,+,+,+)$.
We work in the Heisenberg picture except where otherwise stated.
In order to write concise expressions we use a number
of different summation conventions.
For details, see the discussion below Eq.~\eqref{eq:zeta-twopf-factorization};
the discussion around Eq.~\eqref{eq:extended-summation-example};
and the discussion above Eqs.~\eqref{eq:action-S2}--\eqref{eq:action-S3}.

\section{Why are automated tools necessary?}
\label{sec:why-automated}

\subsection{The standard calculation}
\label{sec:standard-calculation}
Estimates
of the correlation functions
characterizing
inflationary perturbations
have been refined over several decades and are
now very mature.
In particular,
the physical processes contributing to these
correlations are clearly understood.
We focus on equal-time $n$-point functions because
it is these that are observationally relevant.
Each correlator of this type is a function of the
wavenumbers $k_i$ associated with its external legs,
and at the time of evaluation
these are typically either
\emph{subhorizon} in the sense $k/aH \gg 1$
or \emph{superhorizon} in the sense $k/aH \ll 1$.
The short transition period where
$k/aH \sim 1$ is described as the epoch of \emph{horizon exit}
for the mode $k$.

\para{Contributions to correlation functions}
When all wavenumbers are subhorizon, the equivalence principle
means that their correlations must be nearly those of Minkowski space.%
    \footnote{We are assuming that the small-scale fluctuations are
    those of the Minkowski vacuum, described
    in this context as the `Bunch--Davies vacuum'.
    If we allow high-energy fluctuations to have different correlation
    properties, perhaps as a result of nonadiabatic redshifting
    out of a regime controlled by exotic interactions,
    then more general possibilities exist.}
Non-negligible correlations exist only on very short scales and are dominated by
ultraviolet fluctuations.
But as some wavenumbers approach horizon exit and pass to superhorizon
scales
it is possible for
long--short correlations to develop,
because
the equivalence principle cannot prevent
`short' wavelength perturbations
of roughly the Hubble scale from responding to the cosmological background
generated by `long' wavelength modes.
At the same time,
particle production processes
and
interference between
growing and decaying modes
generate a delicate pattern of correlations
between the Hubble-scale modes alone.
Eventually all wavenumbers move into the superhorizon regime.
In this phase there are no interference
effects, and
the correlation
functions describe
an evolving
ensemble of
realizations of the background
with different initial
conditions~\cite{Lyth:1984gv,Bardeen:1983qw,Wands:2000dp,Lyth:2005fi,Malik:2005cy,Seery:2012vj}.

An accurate calculation must capture all these effects.
The difficulty of doing so varies
with the model under discussion,
and
processes that are dominant for one class may be
negligible for others.
But, in general terms, there are two major challenges.

\begin{itemize}
    \item All correlation functions receive `quantum'
    contributions
    from a range of times around horizon exit of each
    momentum $k_i$~\cite{Rigopoulos:2003ak,Lyth:2004gb,Weinberg:2008nf,Weinberg:2008si,Weinberg:2005vy,Weinberg:2006ac}.
    Methods to estimate these contributions are
    well-developed,
    usually yielding
    the result a combination of integrals over
    rapidly-oscillating
    wavefunctions~~\cite{Maldacena:2002vr,Seery:2005wm,Chen:2006nt,Elliston:2012ab}.
    The integrands
    can be
    interpreted as the rate per unit
    volume for $n$-body
    interactions.
    In some cases these integrals can be performed analytically,
    but in others we are already forced to rely on numerical
    methods.
    Evaluation of these rapidly oscillating `Feynman integrals'
    is the first major
    challenge~\cite{Chen:2006xjb,Chen:2008wn,Hazra:2012yn,Funakoshi:2012ms,Adshead:2013zfa,Horner:2013sea,Horner:2015gza}.

    \item In models with `adiabatic' dynamics%
        \footnote{Here, `adiabatic' is a term of art
        meaning that there is only a single independent perturabtion.
        This happens when the calculation is effectively single-field,
        in the sense that all points in spacetime follow the same inflationary
        trajectory through field-space.}
    the calculation is especially simple.
    The interaction rate per unit volume becomes negligible
    when all $k_i$ pass to superhorizon scales,
    meaning that the integrals depend only on a few e-folds around
    horizon-exit of each wavenumber.
    It is only necessary to correctly estimate these contributions.

    If the
    dynamics are not adiabatic then the situation is more complex~\cite{Elliston:2011dr}.
    The interaction rate per unit volume
    need not become negligible even on superhorizon scales, and therefore
    interactions may continue into the indefinite future.
    After subtracting effects already present in Minkowski
    space, it follows that each Feynman integral
    can
    receive contributions from all times when at least one wavenumber
    is of horizon scale or larger.
    (It is the unboundedly growing volume of these integrals
    that is responsible for spoiling
    na\"{\i}ve perturbation theory~\cite{Dias:2012qy}.)
    In this regime the integrand is no longer rapidly oscillatory,
    but
    because some or all wavenumbers are soft in comparison with the Hubble scale
    they may be exquisitely sensitive to the mass spectrum
    and decay channels of the model,
    and therefore to its microphysical details.
    The second major challenge is to obtain
    accurate estimates for all necessary masses and rates
    that enter this calculation,
    including gravitational corrections where appropriate.
\end{itemize}

\para{Analytic approaches}
Whether we adopt analytic or numerical methods,
a practical approach must overcome both these challenges.
In this paper we are going to argue that numerical integration
is essential for evaluation of the rate integrals, even though
it introduces difficulties of its own. In particular,
obtaining accurate results is nontrivial because
the rapidly oscillating integrands
are hard to
accommodate~\cite{Chen:2006xjb,Chen:2008wn,Hazra:2012yn,Horner:2013sea,Horner:2015gza}.
We will show that this problem can be completely removed by writing
`transport' equations for each correlation function
rather than attempting to directly evaluate the Feynman integrals.
To overcome the second challenge we argue that automated methods based
on computer algebra
can be used to include all relevant terms, without
the need for approximations.

In the remainder of this paper we make these arguments in more detail.
But before doing so we pause to describe
how the general challenges identified above
manifest themselves in practical calculations.
Specifically, we describe the standard tools used when
making analytical estimates
and identify those parts of the calculation for which
automated tools offer significant benefits.

\para{Factorization, or the `separate universe' method}
Except for special cases, the only general method
for making analytic estimates
relies on \emph{factorizing} each correlation function
into a sum of terms.
Each term contains
factors capturing a contribution to one of the major
effects described above.
For example, we typically factorize the
equal-time
two-point function of the curvature perturbation $\zeta$
in the form
\begin{equation}
    \langle
        \zeta(\vect{k}_1)
        \zeta(\vect{k}_2)
    \rangle_{t}
    =
    N_a(t,t_\ast)
    N_b(t,t_\ast)
    \langle
        \delta X^a(\vect{k}_1)
        \delta X^b(\vect{k}_2)
    \rangle_{t_\ast} .
    \label{eq:zeta-twopf-factorization}
\end{equation}
The subscript attached to each correlation function denotes the common time
of evaluation for the enclosed operators,
and
we assume $t > t_\ast$.
The result is independent of $t_\ast$
and therefore it can be chosen to suit our
convenience;
normally we set $t_\ast$ to be just after
the horizon-exit epoch for the common scale $k = |\vect{k}_1| = |\vect{k}_2|$.
As we explain below,
this makes the physical interpretation of each factor as simple as possible.

We have collected perturbations in both the fields and their derivatives
into the vector $\delta X^a = (\delta \phi^\alpha, \delta \dot{\phi}^\beta)$.
Here, $\phi^\alpha$ are the fields of the model, labelled by Greek indices
$\alpha$, $\beta$, $\ldots$,
and
$\delta\dot{\phi}^\alpha = \d \delta \phi^\alpha / \d t$.
Latin indices $a$, $b$, {\ldots} range over the combined phase space.
The factorized coefficients
$N_a(t, t_\ast)$ are nearly independent of wavenumber,
and can be computed using any of a number of standard
methods~\cite{Sasaki:1995aw,GarciaBellido:1995qq,Lyth:2005fi,
Vernizzi:2006ve,Yokoyama:2007uu,Yokoyama:2007dw,Mulryne:2009kh,Mulryne:2010rp,Seery:2012vj,Elliston:2011dr}.

In some circumstances the dynamics may satisfy a `slow-roll approximation'
that
makes
the $\dot{\phi}^\alpha$ functions of the field expectation values $\phi^\alpha$.
Where this happens it is possible to collect the contribution from both field and momentum
into a new coefficient $N_\alpha(t, t_\ast)$ and write~\eqref{eq:zeta-twopf-factorization}
as a sum over fields alone. However, this simplification is not necessary for the factorization
property or for our ability to make estimates.

Eq.~\eqref{eq:zeta-twopf-factorization} is the simplest example of a
more general
factorization principle that applies to any correlation function
containing superhorizon wavenumbers~\cite{Sasaki:1995aw,Lyth:2005fi,Dias:2012qy}.
The analogous factorization for the three-point function is
\begin{equation}
\begin{split}
    \langle
        \zeta(\vect{k}_1)
        &
        \zeta(\vect{k}_2)
        \zeta(\vect{k}_3)
    \rangle_{t}
    =
    N_a
    N_b
    N_c
    \langle
        \delta X^a(\vect{k}_1)
        \delta X^b(\vect{k}_2)
        \delta X^c(\vect{k}_3)
    \rangle_{t_\ast}
    \\
    & \mbox{}
    +
    \bigg(
    N_{ab}
    N_c
    N_d
    \int \frac{\d^3 q \, \d^3 r}{(2\pi)^3} \; \delta(\vect{k}_1 - \vect{q} - \vect{r})
    \langle
        \delta X^a(\vect{q})
        \delta X^c(\vect{k}_2)
    \rangle_{t_\ast}
    \langle
        \delta X^b(\vect{r})
        \delta X^d(\vect{k}_3)
    \rangle_{t_\ast}
    \\
    & \mbox{}
    \qquad + \text{2 permutations}
    \bigg).
\end{split}
\label{eq:zeta-threepf-factorization}
\end{equation}
This expression involves a new set of wavenumber-independent
factorization coefficients $N_{ab}(t,t_\ast)$.
To reduce clutter
in Eq.~\eqref{eq:zeta-threepf-factorization}
we have suppressed the time labels associated with both
$N_a$ and $N_{ab}$.
Also, we are temporarily assuming there is no large hierarchy among the $k_i$
so that there is a \emph{single} approximate time of horizon exit $t_\ast$.
For some scenarios, such as those where multiple light fields are active
after horizon exit, almost everything we know about the inflationary bispectrum
comes from applying factorization principles of this kind.

The utility of Eqs.~\eqref{eq:zeta-twopf-factorization}
and~\eqref{eq:zeta-threepf-factorization}
arises from a separation of scales.
With our choice for $t_\ast$
(and continuing to assume there are no large hierarchies among the $k_i$),
the `source' or `initial' factors
$\langle \delta X^a \delta X^b \rangle_{t_\ast}$
and
$\langle \delta X^a \delta X^b \delta X^c \rangle_{t_\ast}$
encode details of correlations between Hubble-scale modes
just after horizon exit.
They represent the oscillatory contribution to the Feynman rate integrals
and are sensitive to particle production and interference effects
around horizon exit.
Meanwhile,
the coefficients $N_a$ and $N_{ab}$
capture growing contributions
associated with the superhorizon
era.
It is these coefficients that exhibit
very strong dependence on the
light part of the
mass spectrum.

Analytic estimates based on factorization have been
used successfully to analyse simple models.
But despite this, the challenges discussed above
mean that such methods do not scale efficiently to more complex
cases.
Expressed in the concrete language of
Eqs.~\eqref{eq:zeta-twopf-factorization}
and~\eqref{eq:zeta-threepf-factorization}
these challenges are:
\begin{itemize}
    \item \semibold{Combinatorics.}
    In a model with $N$ fields, estimates of
    $\langle \zeta \zeta \rangle$
    and
    $\langle \zeta \zeta \zeta \rangle$
    require us to compute
    the quantities
    $N_a$, $N_{ab}$
    and the correlation functions
    $\langle \delta X^a \delta X^b \rangle$,
    $\langle \delta X^a \delta X^b \delta X^c \rangle$.
    Because of symmetries there are $2N$ independent elements of $N_a$,
    and $2N(N+\frac{1}{2})$ independent elements of $N_{ab}$.
    Also, for fixed momenta $\vect{k}_i$ there
    are $2N(N+\frac{1}{2})$ independent elements of
    $\langle \delta X^a \delta X^b \rangle$
    for each $k_i$,
    and
    $8N^3$ independent elements of
    $\langle \delta X^a \delta X^b \delta X^c \rangle$ overall.
    In Table~\ref{table:combinatorics}
    we illustrate how the work required to compute
    $\langle \zeta \zeta \rangle$
    and $\langle \zeta \zeta \zeta \rangle$---%
    for just a \emph{single} wavenumber configuration---%
    scales with $N$.

    In the most favourable cases
    it may be possible to use analytic approximations for
    $\langle \delta X^a \delta X^b \rangle$
    or
    $\langle \delta X^a \delta X^b \delta X^c \rangle$,
    but
    there are already too many terms for hand calculation to be practical
    even at modest $N$. If a subset of these correlation
    functions require special treatment then the situation is much worse.
    The same rapidly growing number of contributions was
    an early catalyst for the development of automated methods in collider
    phenomenology.

    \item \semibold{Estimates of Hubble-scale correlations.}
    Eqs.~\eqref{eq:zeta-twopf-factorization} and~\eqref{eq:zeta-threepf-factorization}
    are useful only if the coefficients $N_a$, $N_{ab}$
    and correlation functions
    $\langle \delta X^a \delta X^b \rangle_{t_\ast}$
    and
    $\langle \delta X^a \delta X^b \delta X^c \rangle_{t_\ast}$ can be estimated
    separately.
    A number of formalisms exist to estimate $N_a$ and $N_{ab}$, but except in special
    cases the equations must be solved
    numerically~\cite{GarciaBellido:1995qq,Vernizzi:2006ve,
    Yokoyama:2007dw,Yokoyama:2007uu,Mulryne:2009kh,Mulryne:2010rp,Seery:2012vj}.
    Therefore we are committed to some numerical
    effort, no matter what other choices
    are made.

    The remaining terms are the $t_\ast$ correlators.
    We can estimate these analytically only in the massless limit.
    If all masses are light compared to the Hubble scale
    then this approximation
    is acceptable, and the estimates
    yield nearly `universal' model-independent expressions
    depending only on the background~\cite{Seery:2005gb}.
    Over the last decade, however, it has been understood that there is a rich
    phenomenology associated with cases where the mass spectrum extends up to the
    Hubble scale, or even above it.
    When such effects must be included the problem becomes much more complicated.
    Universality is almost always lost,
    and specialized calculations using the full machinery of quantum field theory are required.

    Significant efforts have been made to produce estimates
    of the $t_\ast$ correlators in such
    models~\cite{Chen:2009we,Chen:2009zp,Chen:2012ge,Achucarro:2010da,Achucarro:2012sm,Achucarro:2012yr,
    Achucarro:2013cva,Achucarro:2014msa,Gao:2012uq,Gao:2013ota},
    mostly keeping only a subset of the possible effects.
    Nevertheless,
    even with such simplifications the formalism becomes cumbersome
    and accounting for \emph{all} relevant effects is a challenge.

    We briefly review the possible contributing effects in~\S\ref{sec:heavy-influence}
    below.

    \item \semibold{Multiple factorization.}
    The simple three-point function factorization formula~\eqref{eq:zeta-threepf-factorization}
    strictly applies only if there are no large hierarchies among the external wavenumbers $k_i$.
    Where hierarchies exist it is necessary to construct more complex factorizations---for example,
    using the technique of operator product expansions employed in Refs.~\cite{Kenton:2015lxa,Kenton:2016abp,Byrnes:2015dub}.
    The necessity for multiple factorization introduces yet further
    algebraic complexity.

\end{itemize}
\begin{table}

   \begin{center}

        \small
    	\heavyrulewidth=.08em
    	\lightrulewidth=.05em
    	\cmidrulewidth=.03em
    	\belowrulesep=.65ex
    	\belowbottomsep=0pt
    	\aboverulesep=.4ex
    	\abovetopsep=0pt
    	\cmidrulesep=\doublerulesep
    	\cmidrulekern=.5em
    	\defaultaddspace=.5em
    	\renewcommand{\arraystretch}{1.5}

        \rowcolors{2}{gray!25}{white}

        \begin{tabular}{rrrrrrr}

            \toprule
                \semibold{fields}
                & \semibold{$N_a$}
                & \semibold{$N_{ab}$}
                & \semibold{total}
                & \semibold{$\langle \delta X^a \delta X^b \rangle$}
                & \semibold{$\langle \delta X^a \delta X^b \delta X^c \rangle$}
                & \semibold{total} \\
            1 & 2 & 3 & 5 & 9 & 8 & 17 \\
            2 & 4 & 10 & 14 & 30 & 64 & 94 \\
            3 & 6 & 21 & 37 & 63 & 216 & 279 \\
            4 & 8 & 36 & 44 & 108 & 512 & 620 \\
            10 & 20 & 210 & 230 & 630 & 8,000 & 8,630 \\
            50 & 100 & 5,050 & 5,150 & 15,150 & 1,000,000 & 1,015,150 \\
            100 & 200 & 20,100 & 20,300 & 60,300 & 8,000,000 & 8,060,300 \\
            \bottomrule

        \end{tabular}

    \end{center}

    \caption{\label{table:combinatorics}Growth in number of coefficients
    $N_a$, $N_{ab}$ and correlation functions
    $\langle \delta X^a \delta X^b \rangle$,
    $\langle \delta X^a \delta X^b \delta X^c \rangle$
    that must be computed in order to estimate
    $\langle \zeta \zeta \rangle$ and
    $\langle \zeta \zeta \zeta \rangle$
    using~\eqref{eq:zeta-twopf-factorization}--\eqref{eq:zeta-threepf-factorization}.}

\end{table}

\subsection{Influence of heavy fields}
\label{sec:heavy-influence}

In general, one or more of these difficulties will obstruct the use of analytic
methods.
This makes a numerical approach essential,
whether or not we choose to adopt automated
tools to mitigate algebraic complexity.

The most serious difficulties
are encountered
when the massless approximation does not
apply,
and we must
obtain
suitable initial conditions
by some other means.
This normally entails specialized
estimates for the early-time correlation
functions
$\langle \delta X^a \delta X^b \rangle_{t_\ast}$
and
$\langle \delta X^a \delta X^b \delta X^c \rangle_{t_\ast}$.
With this in mind we can divide
multiple-field models into a number of classes:
\begin{itemize}
	\item \semibold{Massless initial conditions apply.}
	This case occurs if all species are light relative to the Hubble
	scale during horizon exit of each relevant mode.
	There is no difficulty in estimating the
	initial conditions
	$\langle \delta X^a \delta X^b \rangle_{t_\ast}$
	and
	$\langle \delta X^a \delta X^b \delta X^c \rangle_{t_\ast}$,
	but calculation of the coefficients $N_a$, $N_{ab}$ will normally
	require numerical methods.%
		\footnote{Factorization methods may be needed
		if there are hierarchies between the external wavenumbers,
		but the time-dependent factors appearing
		in these formulae are related to $N_a$ and $N_{ab}$,
		and therefore as a matter of principle the calculation
		contains no new elements~\cite{Kenton:2015lxa,Kenton:2016abp,Byrnes:2015dub}.}

	For models in this category the advantage of automated methods is
	mostly one of convenience
	in (effectively) computing the mass spectrum
	and setting up the integrals for $N_a$ and $N_{ab}$.%
	   \footnote{In practice, the automated tools described in this paper
	   do not
	   compute correlation functions by factorizing them
	   as in
	   Eqs.~\eqref{eq:zeta-twopf-factorization}--\eqref{eq:zeta-threepf-factorization},
	   and therefore do not explicitly set up integrals for
	   $N_a$ and $N_{ab}$.
	   Instead, each $n$-point function is evolved in its
	   entirety. But
	   if interpreted from the perspective of
	   Eqs.~\eqref{eq:zeta-twopf-factorization}--\eqref{eq:zeta-threepf-factorization}
	   the outcome is the same.}

	\item \semibold{Adiabatic evolution of massive fields.}
	At the opposite extreme, massless initial conditions may fail because
	some modes are very heavy relative to the Hubble scale,
	with masses $M_i \gg H$.
	Based on experience with the decoupling theorem in
	Minkowski space we might expect the effect of such modes to be
	suppressed by inverse powers of $M_i$~\cite{Appelquist:1974tg}.

	During the last decade it has become understood that decoupling is not
	so trivial in a time-dependent background,
	because the time evolution may be associated with scales
	that compensate for the smallness of $1/M_i$.
	A common example is a turn in field space associated with
	an angular velocity $\dot{\theta}$.
	If the turn is sufficiently rapid
	$\dot{\theta}/M_i$ may be order unity or larger even if
	$M_i$ is super-Hubble.
	However,
	because scales such as $\dot{\theta}$
	are generated dynamically from
	the initial conditions they need not be visible merely
	from inspection of the Lagrangian.

	These $\Or(\dot{\theta}/M)$ effects mean that care is needed when
	attempting to integrate-out the heavy fields.
	The simplest case occurs if they are heavy enough to
	`adiabatically' track the minimum of their effective potential,
	and transverse excitations can be neglected.
	Even in this case,
	any kinetic mixing between the light and heavy fields
    will cause the effective potential to differ from the bare potential.
    This scenario was studied by Tolley \& Wyman, who
    demonstrated that the effective theory for the light
    (`gelaton') modes would be
    noncanonical~\cite{Tolley:2009fg}.
    The key feature of such models is that the speed of sound
    of the gelaton modes is renormalized, potentially giving
    a significant bispectrum amplitude on equilateral
    configurations~\cite{Chen:2006nt,Seery:2005wm}.
    The analysis was later refined by other
    authors~\cite{Achucarro:2010jv,Achucarro:2010da,Baumann:2011su,
    Achucarro:2012sm,Achucarro:2012yr,Gao:2012uq}.

	Automated methods offer significant advantages when
	applied to scenarios of this type.
	The first challenge
    is to construct either a suitable effective theory
    for the light modes,
    or to follow the evolution of the full system of light and heavy modes.
    Except in the very simplest
    scenarios it will not be possible to do either analytically,
    because the field trajectories cannot be expressed in closed form.
    Automated methods remove this difficulty by setting up the necessary
    computations and performing them numerically.
    Second, the use of an effective description involving only light
    modes entails an assumption that the evolution is adiabatic throughout,
    which need not be the case (see below).
    Automated calculations involving the full system of light and heavy
    modes automatically take nonadiabatic evolution into account.

    \item \semibold{Nonadiabatic evolution of massive fields.}
    Alternatively, if the time dependence of the background is
    sufficiently rapid then the heavy fields may evolve nonadiabatically,
    resulting in excitation and
    particle production~\cite{Gao:2013ota,Konieczka:2014zja}.
    Excitation of the light fields is always important, and excitation
    of the heavy fields must be taken into account if they couple sufficiently
    strongly that energy transfer can be efficient.

    Scenarios of this type were studied by Gao, Langlois \& Mizuno~\cite{Gao:2012uq,Gao:2013ota}.
    To lowest order in the trajectory bend angle, they estimated that
    the corrections to the power spectrum $\dimlessP$ of the light
    mode could be written
    \begin{equation}
    	\frac{\Delta \dimlessP}{\dimlessP_0}
    	\approx
    	\mathcal{F}_\ell + \mathcal{F}_h + \mathcal{F}_{\ell h}.
    \end{equation}
    In this language, $\mathcal{F}_\ell$ represents particle production
    in the light mode;
    $\mathcal{F}_h$ represents conversion of the heavy fluctuations
    into the light mode;
    and $\mathcal{F}_{\ell h}$ represents the response of the heavy mode
    to fluctuations in the light mode.
	The last of these is an analogue of the gelaton
	mechanism in which a perturbation in the light mode induces
	a heavy perturbation.
	The original
	gelaton behaviour described above
	appears in the limit where this heavy perturbation
	corresponds to an instantaneous adjustment into the minimum of its
	effective potential, preserving adiabatic evolution.

    In some cases it is possible to give analytic estimates
    for $\mathcal{F}_{\ell}$, $\mathcal{F}_h$ and $\mathcal{F}_{\ell h}$.
    Ref.~\cite{Gao:2012uq} estimated $\mathcal{F}_{\ell}$
    assuming particle production to be dominantly
    sourced from the field zero-modes.
    Ref.~\cite{Gao:2013ota} extended this to include contributions
    from coupling to the metric.
	Refs.~\cite{Gao:2012uq,Gao:2013ota} included conversion
	of the vacuum wavefunction in
	$\mathcal{F}_{h}$ but neglected stimulated particle production
	into the heavy mode.
	Later,
	an estimate for the bispectrum induced by production of
	heavy particles and their subsequent decay
	into light modes
	was given
	in Ref.~\cite{Flauger:2016idt},
	at least
	for models with an approximate shift symmetry.

	Where any of these effects are important,
	automated methods dramatically reduce the effort required to produce
	accurate calculations.
	The presence of particle production or strong coupling
	near horizon exit normally invalidates
	any analytic approach, meaning that
	\emph{ab initio} field theory calculations are needed
	to accurately estimate the initial correlation
	functions
	$\langle \delta X^a \delta X^b \rangle_{t_\ast}$
	and
	$\langle \delta X^a \delta X^b \delta X^c \rangle_{t_\ast}$.
	Particle production into both light and heavy modes is
	accounted for automatically,
	and both memory and retardation effects are included
	in $\mathcal{F}_{\ell h}$.%
	   \footnote{By `memory', we mean that $\mathcal{F}_{\ell h}$ contains an
	   integral over time and therefore has memory of conditions at
	   earlier times. Retardation refers to the appearance of the retarded
	   Green's function in $\mathcal{F}_{\ell h}$.}
	(However, as we explain in~\S\S\ref{sec:loop-twopf}--\ref{sec:loop-threepf},
	the subsequent decay of these particles into lighter species
	may not be captured at tree level.)
	Finally,
	if some fields remain active after horizon exit then suitable
	numerical integrals are needed for $N_a$ and $N_{ab}$
	which the automated software can automatically accommodate.

    \item \semibold{Quasi-single field inflation.}
    The `QSFI' scenario is a special case
    which can be regarded
    as a mix of the adiabatic and nonadiabatic cases~\cite{Chen:2009zp,Chen:2009we}.

    We still require the evolution to be nearly adiabatic in the sense that
    the heavy field
    tracks the minimum of its effective potential,
    but we do not allow
    its mass to be much larger than $H$.
    This requires that the angular velocity not
    be too large, perhaps at most $\dot{\theta} \lesssim M \approx H$.
    Under these circumstances
    there will be negligible particle production from coupling to the
    background field configuration.
    At the same time, a Hubble-scale mass is small enough that particle production
    from coupling to the de Sitter metric is not completely suppressed.
    If we choose the cubic self-coupling to be strong,
    say $V''' \gtrsim H$,
    then the heavy field can acquire a sizeable bispectrum.
    This is communicated to the light sector by
    $\Or(\dot{\theta}/M)$ quadratic mixing.
\end{itemize}

\section{Numerical computation of inflationary correlation functions
at tree level}
\label{sec:numerical-computation}

The discussion of~\S\ref{sec:heavy-influence}
makes clear that analytic approximations apply only in restricted
circumstances, and must generally be complemented by numerical methods.
Unfortunately,
as explained in~\S\ref{sec:introduction}, numerical methods
applicable to multiple-field models
have so far been implemented only for the two-point function.
In this paper we go further and explain how to perform numerical
calculations for the three-point function.
Whatever technique we employ, it should allow for arbitrary masses
and couplings, capture all scale-dependent effects, and accurately
track the evolution of all fluctuations both inside and outside the horizon.

\subsection{Evolution equations}
\label{sec:evolution-equations}
Our aim is to evaluate correlation functions of the form
$\langle \Operator \rangle$, where
$\Operator$ is a Heisenberg-picture operator
(possibly composite) and
$\langle \cdots \rangle$ denotes an expectation value
defined by an `in' state,
normally chosen to match
the Minkowski vacuum
for subhorizon oscillators in the range of interest.

In this picture the Hamiltonian is a function $H(Q^\alpha, P_\beta)$
of the Heisenberg-picture field-space coordinates
$Q^\alpha$ and their
canonical momenta $P_\beta$.
When quantized these satisfy the canonical commutation algebra
$[Q^\alpha(\vect{k}_1,t) ,
P_\beta(\vect{k}_2,t') ] = (2\pi)^3
\im \delta^\alpha_\beta \delta(\vect{k}_1 + \vect{k}_2) \delta(t-t')$.
(In this section we are temporarily allowing the theory to be very general
before specializing to the case of interest, where $Q^\alpha = \delta\phi^\alpha$.)

To define an interaction picture we divide $H$ into
`free' and `interacting' parts,
\begin{equation}
	H(Q^\alpha, P_\beta) =
	H_0(Q^\alpha, P_\beta)
	+
	\Hint(Q^\alpha, P_\beta) ,
\end{equation}
which at this stage remain functions of the Heisenberg-picture fields,
and define new interaction-picture operators $q^\alpha$ and $p_\beta$,
\begin{subequations}
\begin{align}
	q^\alpha & = F^\dag Q^\alpha F \\
	p_\beta & = F^\dag P_\beta F ,
\end{align}
\end{subequations}
where $F$ is a unitary operator to be chosen.
At least if $\Operator$ is polynomial in the $Q^\alpha$ and $P_\beta$, the correlation
function $\langle \Operator \rangle$ can be written
\begin{equation}
	\langle \Operator \rangle
	=
	\langle F F^\dag \Operator(Q^\alpha, P_\beta) F F^\dag \rangle
	=
	\langle F \Operator(q^\alpha, p_\beta) F^\dag \rangle .
	\label{eq:intpic-expectation}
\end{equation}
As in~\S\ref{sec:standard-calculation}, we collect the operators $Q^\alpha$, $P_\beta$
and $q^\alpha$, $p_\beta$
into phase-space vectors $X^a = (Q^\alpha, P^\beta)$
and
$x^a = (q^\alpha, p^\beta)$.
If we choose $F$ to satisfy
\begin{equation}
	\frac{\d F}{\d t} = \im \Hint(X) F
	= \im F F^\dag \Hint(X) F
	= \im F \Hint(x)
	\label{eq:f-def}
\end{equation}
then it can be verified that the equation of motion for $x^a$
is the Heisenberg equation for $H_0$,
\begin{equation}
	\frac{\d x_a}{\d t} = - \im [ x^a, H_0(x) ] .
\end{equation}
It follows that $q^\alpha$ and $p_\beta$ can be written in terms of
the wavefunctions for $H_0$,
and the creation--annihilation operators $a_{\vect{k}}$, $a^\dag_{\vect{k}}$
associated with the $H_0$ vacuum state. We denote this state by $| 0 \rangle$.
According to a theorem of Gell-Mann \& Low, the $H_0$ vacuum is related
to the `in' state appearing in the inner product
$\langle \Operator \rangle$
by adiabatic switch-on of the interaction
$\Hint$ in the distant past.

Eq.~\eqref{eq:f-def}
and this adiabatic prescription
show that
the solution for $F$ is an anti-time ordered exponential
when written in terms of the
interaction-picture fields
\begin{equation}
	F = \AntiTimeOrder \exp
	\bigg(
		\im \int^t_{-\infty^+} \Hint[x(t')] \, \d t'
	\bigg) .
	\label{eq:int-pic-f}
\end{equation}
The symbol $\AntiTimeOrder$ is the anti-time ordering operator,
which rewrites its argument in order of \emph{increasing} time.
Its Hermitian conjugate $\TimeOrder = \AntiTimeOrder^{\dag}$
performs the reverse operation, rewriting its argument in
decreasing time order.
The lower limit $-\infty^+$ denotes that the contour of integration
should be deformed above the real axis into the positive imaginary
half-plane at early times,
with the fields appearing in the integrand
defined by analytic continuation.
This accounts for the
adiabatic switching-on required to extract the `in'-state from the $H_0$
vacuum.
Combining Eqs.~\eqref{eq:intpic-expectation} and~\eqref{eq:int-pic-f}
yields
\begin{equation}
	\langle \Operator(X) \rangle
	=
	\bigg\langle
	0
	\bigg|
		\AntiTimeOrder
		\exp
		\Big(
			\im \int^t_{-\infty^+} \Hint[x(t')] \, \d t'
		\Big)
		\Operator(x)
		\TimeOrder
		\exp
		\Big(
			{-\im} \int^t_{-\infty^-} \Hint[x(t'')] \, \d t''
		\Big)
	\bigg|
	0
	\bigg\rangle .
	\label{eq:inin-formula}
\end{equation}
Notice that the expectation value is taken in a different state on the left-
and right-hand sides.
Eq.~\eqref{eq:inin-formula} or equivalent expressions are sometimes
described as the `in--in' or Schwinger formalism for computing expectation
values~\cite{Maldacena:2002vr,Weinberg:2005vy,Weinberg:2006ac}.

\para{Tree-level correlations}
The task of an automated tool is to compute~\eqref{eq:inin-formula}
for suitable choices of operator $\Operator$.
If we wish to compute up to the field-space three-point function
these operators will be
(returning to our original notation)
$\delta X^a \delta X^b$ and $\delta X^a \delta X^b \delta X^c$,
from which we can extract
correlations functions for $\zeta \zeta$ and $\zeta \zeta \zeta$
by making a gauge transformation.

As for any interaction picture
expression, the right-hand side of
Eq.~\eqref{eq:inin-formula} admits a
complex expansion into diagrams
that has been discussed at length in the literature~\cite{Weinberg:2005vy,
Weinberg:2006ac,Seery:2007we,Adshead:2009cb,Arkani-Hamed:2015bza}.
In practical calculations it is almost always necessary
to truncate the expansion to a finite number of terms.
Many such truncations are possible,
but
experience with particle physics has taught us that
it is often a good approximation to restrict
the expansion to a predetermined maximum number of loops.
In calculations of $S$-matrix elements these loops average over
quantum fluctuations that mediate the transition.
For the expectation value in Eq.~\eqref{eq:inin-formula} there are
analogous loops averaging over the effect of high energy fluctuations, but also
new loops that average over intermediate particles
as we describe in~\S\ref{sec:tree-level}.

Diagrams with no loops are described as `tree level' and contain no averaging.
If nonzero, such diagrams often (but not always) constitute the leading contribution
to each transition amplitude or correlation function.
The precise numerical method we are going to describe computes the tree-level
approximation to each correlation function, although
as a matter of principle loop-level contributions could
be retained at the expense of a significant increase in complexity.

\para{Extended summation convention}
To go further we must commit to a particular form for the Hamiltonian.
When writing this and subsequent expressions it is helpful to condense our
notation and include integrals over Fourier modes in the summation convention.
We indicate that this extended interpretation is in use by typesetting
the labels to which it applies in a sans serif face.
An index contraction such as $A_{\ext{a}}B^{\ext{a}}$ therefore implies
\begin{equation}
  \label{eq:extended-summation-example}
  A_{\ext{a}} B^{\ext{a}} =
  \sum_a \int \frac{\d^3 k_a}{(2\pi)^3} \, A_a(\vect{k}_a) B^a(\vect{k}_a) .
\end{equation}
The subscript $a$ on $\vect{k}_a$ indicates that this is the Fourier mode
associated with the $a$ index contraction,
and
because the label $a$ is set in italics
the sum $\sum_a$ on the right-hand side involves only phase-space coordinate labels.
Note that we position indices to respect the normal rules for
covariant expressions, but in this paper we are always using a flat field-space
metric and therefore co- or contravariant index placement has no significance.

With $\vect{k}$-labels included there is an extra complexity
because Fourier-space expressions sometimes produce
the $\delta$-function
$\delta_{\ext{a}\ext{b}} = (2\pi)^3 \delta_{ab} \delta(\vect{k}_a + \vect{k}_b)$.
When integrated out
this reverses the sign of a $\vect{k}$-label, which we indicate
by decorating the label with a bar, as in $B^{\flip{\ext{a}}}$. Hence,
\begin{equation}
  A_{\ext{a}} B^{\flip{\ext{a}}} =
  \sum_a \int \frac{\d^3 k_a}{(2\pi)^3} A_a(\vect{k}_a) B^a(-\vect{k}_a) .
\end{equation}
Note that contraction with ${\delta^{\ext{a}}}_\ext{b}$
will bar an index; for example,
$B^{\flip{\ext{a}}} = \delta^{\ext{a}\ext{b}} B_{\ext{b}}$.

In this notation
the canonical commutation algebra has a compact expression,
\begin{equation}
  [ \delta X^{\ext{a}}, \delta X^{\ext{b}} ] = \im \epsilon^{\ext{a}\ext{b}} ,
\end{equation}
where $\epsilon^{\ext{a}\ext{b}}$ is defined by
\begin{equation}
  \epsilon^{\ext{a}\ext{b}} = (2\pi)^3 \delta(\vect{k}_a + \vect{k}_b) \epsilon^{ab}
\end{equation}
and the matrix on the right-hand side can be written in block form
\begin{equation}
  \epsilon^{ab} =
  \left(
    \begin{array}{cc}
      \vect{0} & \vect{1} \\
      -\vect{1} & \vect{0}
    \end{array}
  \right) .
  \label{eq:epsilon-coeff-matrix}
\end{equation}
In~\eqref{eq:epsilon-coeff-matrix} we are assuming that the
numerical values of the
indices $a$, $b$, \ldots, are organized so that a block of field
labels are followed by a block of momentum labels, in the same order.
The symbol $\epsilon^{\ext{a}\ext{b}}$ obeys the identity
\begin{equation}
  \epsilon^{\ext{a}\ext{b}}{\epsilon_{\ext{b}}}^{\ext{c}}
  = - \delta^{\ext{a}\flip{\ext{c}}} .
\end{equation}
The symbol $\delta^{\ext{a}\flip{\ext{c}}}$
appearing on the right-hand side
behaves as a Kronecker-$\delta$ that does not bar indices;
eg. $B^{\ext{a}} = \delta^{\ext{a}\flip{\ext{b}}} B_{\ext{b}}$.

\para{Equations of motion for Heisenberg- and interaction-picture fields}
With these rules the Hamiltonian can be written
\begin{equation}
  H =
  \frac{1}{2!}
  H_{\ext{a}\ext{b}} \delta X^{\ext{a}} \delta X^{\ext{b}}
  + \frac{1}{3!}
  H_{\ext{a}\ext{b}\ext{c}} \delta X^{\ext{a}} \delta X^{\ext{b}} \delta X^{\ext{c}}
  + \cdots
  \label{eq:hamiltonian}
\end{equation}
where `$\cdots$' denotes terms of higher order in $\delta X$ that have been omitted.
Without loss of generality we can assume the tensors
$H_{\ext{a}\ext{b}}$ and
$H_{\ext{a}\ext{b}\ext{c}}$ to be symmetric under exchange of any index pairs.
In addition, each of these tensors
contains a single momentum-conservation $\delta$-function,
and therefore the quadratic term involves a single momentum integral,
the cubic term involves two momentum integrals, and so on.
The equation of motion for the Heisenberg picture fields $\delta X^{\ext{a}}$
becomes
\begin{equation}
\begin{split}
  \frac{\d \delta X^{\ext{a}}}{\d t}
  & =
  \epsilon^{\ext{a}\ext{c}} H_{\ext{c}\ext{b}}
  \delta X^{\ext{b}}
  +
  \frac{1}{2!} \epsilon^{\ext{a}\ext{d}} H_{\ext{d}\ext{b}\ext{c}}
  \delta X^{\ext{b}}
  \delta X^{\ext{c}}
  + \cdots
  \\
  & =
  {u^\ext{a}}_{\ext{b}} \delta X^{\ext{b}}
  + \frac{1}{2!} {u^\ext{a}}_{\ext{b}\ext{c}} \delta X^{\ext{b}} \delta X^{\ext{c}}
  + \cdots ,
\end{split}
\label{eq:heisenberg-equations}
\end{equation}
which should be regarded as a definition of the tensors
${u^{\ext{a}}}_{\ext{b}}$
and
${u^{\ext{a}}}_{\ext{b}\ext{c}}$.

Our expression~\eqref{eq:inin-formula} for the correlation functions
makes use of the interaction picture, and therefore depends on a
division of $H$ into free and interacting parts.
To make a clean statement about all tree-level graphs we should take $H_0$
to comprise the quadratic terms
$H_{\ext{a}\ext{b}} \delta X^{\ext{a}} \delta X^{\ext{b}}$,
while $\Hint$ contains terms of cubic order or above.%
  \footnote{In analytic calculations we often choose a different division, with
  $H_0$ describing a system of massless modes and masses accommodated
  perturbatively in $\Hint$.
  Although this division is convenient, it is useful only for fields that are
  light compared to the Hubble scale and is not necessary as a point of principle.
  Had we chosen to pursue this approach
  we would encounter extra complexities from the need to resum an infinite
  series of tree-level diagrams containing an arbitrary number of insertions
  of the quadratic part of $\Hint$.
  To avoid this it is preferable to choose $\Hint$ to begin at cubic order.}
The equation of motion for the interaction-picture fields
$\delta x^a$ can be written
\begin{equation}
  \frac{\d \delta x^{\ext{a}}}{\d t} = {u^{\ext{a}}}_{\ext{b}} \delta x^{\ext{b}} .
  \label{eq:int-pic-eom}
\end{equation}

\para{Tree-level expressions for the two- and three-point functions}
We are now in a position to give formulae for the two- and three-point functions
of $\delta X^a$ at tree-level.
Eqs.~\eqref{eq:inin-formula} and~\eqref{eq:hamiltonian} yield
\begin{subequations}
\begin{align}
  \label{eq:twopf-tree}
  \langle \delta X^{\ext{a}} \delta X^{\ext{b}} \rangle_{\text{tree}}
  & =
  \langle 0 | \delta x^{\ext{a}} \delta x^{\ext{b}} | 0 \rangle
  \\
  \label{eq:threepf-tree}
  \langle \delta X^{\ext{a}} \delta X^{\ext{b}} \delta X^{\ext{c}} \rangle_{\text{tree}}
  & =
  \bigg\langle
  0
  \bigg|
  \Big[
    \frac{\im}{3!} \int^t
    H_{\ext{d}\ext{e}\ext{f}} \delta x^{\ext{d}} \delta x^{\ext{e}} \delta x^{\ext{f}}
    \; \d t' ,
    \delta x^{\ext{a}} \delta x^{\ext{b}} \delta x^{\ext{c}}
  \Big]
  \bigg|
  0
  \bigg\rangle .
\end{align}
\end{subequations}
In what follows we drop the subscript `tree', it being understood that
the $\delta X^a$ correlation functions are being computed only to this order.
Since the state is time independent,
differentiating Eq.~\eqref{eq:twopf-tree}
and using~\eqref{eq:int-pic-eom}
yields
\begin{equation}
\begin{split}
  \frac{\d}{\d t} \langle \delta X^{\ext{a}} \delta X^{\ext{b}} \rangle
  & =
  \langle 0 | {u^{\ext{a}}}_{\ext{c}} \delta x^{\ext{c}} \delta x^{\ext{b}} | 0 \rangle
  +
  \langle 0 | \delta x^{\ext{a}} {u^{\ext{b}}}_{\ext{c}} \delta x^{\ext{c}} | 0 \rangle
  \\
  & =
  {u^{\ext{a}}}_{\ext{c}} \langle \delta X^{\ext{c}} \delta X^{\ext{b}} \rangle
  +
  {u^{\ext{b}}}_{\ext{c}} \langle \delta X^{\ext{a}} \delta X^{\ext{c}} \rangle .
\end{split}
\label{eq:twopf-transport}
\end{equation}
Likewise, Eq.~\eqref{eq:threepf-tree} yields
\begin{equation}
\begin{split}
  \frac{\d}{\d t} \langle \delta X^{\ext{a}} \delta X^{\ext{b}} \delta X^{\ext{c}} \rangle
  = \mbox{}
  &
  \bigg\langle
  0
  \bigg|
  \Big[
    \frac{\im}{3!} \int^t
    H_{\ext{d}\ext{e}\ext{f}} \delta x^{\ext{d}} \delta x^{\ext{e}} \delta x^{\ext{f}}
    \; \d t' ,
    {u^{\ext{a}}}_{\ext{g}} \delta x^{\ext{g}} \delta x^{\ext{b}} \delta x^{\ext{c}}
    +
    \text{2 perms}
  \Big]
  \bigg|
  0
  \bigg\rangle
  \\
  & \mbox{} +
  \bigg\langle
  0
  \bigg|
  \Big[
    \frac{\im}{3!}
    H_{\ext{d}\ext{e}\ext{f}} \delta x^{\ext{d}} \delta x^{\ext{e}} \delta x^{\ext{f}}
    ,
    \delta x^{\ext{a}} \delta x^{\ext{b}} \delta x^{\ext{c}}
  \Big]
  \bigg|
  0
  \bigg\rangle
\end{split}
\end{equation}
The terms in the first line produce
${u^{\ext{a}}}_{\ext{b}}$ contracted
with the three-point function in different index combinations.
The second line produces
\begin{equation}
\begin{split}
  \Big[
    H_{\ext{d}\ext{e}\ext{f}} \delta x^{\ext{d}} \delta x^{\ext{e}} \delta x^{\ext{f}}
    , \mbox{}
    &
    \delta x^{\ext{a}} \delta x^{\ext{b}} \delta x^{\ext{c}}
  \Big]
  =
  \Big[
    H_{\ext{d}\ext{e}\ext{f}} \delta x^{\ext{d}} \delta x^{\ext{e}} \delta x^{\ext{f}} ,
    \delta x^{\ext{a}}
  \Big]
  \delta x^{\ext{b}}
  \delta x^{\ext{c}}
  \\
  & \mbox{}
  +
  \delta x^{\ext{a}}
  \Big[
    H_{\ext{d}\ext{e}\ext{f}} \delta x^{\ext{d}} \delta x^{\ext{e}} \delta x^{\ext{f}} ,
    \delta x^{\ext{b}}
  \Big]
  \delta x^{\ext{c}}
  +
  \delta x^{\ext{a}}
  \delta x^{\ext{b}}
  \Big[
    H_{\ext{d}\ext{e}\ext{f}} \delta x^{\ext{d}} \delta x^{\ext{e}} \delta x^{\ext{f}} ,
    \delta x^{\ext{c}}
  \Big] ,
\end{split}
\end{equation}
which can be rewritten
\begin{equation}
\begin{split}
  \Big[
    \frac{1}{3!}
    H_{\ext{d}\ext{e}\ext{f}} \delta x^{\ext{d}} \delta x^{\ext{e}} \delta x^{\ext{f}}
    , \mbox{}
    &
    \delta x^{\ext{a}} \delta x^{\ext{b}} \delta x^{\ext{c}}
  \Big]
  =
  \frac{1}{2!}
  {u^{\ext{a}}}_{\ext{d}\ext{e}}
    \delta x^{\ext{d}} \delta x^{\ext{e}} \delta x^{\ext{b}} \delta x^{\ext{c}}
  \\ & \mbox{}
  +
  \frac{1}{2!}
  {u^{\ext{b}}}_{\ext{d}\ext{e}}
    \delta x^{\ext{a}} \delta x^{\ext{d}} \delta x^{\ext{e}} \delta x^{\ext{c}}
  +
  \frac{1}{2!}
  {u^{\ext{c}}}_{\ext{d}\ext{e}}
    \delta x^{\ext{a}} \delta x^{\ext{b}} \delta x^{\ext{d}} \delta x^{\ext{e}} .
\end{split}
\end{equation}
Retaining only connected contributions gives the evolution equation
\begin{equation}
  \frac{\d}{\d t}
  \langle
    \delta X^{\ext{a}} \delta X^{\ext{b}} \delta X^{\ext{c}}
  \rangle
  =
  {u^{\ext{a}}}_{\ext{d}}
  \langle
    \delta X^{\ext{d}} \delta X^{\ext{b}} \delta X^{\ext{c}}
  \rangle
  +
  {u^{\ext{a}}}_{\ext{d}\ext{e}}
  \langle
    \delta X^{\ext{d}} \delta X^{\ext{b}}
  \rangle
  \langle
    \delta X^{\ext{e}} \delta X^{\ext{c}}
  \rangle
  +
  \text{2 perms} ,
  \label{eq:threepf-transport}
\end{equation}
where the permutations should respect the relative ordering
of $\delta X^{\ext{a}}$, $\delta X^{\ext{b}}$ and $\delta X^{\ext{c}}$.
Although we have derived these evolution equations from the interaction-picture
expresssion~\eqref{eq:inin-formula}, our final expressions are
picture-independent and do not depend on the division of
the Hamiltonian into $H_0$ and $\Hint$.
The argument given here is effectively equivalent to
that given for classical superhorizon evolution in
Ref.~\cite{Seery:2012vj}
and for quantum-mechanical evolution on all scales
in Ref.~\cite{Mulryne:2013uka},
with the new element of explicitly demonstrating equivalence to the
interaction-picture expression for the
tree-level two- and three-point function via Eq.~\eqref{eq:inin-formula}.

\para{No requirement for early-time regulator}
Eqs.~\eqref{eq:twopf-transport} and~\eqref{eq:threepf-transport}
provide an alternative means to compute the two- and three-point
correlation functions.
Rather than deal with the interaction-picture
formula~\eqref{eq:inin-formula} directly,
and its concomitant obligation
to handle the contour deformation
appearing in the time-ordered exponentials,
we can instead integrate these ordinary differential
equations.
No contour deformations appear in these
equations because each one
is derived from a region where
$\int_{-\infty^\pm}^t \Hint[x(t')] \, \d t'$
involves only the real $t'$-axis.
Therefore our method does not require an
explicit regulator
to suppress contributions to the
integral on subhorizon scales.
Such regulators
have been widely
employed in previous attempts to compute
the bispectrum numerically~\cite{Chen:2006xjb,Chen:2008wn,Hazra:2012yn,
Horner:2013sea,Sreenath:2014nca,Horner:2015gza},
but their use necessarily introduces unwanted regulator-dependence
into the numerical results.
Therefore care must be taken to ensure the results
are regulator-independent, as far as possible;
for example, see the discussion in Chen et al.~\cite{Chen:2008wn}.
By comparison our method is unaffected
by such issues.

In practice, a numerical scheme based on these evolution equations requires
two further elements:
first, a set of suitable initial conditions for
$\langle \delta X^{\ext{a}} \delta X^{\ext{b}} \rangle$
and
$\langle \delta X^{\ext{a}} \delta X^{\ext{b}} \delta X^{\ext{c}} \rangle$;
and second,
algebraic expressions for
${u^\ext{a}}_{\ext{b}}$
and
${u^\ext{a}}_{\ext{b}\ext{c}}$.
We return to these issues in
\S\S\ref{sec:u-tensors} and~\ref{sec:initial-conditions},
following a short digression
to understand which physical processes are included in the tree-level
calculation.

\subsection{What is included at tree level?}
\label{sec:tree-level}
Our restriction to tree-level correlation functions
implies that some physical effects will not be modelled
perfectly.
Therefore,
although the tree approximation is sufficient in many scenarios,
it is reasonable to ask where it might fail.
In this section,
without giving a detailed treatment of loop-level processes,
we briefly indicate some common causes of failure.
Care should be taken when using our framework
(or any other tree-level scheme)
to analyse the fluctuations produced
in inflationary scenarios
that exhibit similar phenomenology.

\subsubsection{Background evolution}
When using Feynman diagrams to compute transition amplitudes
we can regard loops as averages over off-shell, virtual particles.
The situation for expectation values is subtly different.
Some of the loops that correct tree-level expectation values
retain an interpretation as averages over virtual
fluctuations, but others
should be interpreted as averages over on-shell particles
that are later converted to the fluctuations of interest.
We will see more details of this interpretation
in~\S\S\ref{sec:loop-twopf}--\ref{sec:loop-threepf} below.

Loop-level terms become important whenever
the averaged contribution of either effect
competes with the tree-level.
We will soon
see that
one way for this to happen is for
individual scattering or decay processes
to make important contributions to a correlation function.
But there is another possibility, which is that the averaged effect of
particle production modifies the inflationary trajectory
by draining energy from the zero-mode.
This is an example of a loop-level effect that is not captured by our
tree-level formalism.
Scenarios where this effect is important include
`warm' inflation~\cite{Berera:1995ie}
and
trapped inflation~\cite{Green:2009ds}.
These are both characterized by an extra friction term in the background
equation of motion.

This backreaction on the zero-mode is not trivial to compute
and we will not give an explicit formula for its contribution.
It is \emph{not} included in the in--in formula~\eqref{eq:inin-formula}
for
the correlation function $\langle \Operator \rangle$,
because before deriving this formula we committed to
a particular choice of background.
If backreaction is important we should adjust our estimate of the background
evolution,
and use this adjusted trajectory in~\eqref{eq:inin-formula}.
As a matter of principle the
proper adjustment
can be obtained by
computing
the Feynman--Vernon influence functional
for the back-reaction of the $k > 0$ fluctuations on the $k=0$
mode~\cite{Fischetti:1979ue,Hartle:1979uf,Hartle:1980nn,
Jordan:1986ug,Jordan:1987we,Calzetta:1986ey,Feynman:1963fq,
Caldeira:1982iu,Caldeira:1982uj}.
This is a loop-level calculation.

In general, the effective equation of motion obtained from the
influence functional will contain both dissipation and noise.
The dissipation tracks energy lost from the zero-mode into a cloud of $k>0$
fluctuations,
whereas the noise represents redeposition of energy from the cloud
due to mutual annihilation of its excitations into long-wavelength modes.
Neither of these effects are accounted for in our tree-level codes,
although it would be straightforward to modify the background equations of
motion to include an \emph{ad hoc} phenomenological description.
With our current technology this would have to be done on a case-by-case basis.

\subsubsection{Two-point function}
\label{sec:loop-twopf}
Now consider loop contributions to the two-point function~\eqref{eq:inin-formula}.
These are generated by higher-order terms from the expansion of
the time-evolution operator $F$, and are present whether or not we choose
to account for back-reaction into the zero-mode.

\para{Mixing and mass effects}
The discussion in~\S\ref{sec:evolution-equations}
used an exact treatment of quadratic terms in the Hamiltonian.
Therefore
the tree-level transport formalism
already contains an exact analysis
of effects from quadratic mixing
or time-dependent masses, including non-adiabatic
particle production.
The conclusion is that our tree-level codes will correctly capture
all physical effects arising from quadratic operators in
the Lagrangian.
For instance, this includes production
of heavy particles
due to a time-varying mass
(including Bose enhancement),
as studied recently
by Flauger et al.~\cite{Flauger:2016idt}.
It also includes the conversion of `isocurvaton' fluctuations
into the adiabatic mode in QSFI-like scenarios~\cite{Chen:2009zp,Chen:2009we}.

\begin{figure}
	\begin{center}
		\begin{fmfgraph*}(80,20)
			\fmfleft{l}
			\fmfright{r}
			\fmf{plain}{l,v}
			\fmf{dashes}{v,r}
			\fmfv{decoration.shape=cross}{v}
			\fmfv{label=$\alpha$}{l}
			\fmfv{label=$\beta$}{r}
		\end{fmfgraph*}
	\end{center}
	\caption{\label{fig:quadratic-vertex}Quadratic
	mixing between species $\alpha$ and $\beta$, distinguished
	by the solid or dashed lines.
	The mixing should be regarded as a perturbative mass
	(possibly time dependent)
	in the case $\alpha = \beta$.
	The cross denotes insertion of a quadratic vertex.}
\end{figure}
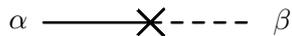

Before moving on to the case of cubic- or higher-order
operators,
we note
that
in an analytic approximation we might have elected
to treat time-varying quadratic terms perturbatively.
This would yield a quadratic
vertex as in Fig.~\ref{fig:quadratic-vertex}.
A quantitatively accurate
treatment of these effects then requires that we resum
all diagrams containing an arbitrary number of insertions of
this vertex in any internal or external line.
Although we are \emph{not} following this strategy,
the terms in~\eqref{eq:twopf-transport}
and~\eqref{eq:threepf-transport} containing the two-index tensor ${u^a}_b$
can be regarded as performing such a
resummation, by incrementally dressing
each
external leg of the correlation function with a copy of the mixing vertex
at each time step.

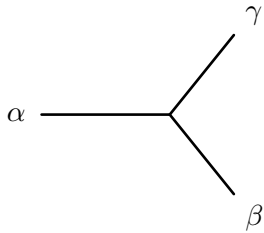
\begin{figure}
    \vspace{5mm}
    \begin{center}
        \begin{fmfgraph*}(80,60)
            \fmfleft{l}
            \fmfright{r1,r2}
            \fmf{plain}{l,v}
            \fmf{plain}{v,r1}
            \fmf{plain}{v,r2}
            \fmfv{label=$\alpha$}{l}
            \fmfv{label=$\beta$}{r1}
            \fmfv{label=$\gamma$}{r2}
        \end{fmfgraph*}
    \end{center}
    \caption{\label{fig:cubic-decay}Tree-level
    S-matrix process in which
    a species $\alpha$ decays into a $\beta$ and $\gamma$.}
\end{figure}

\para{Decays}
Now consider a three-body interaction described by the
S-matrix process in Fig.~\ref{fig:cubic-decay}.
This could be regarded as the decay $\alpha \rightarrow \beta\gamma$
(or likewise for some other permutation of species),
or alternatively as the reverse process $\beta\gamma \rightarrow \alpha$
in which two particles coalesce to produce a third.

A process such as Fig.~\ref{fig:cubic-decay} contributes
to the two-point function
$\langle \delta \phi^\alpha \delta \phi^\beta \rangle$
through a loop correction.
This may be surprising, since the corresponding process
occurs at tree-level in the S-matrix.
To see that the conclusion is correct, insert a complete set of free-particle states
$|m\rangle$
to obtain
\begin{equation}
    \langle
        \delta\phi^\alpha(\vect{k}_1, t)
        \delta\phi^\beta(\vect{k}_2, t)
    \rangle
    =
    \sum_m
    \langle
    0
    |
        F(t)
        \delta\phi^\alpha(\vect{k}_1, t)
    |
    m
    \rangle
    \langle
    m
    |
        \delta\phi^\beta(\vect{k}_2, t)
        F^\dag(t)
    |
    0
    \rangle
    .
    \label{eq:sum-S-matrix}
\end{equation}
The sum $\sum_m$ should be interpreted symbolically;
for a continuous set of states
it should be replaced by an integral with suitable measure.
If we were to compute each factor
appearing on the right-hand side of~\eqref{eq:sum-S-matrix}
by a path integral, we would arrive at the
`closed time path'
representation of the in--in
formalism
and its characteristic doubled
`$+$' and `$-$'
fields~\cite{Hajicek:1979ev,Jordan:1986ug,Jordan:1987we,Calzetta:1986ey,Weinberg:2005vy}.
The assignment of external fields
between the different factors
determines which correlation function
is computed---eg. Wightman functions or time-ordered
correlation functions. For equal-time correlation functions we need
not preserve the distinction between these possibilities.

We can regard~\eqref{eq:sum-S-matrix}
as an average
over the probability
(given by the modulus-square of the corresponding S-matrix
element) for the set of transitions
$|0\rangle \rightarrow |m\rangle$
in which the particles in $|m\rangle$
subsequently coalesce into a $\delta\phi$ fluctuation.
Each of these transitions may be individually corrected
by loops, which have the same interpretation
as loop corrections to an S-matrix element.
When expressed as in--in
diagrams they correspond to the case where all vertices have the
same `$+$' or `$-$' character.
Loops of this kind are genuine quantum corrections
and are not included in our formalism.

\begin{figure}
    \begin{center}
        \parbox{5cm}{
        \begin{fmfgraph*}(120,60)
            \fmfleft{l}
            \fmfright{r}
            \fmftop{t}
            \fmfbottom{b}
            \fmf{plain}{l,v1}
            \fmf{plain}{v1,m1}
            \fmf{plain}{v1,m2}
            \fmf{plain}{r,v2}
            \fmf{plain}{v2,n1}
            \fmf{plain}{v2,n2}
            \fmf{dashes}{t,b}
            \fmfforce{(0,0.5h)}{l}
            \fmfforce{(w,0.5h)}{r}
            \fmfforce{(0.2w,0.5h)}{v1}
            \fmfforce{(0.8w,0.5h)}{v2}
            \fmfforce{(0.4w,0.1h)}{m1}
            \fmfforce{(0.4w,0.9h)}{m2}
            \fmfforce{(0.6w,0.1h)}{n1}
            \fmfforce{(0.6w,0.9h)}{n2}
            \fmfforce{(0.5w,0)}{b}
            \fmfforce{(0.5w,h)}{t}
            \fmfv{label=$\alpha$}{l}
            \fmfv{label=$\lambda$}{r}
            \fmfv{label={\small average over particles}}{t}
        \end{fmfgraph*}
        }
        $\Rightarrow$
        \qquad
        \parbox{5cm}{
        \begin{fmfgraph*}(100,80)
            \fmfleft{l}
            \fmfright{r}
            \fmf{plain,tension=4}{l,v1}
            \fmf{plain,left}{v1,v2}
            \fmf{plain,left}{v2,v1}
            \fmf{plain,tension=4}{v2,r}
            \fmfv{label=$\alpha$}{l}
            \fmfv{label=$\lambda$}{r}
            \fmfv{label=$-$,label.angle=0}{v1}
            \fmfv{label=$+$,label.angle=180}{v2}
        \end{fmfgraph*}
        }
    \end{center}
    \caption{\label{fig:state-average-loop}Loop averaging over final-state
    particles in Schwinger--Keldysh formalism.
    Left: particles in the state $|m\rangle$
    are paired on either side of the `cut', and averaged over by the
    sum $\sum_m$.
    Right: the same process as an in--in diagram. One loop
    vertex is of `$-$' type, and the other is of `$+$' type.}
\end{figure}
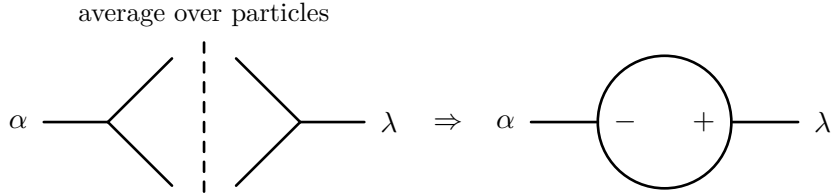

In addition to these S-matrix loops, each factor
in Eq.~\eqref{eq:sum-S-matrix}
contains contributions from
pairing a subset of operators
with particles in the state
$|m\rangle$. The quantum numbers of the
state are then averaged by the sum $\sum_m$.
Diagrammatically, these contributions
are represented by the process of
Fig.~\ref{fig:state-average-loop}.
In the left-hand diagram
we take the S-matrix element of Fig.~\ref{fig:cubic-decay}
and its complex conjugate,
with conjugation represented by crossing the dashed
cut in the centre of the diagram.
The decaying $|m\rangle$-state particles on each side of the cut are paired and averaged,
yielding the loop diagram appearing on the right.%
    \footnote{This interpretation is very similar to the formalism
    of `cut diagrams' used to compute squared matrix elements
    $|\mathcal{M}|^2$ in hadronic physics; see eg.
    Refs.~\cite{Brock:1993sz,Collins:2011zzd}.}
This loop mixes `$+$' and `$-$' vertices.
The conclusion, as advertised above, is that not all loop
corrections in the Schwinger formalism represent averages
over virtual particles.
This is why the contribution
to our expectation value
from tree-level decay
is nonetheless represented
by a loop.

In a simple inflationary model, multiparticle production channels
such as Fig.~\ref{fig:state-average-loop}
are negligible in comparison with the tree-level
because the dominant mechanism for particle creation is gravitational
and therefore weak unless $H$ is very close to the Planck scale.%
    \footnote{It can happen that the loop is enhanced by large infrared
    contributions~\cite{Weinberg:2005vy,Weinberg:2006ac,
    Seery:2007we,Seery:2007wf,Senatore:2009cf,Seery:2009hs,Assassi:2012et},
    in which case multiparticle channels may become
    relevant if inflation is sufficiently long-lasting.
    Unfortunately, obtaining a
    robust prediction in this regime is challenging~\cite{Seery:2010kh}.}
Therefore production of two-
and higher $n$-particle states
is rare compared with production of a single-particle state,
and
the sum $\sum_m$ is dominated by the single-particle term.
But in some models there may exist particle production mechanisms beside
spontaneous creation from the vacuum.
For example, this might happen if the fields roll through an enhanced
symmetry point where some species become
massless~\cite{Chung:1999ve,Elgaroy:2003hp,Romano:2008rr,Green:2009ds}.
In these circumstances it is possible
to generate occupation numbers
that are sufficiently large for multiparticle channels
such as Fig.~\ref{fig:state-average-loop} to be relevant---%
either transiently, or in quasi-steady state if the inflationary trajectory
repeatedly crosses the enhanced symmetry point.
In such scenarios the tree-level prediction may be inadequate.

An explicit example is the scenario recently considered by
Flauger et al.~\cite{Flauger:2016idt},
in which particles of a heavy species $\chi$ are produced
by rapid variation of its mass
and subsequently decay into a lighter species $\phi$.
As explained above, the nonadiabatic production of $\chi$ particles
in their scenario is captured exactly at tree-level.
But the conversion of these $\chi$ fluctuations into $\phi$ particles---and therefore
their contribution to the $\phi$ correlation functions---is \emph{not}
captured at tree-level, precisely because it is represented by
loops of the form appearing in Fig.~\ref{fig:state-average-loop}.

Where multiparticle intermediate states are relevant
it may be difficult to obtain a satisfactory description within perturbation
theory.
In addition to the $2 \rightarrow 1$ process of Fig.~\ref{fig:cubic-decay}
there may be higher-order $n \rightarrow 1$ vertices,
or repeated events,
in which case further loops averaging over these transitions should be
included.
If many of these processes are relevant then we may be faced with failure
of the loop expansion.

\subsubsection{Three-point function}
\label{sec:loop-threepf}
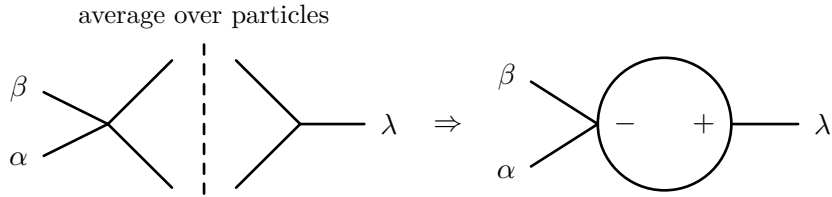
\begin{figure}
    \begin{center}
        \parbox{5cm}{
        \begin{fmfgraph*}(120,60)
            \fmfleft{l1,l2}
            \fmfright{r}
            \fmftop{t}
            \fmfbottom{b}
            \fmf{plain}{l1,v1}
            \fmf{plain}{l2,v1}
            \fmf{plain}{v1,m1}
            \fmf{plain}{v1,m2}
            \fmf{plain}{r,v2}
            \fmf{plain}{v2,n1}
            \fmf{plain}{v2,n2}
            \fmf{dashes}{t,b}
            \fmfforce{(0,0.3h)}{l1}
            \fmfforce{(0,0.7h)}{l2}
            \fmfforce{(w,0.5h)}{r}
            \fmfforce{(0.2w,0.5h)}{v1}
            \fmfforce{(0.8w,0.5h)}{v2}
            \fmfforce{(0.4w,0.1h)}{m1}
            \fmfforce{(0.4w,0.9h)}{m2}
            \fmfforce{(0.6w,0.1h)}{n1}
            \fmfforce{(0.6w,0.9h)}{n2}
            \fmfforce{(0.5w,0)}{b}
            \fmfforce{(0.5w,h)}{t}
            \fmfv{label=$\alpha$}{l1}
            \fmfv{label=$\beta$}{l2}
            \fmfv{label=$\lambda$}{r}
            \fmfv{label={\small average over particles}}{t}
        \end{fmfgraph*}
        }
        $\Rightarrow$
        \qquad
        \parbox{5cm}{
        \begin{fmfgraph*}(100,80)
            \fmfleft{l1,l2}
            \fmfright{r}
            \fmf{plain,tension=2}{l2,v1}
            \fmf{plain,tension=2}{l1,v1}
            \fmf{plain,left}{v1,v2}
            \fmf{plain,left}{v2,v1}
            \fmf{plain,tension=4}{v2,r}
            \fmfforce{(0,0.3h)}{l1}
            \fmfforce{(0,0.7h)}{l2}
            \fmfv{label=$\alpha$}{l1}
            \fmfv{label=$\beta$}{l2}
            \fmfv{label=$\lambda$}{r}
            \fmfv{label=$-$,label.angle=0}{v1}
            \fmfv{label=$+$,label.angle=180}{v2}
        \end{fmfgraph*}
        }
    \end{center}
    \caption{\label{fig:bispectrum-loop}Loop averaging over
    interference between $2 \rightarrow 2$
    scattering (left-hand side) and $2 \rightarrow 1$ decay (right-hand side)
    contributing to the inflationary bispectrum.}
\end{figure}

A similar discussion could be given for the three-point function (or any higher
$n$-point function), although we will not
do so in detail because the analysis is conceptually identical to that for
the two-point function.
For the three-point function,
the three-body interaction of Fig.~\ref{fig:cubic-decay}
is now the leading contribution. However, it can itself be corrected
by $n \rightarrow 1$ or $n \rightarrow 2$
processes.
One new feature is that
the mixed loops may average over scattering events and not just decays;
for example, consider the interference diagram of Fig.~\ref{fig:bispectrum-loop}.
This is a contribution to the three-point function from
interference between $2 \rightarrow 2$ scattering and $2 \rightarrow 1$ decay.

\para{Conclusion}
In summary, the tree-level analysis should be acceptable unless:
\begin{itemize}
    \item quantum corrections to S-matrix elements are already important
    for particle scattering or decays.
    \item copious particle production means that
    multiparticle channels such as nontrivial scattering or decays
    contribute significantly to some $n$-point functions.
    \item back-reaction into the zero-mode significantly affects the inflationary
    trajectory.
\end{itemize}

\section{The tensors ${u^\ext{a}}_{\ext{b}}$
and ${u^{\ext{a}}}_{\ext{b}\ext{c}}$}
\label{sec:u-tensors}

In this section we return to the problem of converting the evolution (or `transport')
equations~\eqref{eq:twopf-transport} and~\eqref{eq:threepf-transport}
into a practical numerical scheme.
Our first task is to obtain explicit expressions for the tensors
${u^{\ext{a}}}_{\ext{b}}$
and
${u^{\ext{a}}}_{\ext{b}\ext{c}}$.

\subsection{Computation of the cubic Hamiltonian}
\para{Background}
We are taking the matter theory to consist of a number of canonically
normalized fields minimally coupled to gravity.
The action is
\begin{equation}
  S = \frac{1}{2} \int \d^4 x \, \sqrt{-g}
  \Big(
    \Mp^2 R
    - \partial_a \phi^\alpha \partial^a \phi_\alpha
    - 2 V(\phi)
  \Big) ,
  \label{eq:model-lagrangian}
\end{equation}
where indices $a$, $b$, $\ldots$ label
the space-time coordinates and $V(\phi)$ is an arbitrary potential
assumed to support the inflationary epoch.
In this paper we mostly use a Hamiltonian formalism
in which the background field equations become
a first-order pair,
\begin{subequations}
\begin{align}
  \frac{\d \phi^\alpha}{\d t} & = \pi^\alpha \\
  \frac{\d \pi^\alpha}{\d t} & = - 3 H \pi^\alpha - \delta^{\alpha\beta} V_\beta ,
\end{align}
\end{subequations}
and $V_\beta = \partial_\beta V$ is the field-space derivative of the potential.
The Hubble rate $H$ is provided by the Friedmann equation
\begin{equation}
  3 H^2 \Mp^2 = \frac{1}{2} \pi^\alpha \pi_\alpha + V .
\end{equation}
Inflation occurs whenever the slow-roll parameter $\epsilon$ is less than unity.
It is defined by
\begin{equation}
  \epsilon \equiv - \frac{\dot{H}}{H^2}
  = \frac{\pi^\alpha \pi_\alpha}{2\Mp^2 H^2} .
\end{equation}

\para{Perturbations}
The discussion in~\S\ref{sec:evolution-equations} shows that
${u^{\ext{a}}}_{\ext{b}}$
and
${u^{\ext{a}}}_{\ext{b}\ext{c}}$
require knowledge of the Hamiltonian to third order in perturbations.
This is a standard calculation
that proceeds by writing the action as a system of coupled
matter and metric fluctuations,
integrating out the lapse and shift,
and finally performing the Legendre transformation
to the Hamiltonian.
The computation was performed by Maldacena in the single field
case~\cite{Maldacena:2002vr} and in Ref.~\cite{Seery:2005gb}
for the more general theory~\eqref{eq:model-lagrangian}.
Transformation to the Hamiltonian was discussed in Ref.~\cite{Seery:2007we}.

In this paper we do not pursue the details of this calculation but merely
quote the result.
We adapt the extended summation convention introduced in~\S\ref{sec:evolution-equations}
for phase space labels,
so that italic field-space labels $\alpha$, $\beta$, {\ldots}
range over the species of scalar
fields but sans serif
versions $\extgreek{α}$, $\extgreek{β}$, {\ldots}
include integration over Fourier modes.
The quadratic and cubic parts of the action
for scalar fluctuations---including coupling to gravity but neglecting the
`tensor' spin-2 modes---can be written
\begin{subequations}
\begin{align}
	\label{eq:action-S2}
	S_2 & = \frac{1}{2} \int \d t \; a^3
	\Big(
		\delta_{\extgreek{α}\extgreek{β}}
		\delta\dot{\phi}^{\extgreek{α}}
		\delta\dot{\phi}^{\extgreek{β}}
		+
		M_{\extgreek{α}\extgreek{β}}
		\delta\phi^{\extgreek{α}}
		\delta\phi^{\extgreek{β}}
	\Big)
	\\
	\label{eq:action-S3}
	S_3 & = \frac{1}{2} \int \d t \; a^3
	\Big(
		A_{\extgreek{α}\extgreek{β}\extgreek{γ}}
		\delta\phi^{\extgreek{α}}
		\delta\phi^{\extgreek{β}}
		\delta\phi^{\extgreek{γ}}
		+
		B_{\extgreek{α}\extgreek{β}\extgreek{γ}}
		\delta\phi^{\extgreek{α}}
		\delta\phi^{\extgreek{β}}
		\delta\dot{\phi}^{\extgreek{γ}}
		+
		C_{\extgreek{α}\extgreek{β}\extgreek{γ}}
		\delta\dot{\phi}^{\extgreek{α}}
		\delta\dot{\phi}^{\extgreek{β}}
		\delta\phi^{\extgreek{γ}}
	\Big)
\end{align}
\end{subequations}
The tensors
$M_{\extgreek{α}\extgreek{β}}$,
$A_{\extgreek{α}\extgreek{β}\extgreek{γ}}$,
$B_{\extgreek{α}\extgreek{β}\extgreek{γ}}$
and
$C_{\extgreek{α}\extgreek{β}\extgreek{γ}}$
can be written in terms of corresponding coefficient tensors
$M_{\alpha\beta}$, $A_{\alpha\beta\gamma}$, $B_{\alpha\beta\gamma}$ and $C_{\alpha\beta\gamma}$,
\begin{subequations}
\begin{align}
	M_{\extgreek{α}\extgreek{β}} & =
	(2\pi)^3 \delta(\vect{k}_\alpha + \vect{k}_\beta) M_{\alpha\beta} \\
	A_{\extgreek{α}\extgreek{β}\extgreek{γ}} & =
	(2\pi)^3 \delta(\vect{k}_\alpha + \vect{k}_\beta + \vect{k}_\gamma) A_{\alpha\beta\gamma} \\
	B_{\extgreek{α}\extgreek{β}\extgreek{γ}} & =
	(2\pi)^3 \delta(\vect{k}_\alpha + \vect{k}_\beta + \vect{k}_\gamma) B_{\alpha\beta\gamma} \\
	C_{\extgreek{α}\extgreek{β}\extgreek{γ}} & =
	(2\pi)^3 \delta(\vect{k}_\alpha + \vect{k}_\beta + \vect{k}_\gamma) C_{\alpha\beta\gamma} .
\end{align}
\end{subequations}
The $M$-tensor satisfies
\begin{subequations}
\begin{equation}
    M_{\alpha\beta} = \frac{\vect{k}_{\alpha} \cdot \vect{k}_{\beta}}{a^2} \delta_{\alpha\beta} - \massmatrix_{\alpha\beta}
    \label{eq:mass}
\end{equation}
where the field-space mass-matrix $\massmatrix_{\alpha\beta}$ is defined by
\begin{equation}
    \massmatrix_{\alpha\beta}
    =
        V_{\alpha\beta}
        - \frac{3 + \epsilon}{\Mp^2} \pi_\alpha \pi_\beta
        - \frac{( \pi_\alpha \dot\pi_\beta
                +  \pi_\beta \dot{\pi}_\alpha)}{H \Mp^2} .
    \label{eq:mass-matrix}
\end{equation}
\end{subequations}
Meanwhile, $A$-, $B$- and $C$-tensors can be written
\begin{subequations}
\begin{align}
\label{eq:reducedTensorsa}
    A_{\alpha\beta\gamma} & =
        - \frac{1}{3} V_{\alpha\beta\gamma}
        - \frac{\pi_{\alpha} V_{\beta\gamma}}{2H\Mp^2}
        + \frac{\pi_{\alpha} \pi_{\beta} \xi_{\gamma}}{8H^2\Mp^4} \nonumber
        \\& \quad \mbox{} + \frac{\pi_{\alpha} \xi_{\beta} \xi_{\gamma}}{32 H^3\Mp^4}
            \bigg(
            	1 - \frac{(\vect{k}_\beta \cdot \vect{k}_\gamma)^2}{k_\beta^2 k_\gamma^2}
            \bigg)
        + \frac{\pi_{\alpha} \pi_{\beta} \pi_{\gamma}}{8H \Mp^4}
            \bigg(
            	6 - \frac{\pi_\sigma \pi_\sigma}{H^2\Mp^2}
            \bigg)
        + \frac{\pi_{\alpha} \delta_{\beta\gamma}}{2H\Mp^2} \frac{\vect{k}_\beta \cdot \vect{k}_\gamma}{a^2}
             \\
\label{eq:reducedTensorsb}
    B_{\alpha\beta\gamma} & =
        \frac{\pi_{\alpha} \pi_{\beta} \pi_{\gamma}}{4H^2 \Mp^4}
        - \frac{\pi_{\alpha} \xi_{\beta} \pi_\gamma}{8 H^3 \Mp^4}
            \bigg(
                1 - \frac{(\vect{k}_\beta \cdot \vect{k}_\gamma)^2}{k_\beta^2 k_\gamma^2}
            \bigg)
        - \frac{\xi_\alpha \delta_{\beta\gamma}}{2H \Mp^2} \frac{\vect{k}_\alpha \cdot \vect{k}_\beta}{k_\alpha^2}
        \\
\label{eq:reducedTensorsc}
    C_{\alpha\beta\gamma} & =
        - \frac{\delta_{\alpha\beta} \pi_\gamma}{2H \Mp^2}
        + \frac{\pi_\alpha \pi_\beta \pi_\gamma}{8 H^3 \Mp^4}
        \bigg(
        	1 - \frac{(\vect{k}_\alpha \cdot \vect{k}_\beta)^2}{k_\alpha^2 k_\beta^2}
        \bigg)
        + \frac{\delta_{\beta\gamma} \pi_\alpha}{H \Mp^2}
            \frac{\vect{k}_\alpha \cdot \vect{k}_\gamma}{k_\alpha^2} .
\end{align}
These expressions should be symmetrized with unit weight.
Specifically, $A_{\alpha\beta\gamma}$ should be symmetrized
over all three indices, with corresponding
exchange of $\vect{k}_\alpha$, $\vect{k}_\beta$ and $\vect{k}_\gamma$;
and $B_{\alpha\beta\gamma}$, $C_{\alpha\beta\gamma}$ should be
symmetrized over $\alpha\beta$, with
corresponding exchange of $\vect{k}_\alpha$ and $\vect{k}_\beta$.
The quantity $\xi_\alpha$ is defined by
\begin{equation}
    \xi_\alpha = 2 \dot\pi_\alpha + \frac{\pi_\alpha}{H} \frac{\pi_\beta \pi_\beta}{\Mp^2} .
    \label{eq:xi-def}
\end{equation}
\end{subequations}
Note that all these expressions are exact. In particular, they do not involve
an expansion or truncation in slow-roll parameters.

Eqs.~\eqref{eq:mass} and~\eqref{eq:reducedTensorsa}--\eqref{eq:reducedTensorsc}
clearly demonstrate the algebraic complexity that makes precise calculations
challenging in the absence of automated tools;
this difficulty of accounting for all relevant microphysical details
was the second major challenge identified in~\S\ref{sec:standard-calculation}.
In an analytic calculation it is often impractical to include
all terms, and we must impose some criterion to select those that are retained.
For example, in Refs.~\cite{Maldacena:2002vr,Seery:2005gb}
the slow-roll approximation was used to discard most of these contributions.
In contrast, in an automated calculation based on the methods of computer algebra
there is no difficulty in keeping \emph{every} contribution
to the Hamiltonian.
This is especially important if we wish to obtain accurate predictions when the
bispectrum amplitude is not large,
because in these circumstances the corrections from
gravitational interactions
are often relevant;
these are the contributions
in $A_{\alpha\beta\gamma}$, $B_{\alpha\beta\gamma}$, $C_{\alpha\beta\gamma}$
and $\xi_\alpha$
that scale like an inverse power of $\Mp$
and therefore vanish in the decoupling limit $\Mp \uparrow \infty$.
The large number of these terms makes it onerous to include them all
in a manual calculation.

\para{Transition to the Hamiltonian}
By definition, the momentum canonically conjugate to the
field perturbations $\delta\phi^{\extgreek{α}}$ satisfies
\begin{equation}
	\delta\pi_{\extgreek{α}}(t) = \frac{\delta S}{\delta [\delta\dot{\phi}^{\extgreek{α}}(t)]} ,
	\label{eq:canonical-momentum}
\end{equation}
where the variational derivative is defined by
\begin{equation}
    \frac{\delta [\delta \phi^\extgreek{α}(t)]}{\delta [\delta\phi^\extgreek{β}(t')]} \equiv \delta^{\extgreek{α}}_{\extgreek{β}}
    \delta(t-t').
    \label{eq:variational-deriv-defn}
\end{equation}
From this definition it follows that $\pi_{\extgreek{β}}$
can be written
\begin{equation}
    \delta\pi_{\extgreek{α}} = a^3 \Big(
        \delta\dot{\phi}_{\extgreek{α}}
        + \frac{1}{2} B_{\extgreek{β} \extgreek{γ} \flip{\extgreek{α}}} \delta\phi^{\extgreek{β}} \delta\phi^{\extgreek{γ}}
        + C_{\flip{\extgreek{α}} \extgreek{β} \extgreek{γ}} \delta\dot{\phi}^{\extgreek{β}} \delta\phi^{\extgreek{γ}}
    \Big) .
    \label{eq:momentum-def}
\end{equation}
The Hamiltonian is
\begin{equation}
    H = \int \d t \;
    \Big(
    	\delta\pi_{\extgreek{α}} \delta\dot{\phi}^{\flip{{\extgreek{α}}}} - L
    \Big) ,
\end{equation}
considered as a functional of $(\delta\phi^{\extgreek{α}}, \delta\pi_{\extgreek{β}})$.
However, to simplify~\eqref{eq:momentum-def}
it is convenient to rescale $\delta\phi_{\extgreek{α}}$
by a factor $a^3$,
$\delta\pi_{\extgreek{α}} \rightarrow a^3 \delta\pi_{\extgreek{α}}$.
This removes the overall factor of $a^3$ from~\eqref{eq:momentum-def}
and means that each term in the Hamiltonian has the same common factor
appearing in the measure.
In terms of the rescaled momentum we find that the time derivative
of the field perturbation can be written
\begin{equation}
\label{eq:newp}
    \delta\dot{\phi}_{\extgreek{α}} =
        \delta\pi_{\extgreek{α}}
        - \frac{1}{2} B_{\extgreek{β} \extgreek{γ} \flip{\extgreek{α}}} \delta\phi^{\extgreek{β}} \delta\phi^{\extgreek{γ}}
        - C_{\flip{\extgreek{α}} \extgreek{β} \extgreek{γ}} \delta\pi^{\extgreek{β}} \delta\phi^{\extgreek{γ}}
        + \cdots ,
\end{equation}
where `$\cdots$' denotes higher-order
interactions that do not need to be retained at the order to which we are working.
Rearranging $H$ and neglecting further higher-order interactions yields
\begin{equation}
\label{eq:hamiltonian-cubic}
\begin{split}
    H = \frac{1}{2} \int \d t \; a^3 \,
    \Big(
    	&
		\delta_{\extgreek{α}\extgreek{β}}
		\delta\pi^{\extgreek{α}}
		\delta\pi^{\extgreek{β}}
		-
		M_{\extgreek{α}\extgreek{β}}
		\delta\phi^{\extgreek{α}}
		\delta\phi^{\extgreek{β}}
	\\ & \mbox{}
		-
		A_{\extgreek{α}\extgreek{β}\extgreek{γ}}
		\delta\phi^{\extgreek{α}}
		\delta\phi^{\extgreek{β}}
		\delta\phi^{\extgreek{γ}}
		-
		B_{\extgreek{α}\extgreek{β}\extgreek{γ}}
		\delta\phi^{\extgreek{α}}
		\delta\phi^{\extgreek{β}}
		\delta\pi^{\extgreek{γ}}
		-
		C_{\extgreek{α}\extgreek{β}\extgreek{γ}}
		\delta\pi^{\extgreek{α}}
		\delta\pi^{\extgreek{β}}
		\delta\phi^{\extgreek{γ}}
	\Big)
\end{split}
\end{equation}
There are no barred indices in this expression.
The first two terms contribute to the quadratic
piece $H_{\ext{a}\ext{b}} \delta X^{\ext{a}} \delta X^{\ext{b}}$
and therefore $H_0$,
while the last three terms contribute to the cubic
piece $H_{\ext{a}\ext{b}\ext{c}} \delta X^{\ext{a}} \delta X^{\ext{b}} \delta X^{\ext{c}}$
and therefore constitute $\Hint$.
In this example $\Hint$ has the appearance of
$-\Lint$ with
$\delta\dot{\phi}^{\extgreek{α}} \rightarrow \delta\pi^{\extgreek{α}}$,
but it should be remembered that
$\delta\pi^{\extgreek{α}} \neq \delta \dot{\phi}^{\extgreek{α}}$
once interactions are included.%
	\footnote{Had our action~\eqref{eq:action-S2}--\eqref{eq:action-S3}
	contained terms cubic in $\delta\dot{\phi}$ then even the
	symbolic appearance $\Hint = -\Lint$ would not have been maintained.}
This will be important in the subsequent discussion.

\subsection{Equations of motion in the Heisenberg picture}
Given the Hamiltonian~\eqref{eq:hamiltonian-cubic}
we can extract the $u$-tensors
${u^{\ext{a}}}_{\ext{b}}$
and
${u^{\ext{a}}}_{\ext{b}\ext{c}}$
by comparison with the Heisenberg equations~\eqref{eq:heisenberg-equations}.
These assume a slightly noncanonical form when using our rescaled momentum
$\delta\pi_{\extgreek{α}}$,
\begin{subequations}
\begin{align}
	\label{eq:heisenberg-q}
    \frac{\d \delta\phi^{\extgreek{α}}}{\d t} & = - \im [ \delta\phi^{\extgreek{α}}, H ] \\
    \label{eq:heisenberg-p}
    \frac{\d \delta\pi_{\extgreek{α}}}{\d t} & = - \im [ \delta\pi_{\extgreek{α}}, H ] - 3H \delta\pi_{\extgreek{α}} .
\end{align}
\end{subequations}
Also, the operators
$\delta\phi^{\extgreek{α}}$
and
$\delta\pi_{\extgreek{α}}$
acquire a modified commutation algebra,
\begin{equation}
	[ \delta\phi^{\extgreek{α}}(t) , \delta\pi_{\extgreek{β}}(t') ]
	=
	\im a^3 \delta^{\extgreek{α}}_{\extgreek{β}} \delta(t-t') .
	\label{eq:adjusted-commutator}
\end{equation}
The noncanonical term $- 3H \delta\pi_{\extgreek{α}}$
appearing in the $\delta\pi_{\extgreek{α}}$ evolution equation can be absorbed
as an extra contribution to ${u^\ext{a}}_{\ext{b}}$.

To proceed we could
use Eqs.~\eqref{eq:heisenberg-q}--\eqref{eq:heisenberg-p}
and
Eq.~\eqref{eq:adjusted-commutator}
in conjunction with~\eqref{eq:hamiltonian-cubic},
or just read off
$H_{\ext{a}\ext{b}}$ and
$H_{\ext{a}\ext{b}\ext{c}}$
from~\eqref{eq:hamiltonian-cubic}
and directly compute
${u^\ext{a}}_{\ext{b}}
= \epsilon^{\ext{a}\ext{c}} H_{\ext{c}\ext{b}}$
and
${u^{\ext{a}}}_{\ext{b}\ext{c}}
= \epsilon^{\ext{a}\ext{d}} H_{\ext{d}\ext{b}\ext{c}}$.
Following the latter route we see that
$H_{\ext{a}\ext{b}}$ can be written
\begin{equation}
	H_{\ext{a}\ext{b}} =
	\left(
		\begin{array}{cc}
			-M_{\extgreek{α}\extgreek{β}} & 0 \\
			0 & \delta_{\extgreek{α}\extgreek{β}}
		\end{array}
	\right)	,
\end{equation}
where we are assuming the same numerical arrangement
for the indices $\ext{a}$, $\ext{b}$
used in Eq.~\eqref{eq:epsilon-coeff-matrix},
and $\alpha$, $\beta$ are the field-space indices
corresponding to the phase-space indices $a$, $b$.
(Explicitly, $\alpha = a \bmod N$, $\beta = b \bmod N$
where $N$ is the number of fields.)
It follows that ${u^{\ext{a}}}_{\ext{b}}$ satisfies
\begin{equation}
	{u^{\ext{a}}}_{\ext{b}}
	=
	\left(
		\begin{array}{cc}
			0 & \vect{1} \\
			-\vect{1} & 0
		\end{array}
	\right)
	\left(
		\begin{array}{cc}
			-{M^{\flip{\extgreek{α}}}}_{\extgreek{β}} & 0 \\
			0 & \delta^{\flip{\extgreek{α}}}_{\extgreek{β}}
		\end{array}
	\right)
	+
	\left(
		\begin{array}{cc}
			0 & 0 \\
			0 & - 3H \delta^{\flip{\extgreek{α}}}_{\extgreek{β}}
		\end{array}
	\right)
	=
	\left(
		\begin{array}{cc}
			0 & \delta^{\flip{\extgreek{α}}}_{\extgreek{β}} \\
			{M^{\flip{\extgreek{α}}}}_{\extgreek{β}} & - 3H \delta^{\flip{\extgreek{α}}}_{\extgreek{β}}
		\end{array}
	\right) .
	\label{eq:u2-general}
\end{equation}
Proceeding to the cubic terms, we find that $H_{\ext{a}\ext{b}\ext{c}}$
satisfies
\begin{equation}
	H_{\ext{a}\ext{b}\ext{c}}
	=
	\left\{
		\begin{array}{c}
			\left(
			\begin{array}{c@{\hspace{3mm}}c}
				- 3 A_{\extgreek{α}\extgreek{β}\extgreek{γ}} & - B_{\extgreek{α}\extgreek{γ}\extgreek{β}} \\
				- B_{\extgreek{β}\extgreek{γ}\extgreek{α}} & -C_{\extgreek{α}\extgreek{β}\extgreek{γ}}
			\end{array}
			\right)
			\\
			\\
			\left(
			\begin{array}{c@{\hspace{3mm}}c}
				- B_{\extgreek{α}\extgreek{β}\extgreek{γ}} & - C_{\extgreek{γ}\extgreek{β}\extgreek{α}} \\
				- C_{\extgreek{α}\extgreek{γ}\extgreek{β}} & 0
			\end{array}
			\right)
		\end{array}
	\right\}
\end{equation}
In this formula, the $\ext{a}$ index labels rows on each $2\times 2$ block matrix; the
$\ext{b}$ index labels corresponding columns;
and the $\ext{c}$ index labels the matrices within the braces $\{ \cdots \}$.
As above, the indices $\alpha$, $\beta$ $\gamma$ are the field-space versions
of $a$, $b$, $c$.
Likewise, $\vect{k}_\alpha$ should be regarded as equal to $\vect{k}_a$,
$\vect{k}_\beta$ should be regarded as equal to $\vect{k}_b$, and so on.

A short calculation yields ${u^{\ext{a}}}_{\ext{b}\ext{c}}$,
\begin{equation}
	{u^\ext{a}}_{\ext{b}\ext{c}}
	=
	\left\{
		\begin{array}{c}
			\left(
			\begin{array}{cc}
				0 & \vect{1} \\
				-\vect{1} & 0
			\end{array}
			\right)
			\left(
			\begin{array}{c@{\hspace{3mm}}c}
				- 3 {A^{\flip{\extgreek{α}}}}_{\extgreek{β}\extgreek{γ}} & - {B^{\flip{\extgreek{α}}}}_{\extgreek{γ}\extgreek{β}} \\
				- {B_{\extgreek{β}\extgreek{γ}}}^{\flip{\extgreek{α}}} & -{C^{\flip{\extgreek{α}}}}_{\extgreek{β}\extgreek{γ}}
			\end{array}
			\right)
			\\
			\\
			\left(
			\begin{array}{cc}
				0 & \vect{1} \\
				-\vect{1} & 0
			\end{array}
			\right)
			\left(
			\begin{array}{c@{\hspace{3mm}}c}
				- {B^{\flip{\extgreek{α}}}}_{\extgreek{β}\extgreek{γ}} & - {C_{\extgreek{γ}\extgreek{β}}}^{\flip{\extgreek{α}}} \\
				- {C^{\flip{\extgreek{α}}}}_{\extgreek{γ}\extgreek{β}} & 0
			\end{array}
			\right)
		\end{array}
	\right\}
	=
	\left\{
		\begin{array}{c}
			\left(
			\begin{array}{c@{\hspace{3mm}}c}
				- {B_{\extgreek{β}\extgreek{γ}}}^{\flip{\extgreek{α}}} & -{C^{\flip{\extgreek{α}}}}_{\extgreek{β}\extgreek{γ}} \\
				3 {A^{\flip{\extgreek{α}}}}_{\extgreek{β}\extgreek{γ}} & {B^{\flip{\extgreek{α}}}}_{\extgreek{γ}\extgreek{β}}
			\end{array}
			\right)
			\\
			\\
			\left(
			\begin{array}{c@{\hspace{3mm}}c}
				- {C^{\flip{\extgreek{α}}}}_{\extgreek{γ}\extgreek{β}} & 0 \\
				{B^{\flip{\extgreek{α}}}}_{\extgreek{β}\extgreek{γ}} & {C_{\extgreek{γ}\extgreek{β}}}^{\flip{\extgreek{α}}}
			\end{array}
			\right)
		\end{array}
	\right\}
	\label{eq:u3-general}
\end{equation}

\section{The transport equations}
\label{sec:transport-equations}
We now have all the elements in place to write an evolution equation for each
configuration of the two- and three-point functions.
First, notice that the arrangement of barred indices in Eqs.~\eqref{eq:u2-general}
and~\eqref{eq:u3-general}
implies that we can define coefficient tensors ${u^a}_{b}$
and ${u^a}_{bc}$
that satisfy
\begin{subequations}
\begin{align}
    \label{eq:u2-coefficient}
	{u^{\ext{a}}}_{\ext{b}} & = (2\pi)^3 \delta(\vect{k}_a - \vect{k}_b) {u^a}_b(\vect{k}_a, \vect{k}_b) \\
	\label{eq:u3-coefficient}
	{u^{\ext{a}}}_{\ext{b}\ext{c}} & = (2\pi)^3 \delta(\vect{k}_a - \vect{k}_b - \vect{k}_c) {u^a}_{bc}(\vect{k}_a, \vect{k}_b, \vect{k}_c) .
\end{align}
\end{subequations}
In these definitions we should still regard
$\vect{k}_a$ as the momentum associated with the index $a$,
$\vect{k}_b$ as the momentum associated with $b$,
\emph{et cetera},
but we have written this dependence explicitly
because our convention of labelling momenta by the corresponding index
will break down
when we come to write an evolution equation
for a single wavenumber configuration in Eq.~\eqref{eq:threepf-transport-raw-ode}
below.

The coefficient tensors ${u^a}_b$ and ${u^a}_{bc}$
are given by expressions
obtained symbolically from Eqs.~\eqref{eq:u2-general} and~\eqref{eq:u3-general},
by replacing extended summation-convention indices ($\extgreek{α}$, $\extgreek{β}$, {\ldots})
with standard indices ($\alpha$, $\beta$, {\ldots}).
Note that because the coefficient tensors
$M_{\alpha\beta}$, $A_{\alpha\beta\gamma}$,
$B_{\alpha\beta\gamma}$ and $C_{\alpha\beta\gamma}$ depend on the
momenta, it is still necessary to keep track of barred labels indicating
that the corresponding sign should be reversed.
However, bars on momenta-independent
tensors such as $\delta^{\alpha}_\beta$ can be safely
discarded.

There is a further simplification that can be made.
In principle, ${u^a}_b$ is a function of the momenta
$\vect{k}_a$ and $\vect{k}_b$.
However, because the $\delta$-function in~\eqref{eq:u2-coefficient}
constrains these to be equal we can replace it with the equivalent form
\begin{equation}
  \tilde{u}^a{}_b(k) =
  \left(
    \begin{array}{c@{\hspace{3mm}}c}
      0 & \delta^\alpha_\beta \\
      \tilde{M}^\alpha{}_\beta & -3H\delta^\alpha_\beta
    \end{array}
  \right)
  \label{eq:u2-simplified}
\end{equation}
and $\tilde{M}_{\alpha\beta}$ is a rewriting of~\eqref{eq:mass},
\begin{equation}
  \tilde{M}_{\alpha\beta} = - \frac{k^2}{a^2} \delta_{\alpha\beta} - m_{\alpha\beta} .
  \label{eq:mass-simplified}
\end{equation}
This accounts for sign reversal from the barred index $\bar{\alpha}$
and uses the common magnitude $k$ of $\vect{k}_a$ and $\vect{k}_b$.

\subsection{Two-point function}
\label{sec:twopf-transport}
It is now straightforward to extract an evolution equation for the two-point
function.
The fully momentum-dependent
transport equation is Eq.~\eqref{eq:twopf-transport}.
To convert this to an ordinary differential equation
we appeal to statistical isotropy and homogeneity and write
the equal-time two-point function in the form
\begin{equation}
  \langle \delta X^{\ext{a}} \delta X^{\ext{b}} \rangle
  =
  (2\pi)^3 \delta(\vect{k}_a + \vect{k}_b) \Sigma^{ab}(k) ,
  \label{eq:sigma-def}
\end{equation}
where as above $k = |\vect{k}_a| = |\vect{k}_b|$ is the common magnitude
of the momenta.
Then Eqs.~\eqref{eq:twopf-transport}, \eqref{eq:u3-coefficient}, \eqref{eq:u2-simplified}
and~\eqref{eq:sigma-def}
combine to give
\begin{equation}
  \frac{\d \Sigma^{ab}(k)}{\d t}
  =
  \tilde{u}^a{}_c(k) \Sigma^{cb}(k) + \tilde{u}^b{}_c(k) \Sigma^{ac}(k) .
  \label{eq:twopf-transport-ode}
\end{equation}

\para{Reality and symmetry properties}
The two-point function $\Sigma^{ab}(k)$ will generally be complex,
and
a practical numerical implementation must evolve its real and imaginary
parts separately. Fortunately the linearity of Eq.~\eqref{eq:twopf-transport-ode}
makes this straightforward.
We break $\Sigma^{ab}$ into real and imaginary parts
\begin{equation}
  \Sigma^{ab} = \Sigma_{\Re}^{ab} + \im \Sigma_{\Im}^{ab} ,
\end{equation}
where both $\Sigma_{\Re}$ and $\Sigma_{\Im}$ are real.
Then, because $\tilde{u}^a{}_b$ is real,
each of $\Sigma_{\Re}$ and $\Sigma_{\Im}$ obey Eq.~\eqref{eq:twopf-transport-ode}
separately.
We only require the real part to compute
the $\zeta$ two-point function
$\langle \zeta \zeta \rangle$,
but we will see below that computation of
$\langle \delta X^{\ext{a}} \delta X^{\ext{b}} \delta X^{\ext{c}} \rangle$
or
$\langle \zeta \zeta \zeta \rangle$ requires knowledge of both the
real and imaginary parts.

It follows from the relation
$\langle \Operator \rangle^\ast
= \langle \Operator^\dag \rangle$
(where $\ast$ denotes conjugation of a complex number, and
$\dag$ denotes Hermitian conjugation of an operator)
that
$\Sigma_{\Re}$ is symmetric and $\Sigma_{\Im}$ is antisymmetric,
\begin{subequations}
\begin{align}
  \Sigma_{\Re}^{ab} & = \Sigma_{\Re}^{ba} \\
  \Sigma_{\Im}^{ab} & = -\Sigma_{\Im}^{ba} .
\end{align}
\end{subequations}
The imaginary part normally decays outside the horizon.

\subsection{Tensor modes}%
\label{sec:tensor-transport}
We can compute a transport equation for the components of the
tensor 2-point function in a similar way; for details see
Ref.~\cite{Dias:2015rca}.
Here we summarize the calculation.

Tensor perturbations $\bm{\gamma}$ are transverse, traceless
perturbations of the spatial metric.
They represent gravitational waves excited by the inflationary epoch.
Up to quadratic order their evolution is controlled by the action
\begin{equation}
  S \supseteq \frac{\Mp^2 }{8} \int \d^3 x\, \d t \; a^3
    \bigg\{
        \dot{\gamma}_{ij} \dot{\gamma}^{ij}
        -
        \partial_k \gamma_{ij} \partial^k \gamma^{ij}
    \bigg\}.
\end{equation}
The indices $i$, $j$, $k$, $\ldots$ label spatial coordinates.
To write a scalar transport equation it is convenient to decompose $\bm{\gamma}$
into a basis of polarization tensors.
Working in Fourier space, we write
\begin{equation}
  {\bm \gamma}(\vect{x}) =
    \int
    \frac{\d^3 k}{(2\pi)^3} \, \sum_{s}
    \gamma_s(\vect{k}) \bm{e}^s(\vect{k}) \, \e{i\vect{k} \cdot \vect{x}}
\end{equation}
where $s = \{ +, \times \}$ labels the available polarization states,
and the polarization tensors $\bm{e}(\vect{k})$ satisfy the relations
$\vect{k} \cdot \bm{e}^s(\vect{k}) = 0$
and
$\tr (\bm{e}^s \cdot \bm{e}^{s'}) = 2 \delta^{s s'}$.
Each polarization $\Mp \gamma_s(\vect{k}) / \sqrt{2}$
behaves as a
canonically-normalized
massless scalar field.

In analogy with the case of field-space fluctuations we
define a tensor momentum $\pi_s = \d \gamma_s / \d t$
and collect $\gamma_s$ and $\pi_s$
into a two-component vector $Y_s^{\tensoridx{a}} = (\gamma_s, \pi_s)$.
Gothic indices $\tensoridx{a}$, $\tensoridx{b}$, $\ldots$
label the components of this vector.
The two-point function can be written
\begin{equation}
  \langle
    Y_s^{\tensoridx{a}}(\vect{k}_1)
    Y_{s'}^{\tensoridx{b}}(\vect{k}_2)
  \rangle
  =
  (2\pi)^3 \delta_{ss'}
  \delta(\vect{k} + \vect{k}')
  \Gamma^{\tensoridx{a}\tensoridx{b}}(k)
  \label{eq:Gamma-def}
\end{equation}
Then it follows
from the same argument leading to~\eqref{eq:twopf-transport-ode}
that
the transport equation for
$\Gamma^{\tensoridx{a}\tensoridx{b}}$ can be written
\begin{equation}
  \frac{\d \Gamma^{\tensoridx{a}\tensoridx{b}}(k)}{\d t} =
    {w^\tensoridx{a}}_{\tensoridx{c}}(k) \Gamma^{\tensoridx{c}\tensoridx{b}}(k)
    +
    {w^\tensoridx{b}}_{\tensoridx{c}}(k) \Gamma^{\tensoridx{a}\tensoridx{c}}(k)
    +
    \cdots
    .
\end{equation}
The functional form of the matrices ${w^\tensoridx{a}}_{\tensoridx{b}}$
follows directly from the massless limit of Eqs.~\eqref{eq:u2-simplified}
and \eqref{eq:mass-simplified}, that is,
\begin{equation}
    {w^{\tensoridx{a}}}_{\tensoridx{b}}(k)
    =
    \left(
        \begin{array}{c@{\hspace{3mm}}c}
            0 & 1 \\
            -k^2/a^2 & - 3H
        \end{array}
    \right)
\end{equation}
The coefficient matrix $\Gamma^{\tensoridx{a}\tensoridx{b}}$
will generally be complex on subhorizon scales,
as in~\S\ref{sec:twopf-transport}.
For the same reasons
explained above the real and imaginary components evolve independently
at tree-level.
The imaginary part decays outside the horizon and does not
affect observables.
Therefore for practical purposes we can restrict our attention
to the real part.

\subsection{Three-point function}
\label{sec:threepf-transport}
Now consider the three-point function.
Statistical isotropy and homogeneity imply that this takes the form
\begin{equation}
  \langle \delta X^{\ext{a}} \delta X^{\ext{b}} \delta X^{\ext{c}} \rangle
  =
  (2\pi)^3 \delta(\vect{k}_a + \vect{k}_b + \vect{k}_c) B^{abc}(k_a, k_b, k_c) .
  \label{eq:B-def}
\end{equation}
The momentum-dependent transport equation is now~\eqref{eq:threepf-transport}.
Applied to Eq.~\eqref{eq:B-def}
we find
\begin{equation}
\begin{split}
  \frac{\d B^{abc}(k_a, k_b, k_c)}{\d t}
  = \mbox{} &
  \tilde{u}^a{}_d(k_a) B^{dbc}(k_a, k_b, k_c)
  +
  \tilde{u}^b{}_d(k_b) B^{adc}(k_a, k_b, k_c)
  +
  \tilde{u}^c{}_d(k_c) B^{abd}(k_a, k_b, k_c)
  \\
  & \mbox{}
  +
  {u^a}_{de}(\vect{k}_a, -\vect{k}_b, -\vect{k}_c)
  \Sigma^{db}(k_b) \Sigma^{ec}(k_c)
  \\
  & \mbox{}
  +
  {u^b}_{de}(\vect{k}_b, -\vect{k}_a, -\vect{k}_c)
  \Sigma^{ad}(k_a) \Sigma^{ec}(k_c)
  \\
  & \mbox{}
  +
  {u^c}_{de}(\vect{k}_c, -\vect{k}_a, -\vect{k}_b)
  \Sigma^{ad}(k_a) \Sigma^{be}(k_c) .
  \label{eq:threepf-transport-raw-ode}
\end{split}
\end{equation}
We have written out all terms explicitly in order
to avoid ambiguity about the index ordering.
This is important in the $\Sigma\Sigma$ terms because,
as explained above,
$\Sigma^{ab}$ is not symmetric
if it contains a significant imaginary part.

Comparison with Eq.~\eqref{eq:u3-general}
shows that
the pattern of sign reversals in these terms
corresponds to reversal of
\emph{all} momenta in Eqs.~\eqref{eq:reducedTensorsa}--\eqref{eq:reducedTensorsc}.
However, because the momenta appear only in inner products with each
other (and therefore always occur in pairs),
this is the same as reversing \emph{none} of them.
Also, like the two-point function,
$B^{abc}$ will normally be complex
and should be broken into its real and imaginary parts,
\begin{equation}
  B^{abc}(k_1, k_2, k_3) = B_{\Re}^{abc}(k_1, k_2, k_3) + \im B_{\Im}^{abc}(k_1, k_2, k_3) .
\end{equation}
We will see below that the imaginary part is zero at tree-level
and therefore plays no role in the calculation.
Putting all this together,
the ordinary differential equation we are seeking
for $B_{\Re}^{abc}$
can be written
\begin{equation}
\begin{split}
  \frac{\d B^{abc}_{\Re}(k_a, k_b, k_c)}{\d t}
  = \mbox{} &
  \tilde{u}^a{}_d(k_a) B^{dbc}_{\Re}(k_a, k_b, k_c)
  +
  \tilde{u}^b{}_d(k_b) B^{adc}_{\Re}(k_a, k_b, k_c)
  +
  \tilde{u}^c{}_d(k_c) B^{abd}_{\Re}(k_a, k_b, k_c)
  \\
  & \mbox{}
  +
  \tilde{u}^a{}_{de}(\vect{k}_a, \vect{k}_b, \vect{k}_c)
  \Sigma_{\Re}^{db}(k_b) \Sigma_{\Re}^{ec}(k_c)
  -
  \tilde{u}^a{}_{de}(\vect{k}_a, \vect{k}_b, \vect{k}_c)
  \Sigma_{\Im}^{db}(k_b) \Sigma_{\Im}^{ec}(k_c)
  \\
  & \mbox{}
  +
  \tilde{u}^b{}_{de}(\vect{k}_b, \vect{k}_a, \vect{k}_c)
  \Sigma_{\Re}^{ad}(k_a) \Sigma_{\Re}^{ec}(k_c)
  -
  \tilde{u}^b{}_{de}(\vect{k}_b, \vect{k}_a, \vect{k}_c)
  \Sigma_{\Im}^{ad}(k_a) \Sigma_{\Im}^{ec}(k_c)
  \\
  & \mbox{}
  +
  \tilde{u}^c{}_{de}(\vect{k}_c, \vect{k}_a, \vect{k}_b)
  \Sigma_{\Re}^{ad}(k_a) \Sigma_{\Re}^{be}(k_b) ,
  -
  \tilde{u}^c{}_{de}(\vect{k}_c, \vect{k}_a, \vect{k}_b)
  \Sigma_{\Im}^{ad}(k_a) \Sigma_{\Im}^{be}(k_b) ,
  \label{eq:threepf-transport-ode}
\end{split}
\end{equation}
where
$\tilde{u}^a{}_{bc}$ is defined by
the plain summation-convention version of
Eq.~\eqref{eq:u3-general} but with sign reversal discarded,
\begin{equation}
	\tilde{u}^a{}_{bc}
	=
	\left\{
		\begin{array}{c}
			\left(
			\begin{array}{c@{\hspace{3mm}}c}
				- {B_{\beta\gamma}}^{\alpha} & -{C^{\alpha}}_{\beta\gamma} \\
				3 {A^{\alpha}}_{\beta\gamma} & {B^{\alpha}}_{\gamma\beta}
			\end{array}
			\right)
			\\
			\\
			\left(
			\begin{array}{c@{\hspace{3mm}}c}
				- {C^{\alpha}}_{\gamma\beta} & 0 \\
				{B^{\alpha}}_{\beta\gamma} & {C_{\gamma\beta}}^{\alpha}
			\end{array}
			\right)
		\end{array}
	\right\}
	\label{eq:u3-tilde} .
\end{equation}
We have not written the $\vect{k}$ dependence explicitly, but the
left-hand side should be regarded as a function of
the momenta $\vect{k}_a$, $\vect{k}_b$, $\vect{k}_c$
associated with each index.
On the right-hand side, the
tensors $A_{\alpha\beta\gamma}$,
$B_{\alpha\beta\gamma}$
and $C_{\alpha\beta\gamma}$
are functions of momenta
$\vect{k}_\alpha$, $\vect{k}_\beta$, $\vect{k}_\gamma$
that should be obtained from the assignments
$\vect{k}_a \rightarrow \vect{k}_\alpha$,
$\vect{k}_b \rightarrow \vect{k}_\beta$,
$\vect{k}_c \rightarrow \vect{k}_\gamma$.

\section{Initial conditions}
\label{sec:initial-conditions}

Finally, we must provide initial conditions for
$\Sigma_{\Re}^{ab}$,
$\Sigma_{\Im}^{ab}$
and
$B_{\Re}^{abc}$.
In our framework these are
computed analytically from
the interaction-picture expression~\eqref{eq:inin-formula}.

This proposition
might appear hopeless because the premise of our
argument is that analytic approximations to Eq.~\eqref{eq:inin-formula}
are either prohibitively complex or cannot capture
the relevant physical processes such as particle creation and energy
exchange.
However, as we now explain, it is normally possible to
use~\eqref{eq:inin-formula} to obtain accurate initial
conditions provided the initial time is taken
sufficiently early that the relevant modes are deep in the subhorizon
era.

Inspection of Eq.~\eqref{eq:mass} shows that
the mass-matrix $m_{\alpha\beta}$
defined in~\eqref{eq:mass-matrix}
models the influence of
complex processes such as energy exchange
in the two-point function.
The three-point function inherits all of these effects,
and also includes others from the three-body
vertices~\eqref{eq:reducedTensorsa}--\eqref{eq:reducedTensorsc}.
But at very early times
the term $(k/a)^2$ in $M_{\alpha\beta}$
will dominate $m_{\alpha\beta}$
if all eigenvalues of the mass matrix remain bounded.
Therefore,
at sufficiently early times,
the two-point function will approach that for a set of
\emph{massless} uncoupled scalar fields.
This is much easier to handle analytically.
Meanwhile, for the three-point function,
the $\vect{k}$-dependent terms in
Eqs.~\eqref{eq:reducedTensorsa}--\eqref{eq:reducedTensorsc}
will dominate.
These significant simplifications mean that analytic
approximations become both tractable and accurate.

\subsection{Two-point function}
We define $m^2$ to be the largest eigenvalue of the mass matrix~\eqref{eq:mass-matrix}.
Provided this eigenvalue
remains bounded
the kinetic term $(k/a)^2$ becomes increasingly important at early times,
while contributions from the mass matrix become irrelevant.
If we choose the initial time so that
$(k/a)^2 \gg m^2$
then the fields can be approximated as massless
and non-interacting with very good
accuracy.
The
corresponding time-ordered unequal-time two-point function
in a nearly de Sitter spacetime
assumes a standard form
in terms of conformal time
$\d \tau = \d t / a(t)$,
\begin{equation}
  \langle
    \TimeOrder
    \delta \phi^{\alpha}(\vect{k}, \tau)
    \delta \phi^{\beta}(\vect{k}', \tau')
  \rangle
  =
  (2\pi)^3
  \delta(\vect{k} + \vect{k}')
  \delta^{\alpha\beta}
  \times
  \left\{
  \begin{array}{l@{\hspace{4mm}}l}
      f_k(\tau) f_{k'}^\ast(\tau') & \tau < \tau' \\
      f_k^\ast(\tau) f_{k'}(\tau') & \tau > \tau'
  \end{array}
  \right.
  \label{eq:2ptdeSitter}
\end{equation}
where $f_k(\tau) = H(\tau) (1-\im k \tau) \e{\im k \tau} / \sqrt{2k^3}$
is the elementary wavefunction for a mode of wavenumber $k$.
The time-ordering operator $\TimeOrder$ was defined below
Eq.~\eqref{eq:int-pic-f},
and
we have evaluated the correlation function in the Bunch--Davies
vacuum state.

In addition to the condition $(k/a)^2 \gg m^2$,
Eq.~\eqref{eq:2ptdeSitter}
requires that the expansion rate $H(t)$
is sufficiently slowly varying that
$\epsilon \equiv -\dot{H}/H^2 \ll 1$,
where the overdot denotes a derivative with respect to cosmic time $t$.
Therefore, although the rest of our numerical scheme
makes no use of the slow-roll approximation,
our use of~\eqref{eq:2ptdeSitter}
to compute initial conditions
\emph{does} require
that $\epsilon$ be at least modestly small
at the initial time---and for at least a few prior e-folds
in order that each mode has time to relax safely into its vacuum state.%
    \footnote{In general, this means it is not possible to
    use the initial conditions obtained in this
    section to study effects very close to the
    beginning of inflation.
    A separate prescription is required to study modes that exit
    the horizon
    very early in the inflationary phase.}
To successfully estimate initial values for the three-point
correlation functions we must impose a similar slow-variation
condition on other background quantities such as the
field derivatives $\dot{\phi}^\alpha$.
Two comments can be made about this restriction.
First, while it necessarily involves some loss of flexibility,
the requirement that $\epsilon < 1$
and $\dot{\phi}^\alpha$ is slowly varying
is less restrictive than it may appear
because
we would normally expect to place the initial time during
a quiescent phase.
Second, although $\epsilon < 1$ should be approximately
satisfied, we normally do not require the strong condition
$\epsilon \ll 1$ \emph{provided} the
slow-roll approximation is close enough to the true solution
that it lies within its basin of attraction,
\emph{and}
there is enough time
for the numerical solution to relax to this `true' value.

On subhorizon scales
$k/aH$ is exponentially large
and therefore
the highest power of $k\tauinit \approx
k/(aH)_{\text{init}}$ is dominant,
where the subscript `init' denotes evaluation at the
initial cosmic time $\tinit$.
From Eq.~\eqref{eq:2ptdeSitter}
we can deduce all two-point functions containing any combination
of fields and momenta.
At highest order in $k\tauinit$
and lowest order in $\epsilon$, and neglecting any mass terms,
this gives
\begin{subequations}
\begin{align}
  \label{eq:sigma-re-ics}
  \Sigma_{\Re}^{ab}(\tinit)
  & =
  \frac{1}{2a^3 k}
  \left(
    \begin{array}{c@{\hspace{3mm}}c}
      a\delta^{\alpha\beta} & - aH \delta^{\alpha\beta} \\
      - aH \delta^{\alpha\beta} & (k^2 / a) \delta^{\alpha\beta}
    \end{array}
  \right)
  \\
  \label{eq:sigma-im-ics}
  \Sigma_{\Im}^{ab}(\tinit)
  & =
  \frac{1}{2a^3 k}
  \left(
    \begin{array}{c@{\hspace{3mm}}c}
      0 & k \delta^{\alpha\beta} \\
      k \delta^{\alpha\beta} & 0
    \end{array}
  \right)
  ,
\end{align}
\end{subequations}
where all quantities on the right-hand side are to be evaluated
at $\tinit$.

\subsection{Tensor modes}
Since tensor modes behave like free scalar fields
(aside from normalization)
their initial conditions follow directly from Eq.~\eqref{eq:sigma-re-ics}.
Remembering that we need work only with the real part, the required
initial condition is
\begin{align}
  \Gamma^{\tensoridx{a}\tensoridx{b}}(\tinit)
  & =
  \frac{1}{a^3 k \Mp^2}
  \left(
    \begin{array}{c@{\hspace{3mm}}c}
      a & - aH \\
      - aH & k^2 / a
  \end{array}
  \right) ,
\end{align}
where $\Gamma^{\tensoridx{a}\tensoridx{b}}$
was defined in Eq.~\eqref{eq:Gamma-def}.
As above, all quantities on the right-hand side
are to be evaluated at $\tinit$.

\subsection{Three-point function}
To compute an initial condition for the three-point function
we use the in--in formula~\eqref{eq:inin-formula}
at tree-level
in conjunction with
the Bunch--Davies unequal-time
two-point function~\eqref{eq:2ptdeSitter}.
The interaction Hamiltonian
$\Hint$ has already been computed in
Eqs.~\eqref{eq:reducedTensorsa}--\eqref{eq:reducedTensorsc}.
As above, it is sufficient to perform the calculation to highest
order in $k\tauinit$.
This means that
we may neglect mass terms in the propagator.
The method is the same used in
standard calculations of the three-point
function~\cite{Maldacena:2002vr,Seery:2005wm,Seery:2005gb,Chen:2006nt},
but here
we extract a limit long before horizon exit
rather than long after.

\para{Example}
Consider
the equal-time
three-point function
$\langle \delta\phi^\alpha(\vect{k}_1)
\delta\phi^\beta(\vect{k}_2)
\delta\phi^\gamma(\vect{k}_3) \rangle_{\tinit}$.
The tree-level calculation involves the third-order
vertex functions
$A_{\alpha\beta\gamma}$,
$B_{\alpha\beta\gamma}$
and
$C_{\alpha\beta\gamma}$
defined in Eqs.~\eqref{eq:reducedTensorsa}--\eqref{eq:reducedTensorsc}.
Of these, $A_{\alpha\beta\gamma}$ contains a `fast' term of order
$(k/a)^2 \sim (k\tau)^2 $ that grows exponentially on subhorizon scales,
whereas the
remaining terms in
$A_{\alpha\beta\gamma}$,
and all terms in
$B_{\alpha\beta\gamma}$ and $C_{\alpha\beta\gamma}$,
are `slow' and do not grow exponentially.
Meanwhile, Eqs.~\eqref{eq:2ptdeSitter} and
\eqref{eq:sigma-re-ics}--\eqref{eq:sigma-im-ics}
show that on subhorizon scales the wavefunction for a field
perturbation grows like $(1-\im k \tau) \e{\im k\tau} \sim k\tau$,
whereas the wavefunction for a momentum perturbation
behaves like its $t$ derivative $\sim H k^2 \tau^2 \e{\im k\tau}$.%
  \footnote{Recall that in the interaction picture we should use
  the relation $\delta \pi^\alpha = \delta{\dot{\phi}}^\alpha$
  to express the momentum perturbation in terms of the creation and
  annihilation operators for $\delta\phi^\alpha$,
  and not Eq.~\eqref{eq:newp}.}

These estimates enable us to decide which terms should be retained
when computing correlations at $\tinit$.
For example,
on subhorizon scales,
the fast part of
the $A_{\alpha\beta\gamma}$ vertex integral will scale like
\begin{equation}
  \text{fast $A_{\alpha\beta\gamma}$}
  \sim
  \int_{-\infty^+}^{\tauinit}
  \d \eta \; a(\eta)^2
  f_{k_1}(\eta)
  f_{k_2}(\eta)
  f_{k_3}(\eta)
  \sim
  \int_{-\infty^+}^{\tauinit}
  \frac{\d \eta}{H^2 \eta^2}
  k_1 k_2 k_3 \eta^3 \e{\im k_t \eta} .
\end{equation}
The $\eta$ integral can be estimated by the usual method of asymptotic
expansion~\cite{Maldacena:2002vr,Seery:2005gb,Burrage:2011hd,Dias:2012qy},
yielding
\begin{equation}
  \int_{-\infty^+}^{\tauinit}
  \d \eta \; \eta \e{\im k_t \eta}
  =
  -\frac{\im}{k_t^3}
  \Big(
    k_t \tauinit \e{\im k_t \tauinit} + \Or(k_t \tauinit)^0
  \Big) .
  \label{eq:time-integral-estimate}
\end{equation}
There is no contribution from the lower limit of integration
because it is killed by the contour deformation prescription that
implements adiabatic switch-on of the interaction.
For sufficiently early $\tinit$
we need retain only
the term proportional
to $\tauinit$ in Eq.~\eqref{eq:time-integral-estimate}.
By comparison any lower-order terms are exponentially suppressed.

The vertex integral for the
slow part of $A_{\alpha\beta\gamma}$
yields terms of order
$\Or(k_t \tauinit)^{-1}$,
making it negligible in many models.
However, these slow terms must be included
in QSFI-like scenarios where $V_{\alpha\beta\gamma}/H \sim 1$
and therefore $A_{\alpha\beta\gamma} \sim H$.
This generates a
large contribution that can compensate for suppression by
$(k_t \tauinit)^2$,
at least if $\tauinit$ is selected to lie a modest number of e-folds before
horizon exit.
We will
see in~\S\ref{sec:massless-efolds}
that
this is usually the case for practical numerical work.
The conclusion is that, in QSFI-like scenarios, both fast and slow
contributions to the initial
condition from $A_{\alpha\beta\gamma}$ are comparable.

To accommodate this, it is safest to retain the highest
power of $k_t \tauinit$
from each of the four distinct contributions
(fast-$A$, slow-$A$, $B$, $C$).
In simple models the highest dimension operators are typically
most relevant, and therefore the fast-$A$ and $C$-type
contributions dominate the initial condition.
These are dimension-5 operators.
In QSFI-type models the
slow-$A$ (dimension-3) contributions are also relevant
due to their large coupling,
as explained above.
In most models the
$B$-type (dimension-4) contributions are irrelevant,
but they are easily kept
and therefore
we retain them as a precaution.
This allows us to
successfully handle
models where a
tuned combination of couplings makes them competitive with the
dimension-5 contributions.

To illustrate the method we continue with dimension-5 operators
only,
although our final results~\eqref{eq:bsp-re-fff}--\eqref{eq:bsp-re-ppp}
contain contributions from
operators of all types.
When
$|k_i \tau| \gg 1$ for all relevant scales $k_i$,
the dimension-5 contribution
to the initial condition is
\begin{equation}
\begin{split}
  B_{\Re,\text{dim-5}}^{\alpha\beta\gamma}
  \approx
  \mbox{}
  &
  \frac{\im \Hinit^3}{8 (k_1 k_2 k_3)}
  \tauinit^3 \e{-\im k_t \tauinit}
  \int_{-\infty^+}^{\tauinit}
  \d \eta \; H\eta  \e{\im k_t \eta}
  \Big(
    \frac{\pi^\alpha}{4H \Mp^2}
    \delta^{\beta\gamma}
    (\vect{k}_2 \cdot \vect{k}_3)
    -
    \frac{C^{\alpha\beta\gamma}}{2} k_1 k_2
  \Big)
  \\
  &
  \mbox{} +
  \text{5 perms}
  +
  \text{c.c.} ,
\end{split}
\end{equation}
where the permutations are produced by simultaneous
cyclic exchange
of the indices $\alpha$, $\beta$, $\gamma$
and their corresponding wavevectors $\vect{k}_1$, $\vect{k}_2$, $\vect{k}_3$,
and `c.c.' denotes the complex conjugate of the entire
expression.%
    \footnote{In this section we are not making use of the
    extended summation convention described on p.\pageref{eq:extended-summation-example},
    and therefore we revert to labelling the momenta
    as $\vect{k}_1$, $\vect{k}_2$, $\vect{k}_3$
    to avoid ambiguity in combinations such as
    $C^{\alpha\beta\gamma} k_1 k_2$.}
Note that this makes the tree-level three-point function automatically real.

If the time-dependent background quantities
$H$, $\dot{\phi}^\alpha$ (and so on)
are slowly varying near $\tauinit$---as we are assuming---then
we may expand them in a Taylor series.
The exponential decay of the integrand
produced by the contour deformation at large $|\eta|$
implies that high-order
terms in the Taylor expansion make small contributions to the
integral~\cite{Burrage:2011hd,Dias:2012qy},
and to leading order we need only the first term.
In practice this means we can evaluate all
background terms at $\tauinit$ and take them outside
the $\eta$ integral.

\para{Results}
Repeating this exercise for each correlation function
we find the following results.
In each case we quote the correlation function as a initial
term plus a number of permutations
over simultaneous cyclic exchange of
$\alpha$, $\beta$, $\gamma$
and $\vect{k}_1$, $\vect{k}_2$, $\vect{k}_3$, as above.
Notice that the permutations should only be made
\emph{within the bracket} in which the sum appears;
not all terms should be symmetrized, because momentum factors from
the external wavefunctions may be asymmetric in mixed
field--momentum correlation functions.
The species labels $\alpha$, $\beta$, $\gamma$
correspond to the phase-space labels $a$, $b$, $c$,
and $\Aslow^{\alpha\beta\gamma}$ is defined to contain only the
slow terms in $A^{\alpha\beta\gamma}$.
We set
$k_t \equiv k_1 + k_2 + k_3$ equal to the perimeter of the momentum
triangle
and use the abbreviation
$K^2 \equiv k_1 k_2 + k_1 k_3 + k_2 k_3$.
\begin{subequations}
\begin{itemize}
  \item \semibold{$a$, $b$, $c$ all fields} \\
  \begin{equation}
  \begin{split}
    \label{eq:bsp-re-fff}
    B_{\Re}^{abc}(\tinit) =
    \frac{1}{4a(\tinit)^4} \frac{1}{k_1 k_2 k_3 k_t}
    \bigg(
      &
      \frac{\pi^\alpha}{4H \Mp^2} \delta^{\beta\gamma}
      \vect{k}_2 \cdot \vect{k}_3
      -
      \frac{1}{2} C^{\alpha\beta\gamma} k_1 k_2
      + \frac{a^2}{2} \Aslow^{\alpha\beta\gamma}
    \\
      &
      \mbox{}
      + \frac{a^2 H}{2} B^{\alpha\beta\gamma}
      \bigg[
        \frac{(k_1+k_2)k_3}{k_1 k_2} - \frac{K^2}{k_1 k_2}
      \bigg]
    + \text{5 perms}
    \bigg)_{t=\tinit}
  \end{split}
  \end{equation}
  We can
  estimate the size of
  the dimension-5 contributions from
  $C^{\alpha\beta\gamma} k_1 k_2$,
  which is
  roughly $k^2/\Mp$ assuming all $k_i$
  are close to a common scale $k$.
  Meanwhile, the slow-$A$ contribution
  is $a^2 A^{\alpha\beta\gamma}$.
  In a QSFI-like model this is $\sim a^2 H$.
  The relative importance of the slow-$A$ terms is then roughly
  $(a^2 H^2 / k^2) (\Mp/H) \sim (k\tauinit)^{-2} (\Mp/H)$,
  confirming the conclusion above that these contributions can be
  comparable if $\tinit$ is not too early.

  \item \semibold{$a$ momentum, $b$, $c$ fields} \\
  \begin{equation}
  \begin{split}
    \label{eq:bsp-re-ffp}
    B_{\Re}^{abc}&(\tinit) =
    \frac{1}{4a(\tinit)^4} \frac{1}{(k_1 k_2 k_3)^2 k_t}
    \\
    & \times
    \bigg(
      k_1^2 (k_2 + k_3)
      \bigg[
        \frac{\pi^\alpha}{4H\Mp^2} \delta^{\beta\gamma} \vect{k}_2 \cdot \vect{k}_3
        + \frac{a^2}{2} \Aslow^{\alpha\beta\gamma}
        -
        \frac{1}{2} C_{\alpha\beta\gamma} k_1 k_2
        + \text{5 perms}
      \bigg]
    \\
    & \qquad
    \mbox{} +
      k_1
      \bigg[
        \frac{k_1^2 k_2^2}{2} \Big( 1 + \frac{k_3}{k_t} \Big) C^{\alpha\beta\gamma}
        +
        \frac{1}{2H} B^{\alpha\beta\gamma} k_1 k_2 k_3^2
        -
        \frac{\pi^\alpha}{4H\Mp^2} \delta^{\beta\gamma}
        \Big(
          K^2 + \frac{k_1 k_2 k_3}{k_t}
        \Big)
        \vect{k}_2 \cdot \vect{k}_3
        \\
        &
        \qquad\qquad\quad
        \mbox{} -
        \frac{a^2}{2} \Aslow^{\alpha\beta\gamma}
        \Big(
            K^2 - \frac{k_1 k_2 k_3}{k_t}
        \Big)
        + \text{5 perms}
      \bigg]
    \bigg)_{t=\tinit}
  \end{split}
  \end{equation}

  \item \semibold{$a$, $b$ momenta, $c$ field} \\
  \begin{equation}
  \begin{split}
    \label{eq:bsp-re-fpp}
    B_{\Re}^{abc}&(\tinit) =
    \frac{1}{4a(\tinit)^6} \frac{1}{(k_1 k_2 k_3)^2 k_t H(\tinit)^2}
    \\
    & \times
    \bigg(
      k_1^2 k_2^2 k_3
      \bigg[
          \frac{1}{2}
          C^{\alpha\beta\gamma} k_1 k_2
          -
          \frac{\pi^\alpha}{4H \Mp^2} \delta^{\beta\gamma}
          \vect{k}_2 \cdot \vect{k}_3
          -
          \frac{a^2}{2} \Aslow^{\alpha\beta\gamma}
          -
          \frac{a^2 H}{2} B^{\alpha\beta\gamma}
          \frac{(k_1 + k_2)k_3}{k_1 k_2}
          \\
          &
          \qquad\qquad\quad
          \mbox{} +
          \text{5 perms}
      \bigg]
      +
      k_1^2 k_3^2
      \bigg[
          \frac{a^2 H}{2} B^{\alpha\beta\gamma}
          k_3
          +
          \text{5 perms}
      \bigg]
    \bigg)_{t=\tinit}
  \end{split}
  \end{equation}

  \item \semibold{$a$, $b$, $c$ all momenta} \\
  \begin{equation}
  \begin{split}
    \label{eq:bsp-re-ppp}
    B_{\Re}^{abc}(\tinit) =
    \frac{1}{4a(\tinit)^6} \frac{1}{k_1 k_2 k_3 k_t H(\tinit)^2}
    \bigg(
      &
        \frac{\pi^\alpha}{4H\Mp^2} \delta^{\beta\gamma}
        \Big[
          K^2 + \frac{k_1 k_2 k_3}{k_t}
        \Big]
        \vect{k}_2 \cdot \vect{k}_3
        -
        \frac{1}{2H} B_{\beta\gamma\alpha} k_1^2 k_2 k_3
        \\
        &
        \mbox{} -
        \frac{k_1^2 k_2^2}{2}
        \Big[
            1 + \frac{k_3}{k_t}
        \Big]
        C^{\alpha\beta\gamma}
        +
        \frac{a^2}{2} \Aslow^{\alpha\beta\gamma}
        \Big[
            K^2 - \frac{k_1 k_2 k_3}{k_t}
        \Big]
        \\
        &
        \mbox{} +
        \text{5 perms}
    \bigg)_{t=\tinit}
  \end{split}
  \end{equation}
\end{itemize}
\end{subequations}

Once these initial conditions have been obtained,
their subsequent evolution takes place on the real time axis.
As explained above,
the evolution can therefore be implemented by ordinary differential equations
without any requirement to account for contour rotation.

\section{Gauge transformation to the curvature perturbation}
\label{sec:gauge-transform}

The formalism developed in~\S\S\ref{sec:numerical-computation}--\ref{sec:initial-conditions}
yields a numerical scheme suitable to compute
the phase-space correlation functions
$\langle \delta X^{\ext{a}} \delta X^{\ext{b}} \rangle$
and
$\langle \delta X^{\ext{a}} \delta X^{\ext{b}} \delta X^{\ext{c}} \rangle$.
However, to construct inflationary observables we are
normally interested in the correlation functions of the curvature
perturbation
$\langle \zeta \zeta \rangle$
and
$\langle \zeta \zeta \zeta \rangle$.%
    \footnote{There are other possible observables of interest,
    such as those related to isocurvature perturbations. In this paper
    we focus on $\zeta$, but it is straightforward to compute correlation functions
    for isocurvature modes using the transport framework. In the same way that
    $\zeta$ can be regarded as a scaled projection
    along the instantaneous background trajectory, isocurvature perturbations
    correspond to projections onto directions orthogonal to it.
    We do not pursue the details here; see the discussion in Ref.~\cite{Seery:2012vj}.}
Since $\zeta$ and the $\delta X^a$ are related by a nonlinear gauge
transformation
the same is true for their correlation functions.
To compute the two- and three-point functions of $\zeta$
we need the details of this gauge transformation up to second order
in $\delta X^{\ext{a}}$.
Suitable expressions were obtained in Ref.~\cite{Dias:2014msa}.

It is possible to present the result in a number of different ways,
each of which are related by constraint equations.
For our purposes, a convenient real-space form is
\begin{equation}
\begin{split}
    \zeta =
        - \frac{\dot{\phi}^\alpha \delta\phi_\alpha}{2H \Mp^2 \epsilon}
    +
        \frac{1}{6H^2 \Mp^2 \epsilon}
        \bigg(
            &
            \frac{\dot{\phi}_\alpha \dot{\phi}_\beta}{\Mp^2}
                \Big[
                    - \frac{3}{2}
                    + \frac{9}{2\epsilon}
                    + \frac{3}{4\epsilon^2} \frac{V_\gamma \dot{\phi}^\gamma}{\Mp^2 H^3}
                \Big] \delta\phi^\alpha \delta\phi^\beta
    \\ & \mbox{}
            + \frac{3}{H \epsilon}
                \frac{\dot{\phi}_\alpha \dot{\phi}_\beta}{\Mp^2}
                \delta\phi^\alpha \delta\dot{\phi}^\beta
        - 3 H \partial^{-2}
            \Big[
                \partial_j \delta\dot{\phi}^\alpha \partial_j \delta\phi_\alpha
                + \delta\dot{\phi}^\alpha \partial^2 \delta\phi_\alpha
            \Big]
        \bigg) .
    \label{eq:zeta-simple-two}
\end{split}
\end{equation}
All quantities in this relation should be evaluated at the same time,
corresponding to the time of evaluation for $\zeta$.

The transformation~\eqref{eq:zeta-simple-two}
is expressed in terms of time derivatives
$\delta \dot{\phi}^\alpha$ rather than momenta $\delta \pi^\alpha$.
To account for this we must either exchange
the phase-space correlation functions
$\langle \delta X^{\ext{a}} \delta X^{\ext{b}} \rangle$,
$\langle \delta X^{\ext{a}} \delta X^{\ext{b}} \delta X^{\ext{c}} \rangle$
for correlation functions of $\delta \phi^\alpha$, $\delta\dot{\phi}^\alpha$,
\emph{or}
use Eq.~\eqref{eq:newp}
to rewrite~\eqref{eq:zeta-simple-two}
in terms of the $\delta \pi^\alpha$.
If we adopt this second strategy then
Eq.~\eqref{eq:zeta-simple-two}
is an especially simple representation because the absence of
a linear term in $\delta\dot{\phi}^\alpha$ means that the second-order
parts of~\eqref{eq:newp} are not needed.

Following this procedure and
reverting to our extended summation convention, the gauge transformation
can be written
\begin{equation}
  \zeta(\vect{k}) = N_{\ext{a}}(\vect{k}) \delta X^{\ext{a}}
    + \frac{1}{2} N_{\ext{a}\ext{b}}(\vect{k}) \delta X^{\ext{a}} \delta X^{\ext{b}} .
\end{equation}
In order to give a complete set of equations we quote results for
$N_{\ext{a}}$ and $N_{\ext{a}\ext{b}}$ explicitly.
We write
\begin{subequations}
\begin{align}
  N_{\ext{a}}(\vect{k}) & = (2\pi)^3 \delta(\vect{k} - \vect{k}_a) N_a \\
  N_{\ext{a}\ext{b}}(\vect{k}) & = (2\pi)^3 \delta(\vect{k} - \vect{k}_a - \vect{k}_b)
  N_{ab}(\vect{k}, \vect{k}_a, \vect{k}_b) .
\end{align}
\end{subequations}
The coefficient matrices $N_a$ and $N_{ab}$ satisfy
\begin{subequations}
\begin{align}
  \label{eq:Na-expr}
  N_a & =
  - \frac{\pi_\alpha}{2H \Mp^2 \epsilon}
  \left(
    \begin{array}{c}
      \displaystyle
      1 \\
      0
    \end{array}
  \right)
  \\
  \label{eq:Nab-expr}
  N_{ab} & =
  \frac{1}{3H^2 \Mp^2 \epsilon}
  \left(
    \begin{array}{c@{\hspace{3mm}}c}
      \displaystyle
      \frac{\pi_\alpha \pi_\beta}{\Mp^2}
      \Big[
        -\frac{3}{2} + \frac{9}{2\epsilon} + \frac{3}{4\epsilon^2} \frac{V_\gamma \pi^\gamma}{\Mp^2 H^3}
      \Big]
      &
      \displaystyle
      \frac{3}{H\epsilon} \frac{\pi_\alpha \pi_\beta}{\Mp^2}
      - \frac{3H}{k^2}
      \Big[
        \vect{k}_a \cdot \vect{k}_b + k_a^2
      \Big]
      \delta_{\alpha\beta} \\
      \displaystyle
      \frac{3}{H\epsilon} \frac{\pi_\alpha \pi_\beta}{\Mp^2}
      - \frac{3H}{k^2}
      \Big[
        \vect{k}_a \cdot \vect{k}_b + k_b^2
      \Big]
      \delta_{\alpha\beta}
      &
      0
    \end{array}
  \right)
\end{align}
\end{subequations}
Notice that, in this representation, $N_a$ is independent of wavenumber.
We define the spectrum and bispectrum of $\zeta$ by the rules
\begin{subequations}
\begin{align}
  \langle \zeta(\vect{k}_1) \zeta(\vect{k}_2) \rangle
  & =
  (2\pi)^3 \delta(\vect{k}_1 + \vect{k}_2) P(k) \\
  \langle \zeta(\vect{k}_1) \zeta(\vect{k}_2) \zeta(\vect{k}_3) \rangle
  & =
  (2\pi)^3 \delta(\vect{k}_1 + \vect{k}_2 + \vect{k}_3) B(k_1, k_2, k_3) .
\end{align}
\end{subequations}
To compute these from the phase-space correlation functions we should use
(assuming the time of evaluation to be sufficiently late that any
imaginary components of the two-point function have decayed)
\begin{subequations}
\begin{align}
  \label{eq:zeta-spectrum}
  P(k) & = N_a N_b \Sigma^{ab}_{\Re}(k) \\
  \label{eq:zeta-bispectrum}
  B(k_1, k_2, k_3) & = N_a N_b N_c B_{\Re}^{abc}(k_1, k_2, k_3)
  +
  \Big(
    N_a N_b N_{cd}(\vect{k}_3, \vect{k}_1, \vect{k}_2)
    \Sigma^{ac}_{\Re}(k_1) \Sigma^{bd}_{\Re}(k_2)
    + \text{2 cyclic}
  \Big) .
\end{align}
\end{subequations}

\para{Tensor fraction}%
The tensor power spectrum is defined by analogy with the scalar spectrum,
\begin{equation}
    \langle
        \gamma_{ij}(\vect{k}_1)
        \gamma^{ij}(\vect{k}_2)
    \rangle
    \equiv
    (2\pi)^3
    \delta(\vect{k}_1+\vect{k}_2)
    P_\gamma .
\end{equation}
Using the normalization condition
$\tr (\bm{e}^s \cdot \bm{e}^{s'}) = 2\delta^{ss'}$
it follows that
\begin{equation}
    \langle
        \gamma_{ij}(\vect{k}_1)
        \gamma^{ij}(\vect{k}_2)
    \rangle = \sum_s \sum_{s'}  \langle
        \gamma_s(\vect{k}_1)
        \gamma_s(\vect{k}_2)
    \rangle
    \tr (\bm{e}^s \cdot \bm{e}^{s'} )
    =
    2 \sum_s \langle
        \gamma_{s}(\vect{k}_1)
        \gamma_{s}(\vect{k}_2) \rangle .
\end{equation}
Each polarization has the same amplitude and therefore the final
result is $P_\gamma = 4 \langle \gamma_+ \gamma_+ \rangle =
4 \langle \gamma_\times \gamma_\times \rangle$.
The tensor-to-scalar ratio $r$ is
defined by
\begin{equation}
  r \equiv \frac{P_\gamma}{P_\zeta} = \frac{4 \Gamma^{\gamma\gamma}}{P_{\zeta}} ,
\end{equation}
where $\Gamma^{\tensoridx{a}\tensoridx{b}}$ is the tensor two-point function
defined in Eq.~\eqref{eq:Gamma-def}.

\section{The {\PyTransport} and {\CppTransport} codes}
\label{sec:automated-codes}

We have now assembled the formalism
necessary
to build automated
tools
that compute the
two- and three-point functions
of phase-space perturbations $\delta X^a$
or the curvature perturbation $\zeta$.
In summary, the steps involved are:
\begin{enumerate}
  \item Information about the model must be provided---specifically, the number
  of fields, the form of the potential, the background initial
  conditions and the total integration time.
  In {\PyTransport} this is done by writing a Python script that builds the
  potential as a {\SymPy} expression.
  In {\CppTransport} it is achieved by writing a separate `model description file'
  and passing it to a specialized translation tool.

  \item In the case of {\CppTransport}, the automated tool
  uses this information to symbolically compute the
  tensors $A_{\alpha\beta\gamma}$, $B_{\alpha\beta\gamma}$
  and $C_{\alpha\beta\gamma}$. These are used to
  build expressions for ${u^{\ext{a}}}_{\ext{b}}$
  and ${u^\ext{a}}_{\ext{b}\ext{c}}$.
  In the case of {\PyTransport} the automated tool symbolically computes only the
  derivatives of the  potential.

  \item In {\CppTransport} these symbolic expressions are used to construct
  specialized {\CC}
  code that solves the ordinary differential
  equations~\eqref{eq:twopf-transport-ode}
  and~\eqref{eq:threepf-transport-ode}.
  Further specialized code is generated
  to compute the initial conditions~\eqref{eq:sigma-re-ics}--\eqref{eq:sigma-im-ics}
  and~\eqref{eq:bsp-re-fff}--\eqref{eq:bsp-re-ppp},
  and the gauge transformation matrices~\eqref{eq:Na-expr}--\eqref{eq:Nab-expr}.

  This specialized code is generated by the translation tool.
  The platform provides an object-based interface that is used to define
  specific tasks, such as the computation of an $n$-point function
  over a fixed range of wavenumber configurations.
  These tasks are written into a database,
  after which they can be executed. No specific logic is needed to control how
  the calculation is performed.

  In {\PyTransport} specialized {\CC} code
  to compute the derivatives of the potential
  is generated by a Python script.
  The tensors $A_{\alpha\beta\gamma}$, $B_{\alpha\beta\gamma}$
  and $C_{\alpha\beta\gamma}$ and all other quantities
  are hard-coded in terms of the potential
  and its derivatives. Finally, the {\CC} code
  is compiled into a Python module
  and thereafter can be included in further Python scripts that use its
  functionality to compute $n$-point functions.

  \item To solve for a specific $n$-point function
  we should determine how early it is necessary to set
  the initial conditions in order that the massless approximation
  is sufficiently good.
  In {\CppTransport} this calculation is automated, whereas in
  {\PyTransport} Python scripts are provided
  to perform this task.
  The compiled {\CC} code is used to produce suitable initial
  conditions
  and then to evolve the correlation functions
  until some specified final time.

  Computation of the three-point function requires knowledge
  of the background
  inflationary trajectory
  and the two-point functions
  that act as source terms in Eq.~\eqref{eq:threepf-transport-ode}.
  In principle both of these
  could be pre-computed over the required range of
  scales,
  but in practice the integration time is normally dominated by evolution of the
  three-point function.
  Therefore we simultaneously evolve the
  background together with the
  real and imaginary parts of the two-point function
  for each of the scales $k_1$, $k_2$, $k_3$.

  \item Eqs.~\eqref{eq:Na-expr}--\eqref{eq:Nab-expr} are used to convert the phase-space
  correlation functions into $\zeta$ correlators.
  At this stage it is possible to extract derived quantities such as the spectral
  index of the power spectrum $P(k)$
  or the reduced bispectrum
  $(6/5) \fNL(k_1, k_2, k_3) = B(k_1, k_2, k_3) / [ P(k_1)P(k_2) + P(k_1)P(k_3) + P(k_2)P(k_3) ]$.
  Alternatively the $\zeta$ bispectrum can be used to compute
  an amplitude associated with one of the standard bispectrum shapes
  such as the `local' or `equilateral' templates.
\end{enumerate}

\para{Obtaining the codes}
{\PyTransport} and {\CppTransport}
can be downloaded
from the website \href{http://transportmethod.com}{transportmethod.com}.
Additionally,
releases of both platforms are archived
at the \href{https://zenodo.org}{zenodo.org} repository.
At the time of writing the current version
of {\PyTransport} is
\href{https://zenodo.org/record/61265}{v1.0}
and the current version of {\CppTransport} is
\href{https://zenodo.org/record/61237}{2016.3}.

{\CppTransport} can alternatively be downloaded
from the development repository
hosted at
\href{https://github.com/ds283/CppTransport}{GitHub}.
This makes it possible to obtain pre-release versions of the software.

\para{Plots included in this paper}
In~\S\ref{sec:examples} we give a number of examples that illustrate
the usefulness of our framework,
and the capabilities of the {\PyTransport} and {\CppTransport}
platforms specifically.
In each case we have verified that both platforms give equivalent
results, but the plots we include were generated
by {\CppTransport}.

\begin{itemize}
    \item For {\CppTransport} the plots were generated using
    the 2016.3 release,
    or platform revision
    \GitRevision{36ac30da}{https://github.com/ds283/CppTransport/commit/36ac30da15035df144d92a803b3f5d48d10bbf26}.
    Source code for the plots can be obtained from a separate
    GitHub repository
    using revision
    \GitRevision{7ef3852f}{https://github.com/ds283/transport-paper/commit/7ef3852ff059325d5d15bb23a5902fbfcfc087ed}.

    \item For {\PyTransport} the plots were generated using platform
    revision release version 1.0, and are stored in a folder
    that accompanies the release.
    The code used to generate the plots is contained in
    subfolders of the \emph{Examples} folder that accompanies
    the release.
\end{itemize}

\section{Numerical Examples}
\label{sec:examples}

In this section we illustrate the utility of our method by computing the
two- and three-point functions for a collection of example theories.

There are several ways to present numerical solutions
for the spectrum and bispectrum.
For the two-point function
we normally plot the `dimensionless' spectrum
$\dimlessP$, defined by
\begin{equation}
    \dimlessP(k) = \frac{k^3}{2\pi^2} P(k) ,
\end{equation}
where $P(k)$ is the $\zeta$ power spectrum defined in
Eq.~\eqref{eq:zeta-spectrum}.
For the three-point function we plot
either the dimensionless bispectrum
\begin{equation}
    \dimlessB(k_1, k_2, k_3) = (k_1 k_2 k_3)^2 B(k_1, k_2, k_3) ,
    \label{eq:dimensionless-B-def}
\end{equation}
or the `reduced bispectrum',
\begin{equation}
    \frac{6}{5} \fNL(k_1, k_2, k_3) = \frac{B(k_1, k_2, k_3)}{P(k_1)P(k_2) + P(k_1)P(k_3) + P(k_2)P(k_3)} .
    \label{eq:reduced-B-def}
\end{equation}
The dimensionless bispectrum is sometimes called the
`shape function' (up to an irrelevant normalization)
and denoted $S$~\cite{Fergusson:2008ra}.
In both~\eqref{eq:dimensionless-B-def} and~\eqref{eq:reduced-B-def}
$B(k_1, k_2, k_3)$ is the $\zeta$
bispectrum of Eq.~\eqref{eq:zeta-bispectrum}.
The dimensionless bispectrum is a measure
of the three-point function alone, whereas
the reduced bispectrum
measures the \emph{relative} nonlinearity
between the two- and three-point functions.
If the bispectrum is generated
by a quadratic contribution
$\zeta = \zeta_g + (3/5) \fNLlocal \zeta_g^2$
(where $\zeta_g$ is a Gaussian random field)
then $\fNL(k_1, k_2, k_3)$ is equal to $\fNLlocal$,
but more generally it will depend on the wavenumber configuration.

To express this configuration dependence it is convenient to
distinguish between the \emph{scale} and \emph{shape}
of the triangle $\vect{k}_1 + \vect{k}_2 + \vect{k}_3$ formed from
the individual wavenumbers.
The scale can be measured by the perimeter $k_t = k_1 + k_2 + k_3$ of the
triangle.
To measure the shape we use the parameters
$\alpha$ and $\beta$ introduced by Fergusson \& Shellard~\cite{Fergusson:2006pr},
\begin{subequations}
\begin{align}
    k_1 & = \frac{k_t}{4}(1 + \alpha + \beta) \\
    k_2 & = \frac{k_t}{4}(1 - \alpha + \beta) \\
    k_3 & = \frac{k_t}{2}(1-\beta)
\end{align}
\end{subequations}
The allowed values of $(\alpha, \beta)$
fall inside the triangle
with vertices
$(-1,0)$, $(1,0)$
and $(0,1)$.
At fixed $k_t$
isosceles configurations have $\alpha=0$
and the equilateral configuration is $\alpha=0$, $\beta=1/3$.
Finally, the vertices
correspond to `squeezed limits'
where one $k_i$ becomes much smaller than the other two.

\subsection{Axion-quartic model: local-mode bispectrum}
\label{sec:axion-quartic}
The axion-quartic model was introduced by Elliston
et al.~\cite{Elliston:2012wm}
as an analytically tractable proxy for
a large-$N$ $N$-flation model where some
fields have initial conditions close to the
hilltop of an axionic potential~\cite{Kim:2010ud,Kim:2011jea}.
In this region the potential has a large
negative $\eta$ parameter
that is communicated to the final bispectrum,
even though the slow-roll parameter $\epsilon = - \dot{H}/H^2$
and its derivatives are small.
The model produces a mostly local-mode bispectrum
with $\fNLlocal \sim \Or(10)$.

This is a two-field model
with potential
\begin{equation}
    V = \frac{1}{4}g \phi^4 +
    \Lambda^2
    \Big(
        1 - \cos \frac{2\pi \chi}{f}
    \Big) .
    \label{eq:axion-quartic}
\end{equation}
The field $\chi$ represents the degree of freedom whose initial
position lies closest to the cosine hilltop,
whereas $\phi$ represents the aggregate effect of
the other fields~\cite{Elliston:2012wm}.
A similar potential had been studied earlier in
Ref.~\cite{Elliston:2011dr}, but Eq.~\eqref{eq:axion-quartic}
has the advantage of producing a spectrum with
acceptable tilt.
In our numerical results we choose
$g = 10^{-10}$,
$\Lambda^4 = (25/2\pi)^2 g \Mp^4$
and
$f = \Mp$.
The initial conditions are
$\phi = 23.5 \Mp$ and
$\chi = f/2 - 10^{-3}\Mp$,
and
the initial field derivatives are determined by the
slow-roll approximation.

\begin{figure}
    \begin{center}
        \includegraphics[scale=0.3]{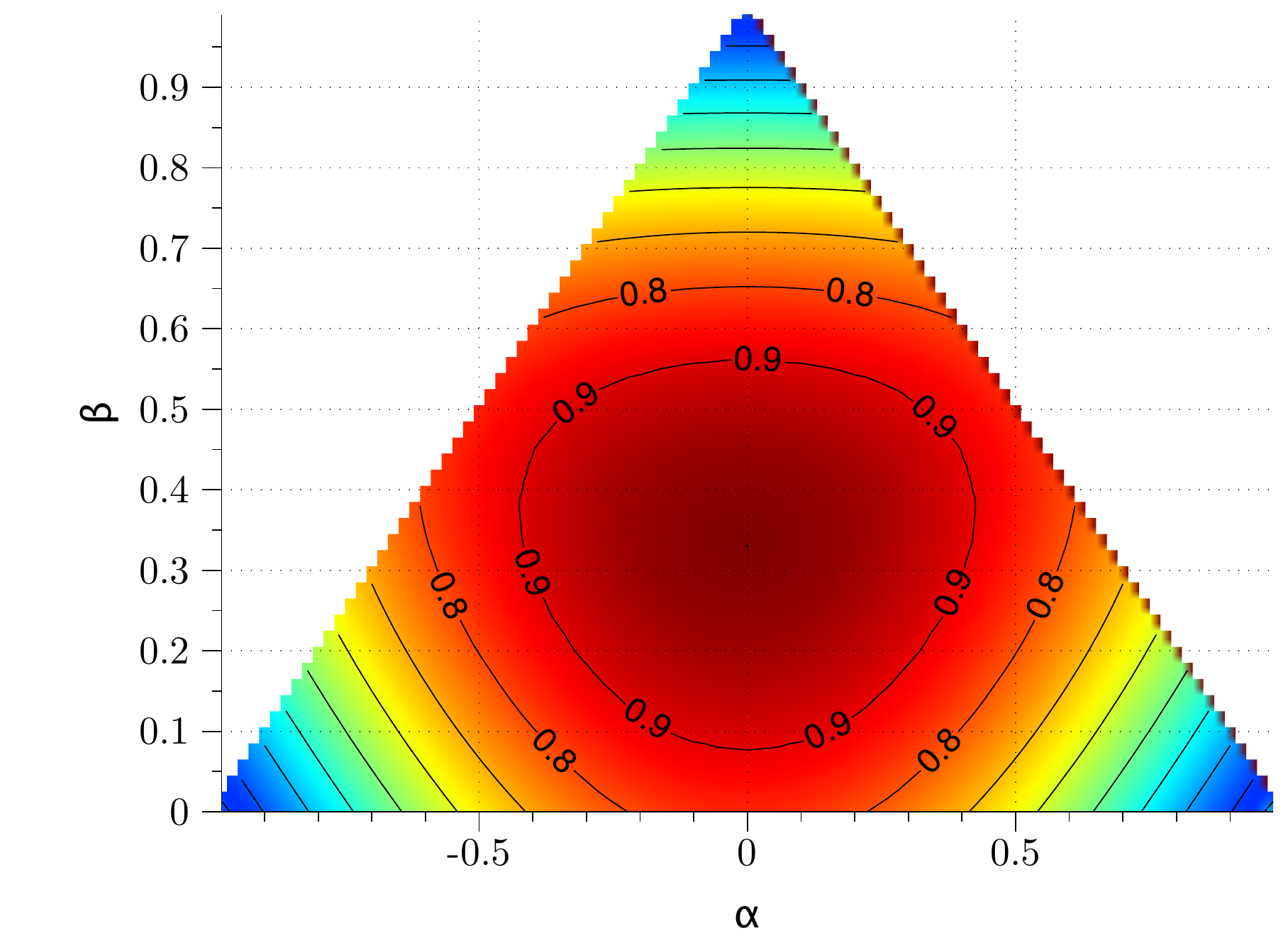}
        \qquad
        \includegraphics[scale=0.2]{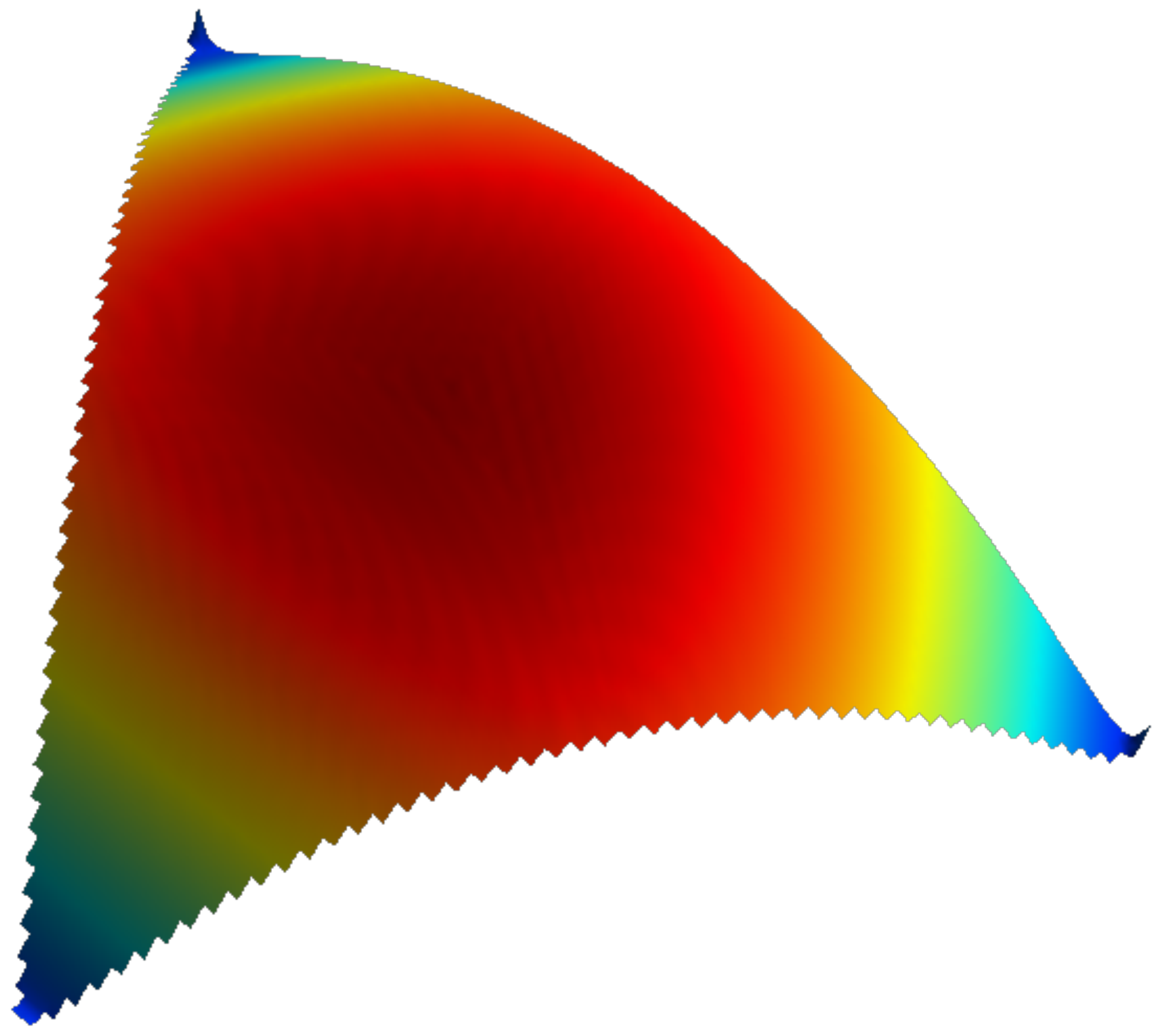}
    \end{center}
    \vspace{2mm}
    \begin{center}
        \includegraphics[scale=0.3]{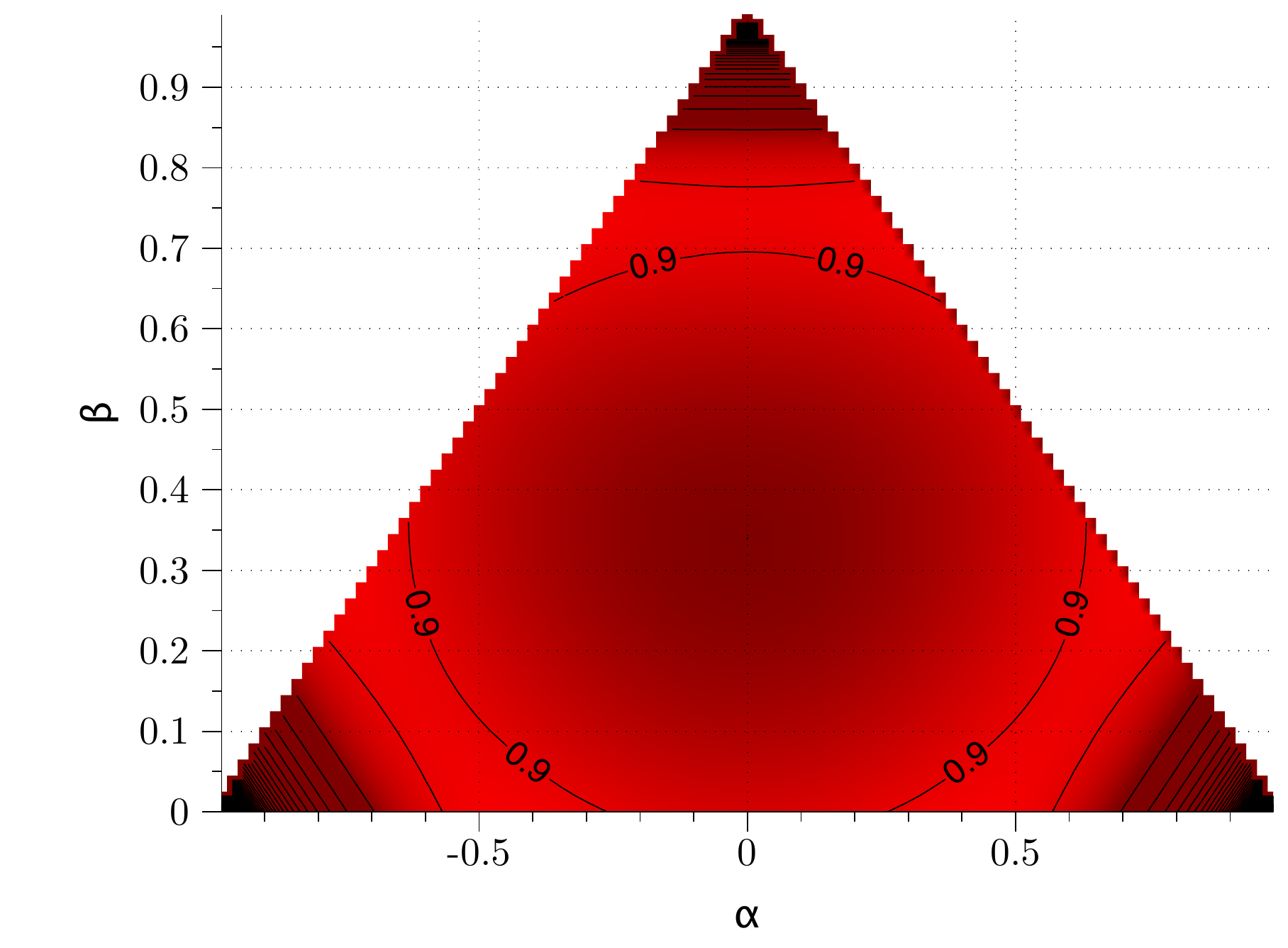}
        \qquad
        \includegraphics[scale=0.2]{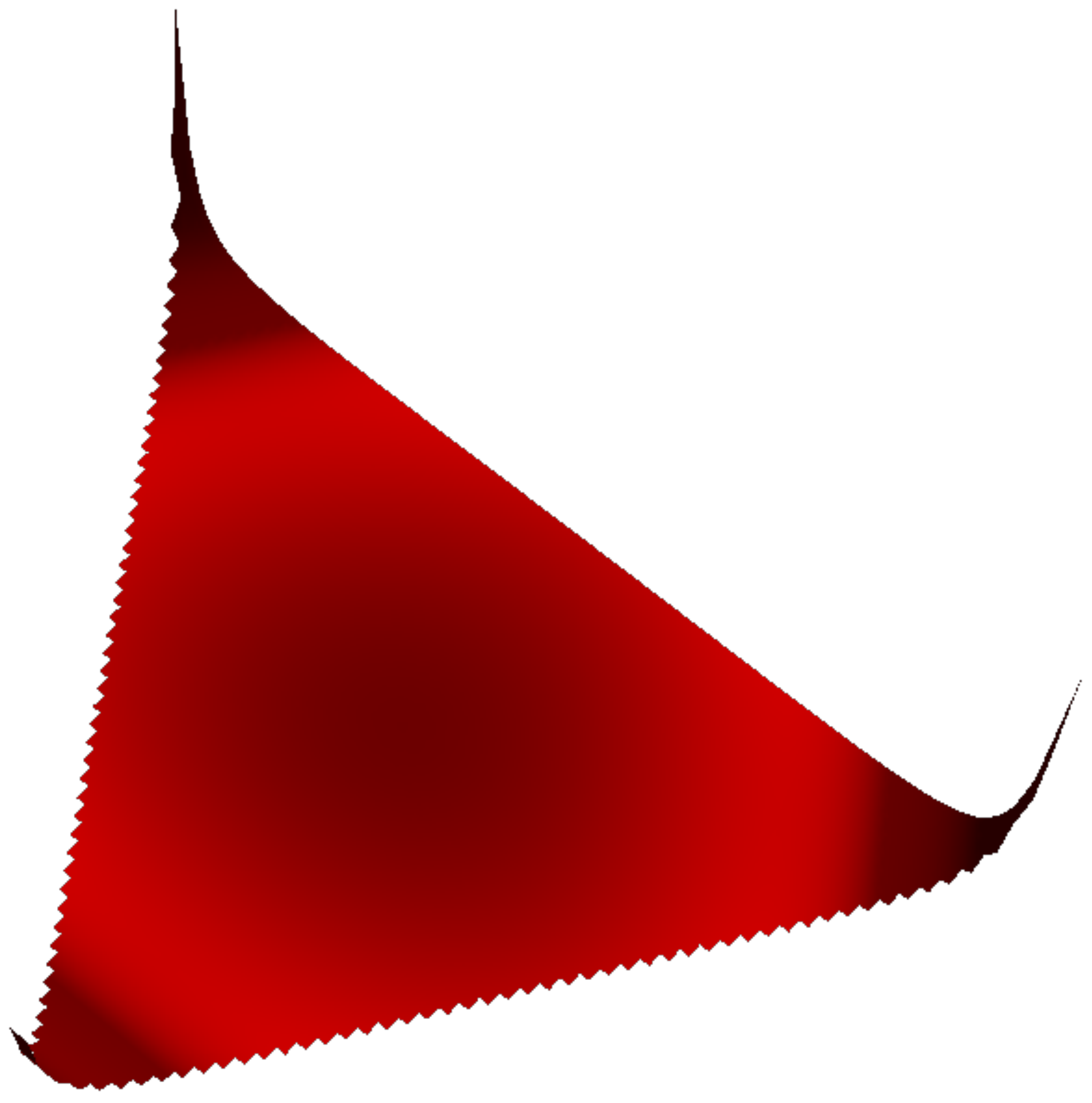}
    \end{center}
    \vspace{2mm}
    \begin{center}
        \includegraphics[scale=0.3]{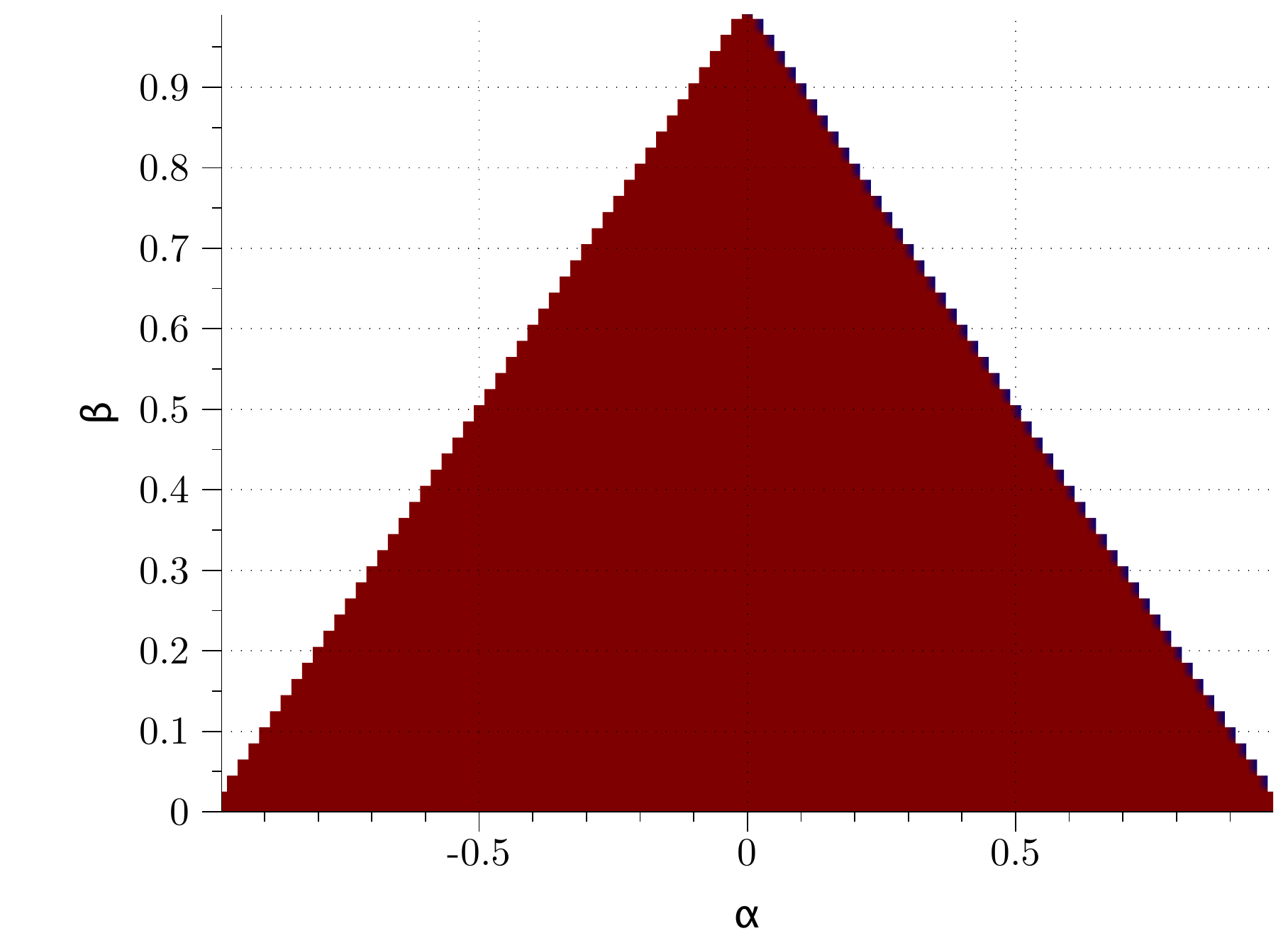}
        \qquad
        \includegraphics[scale=0.2]{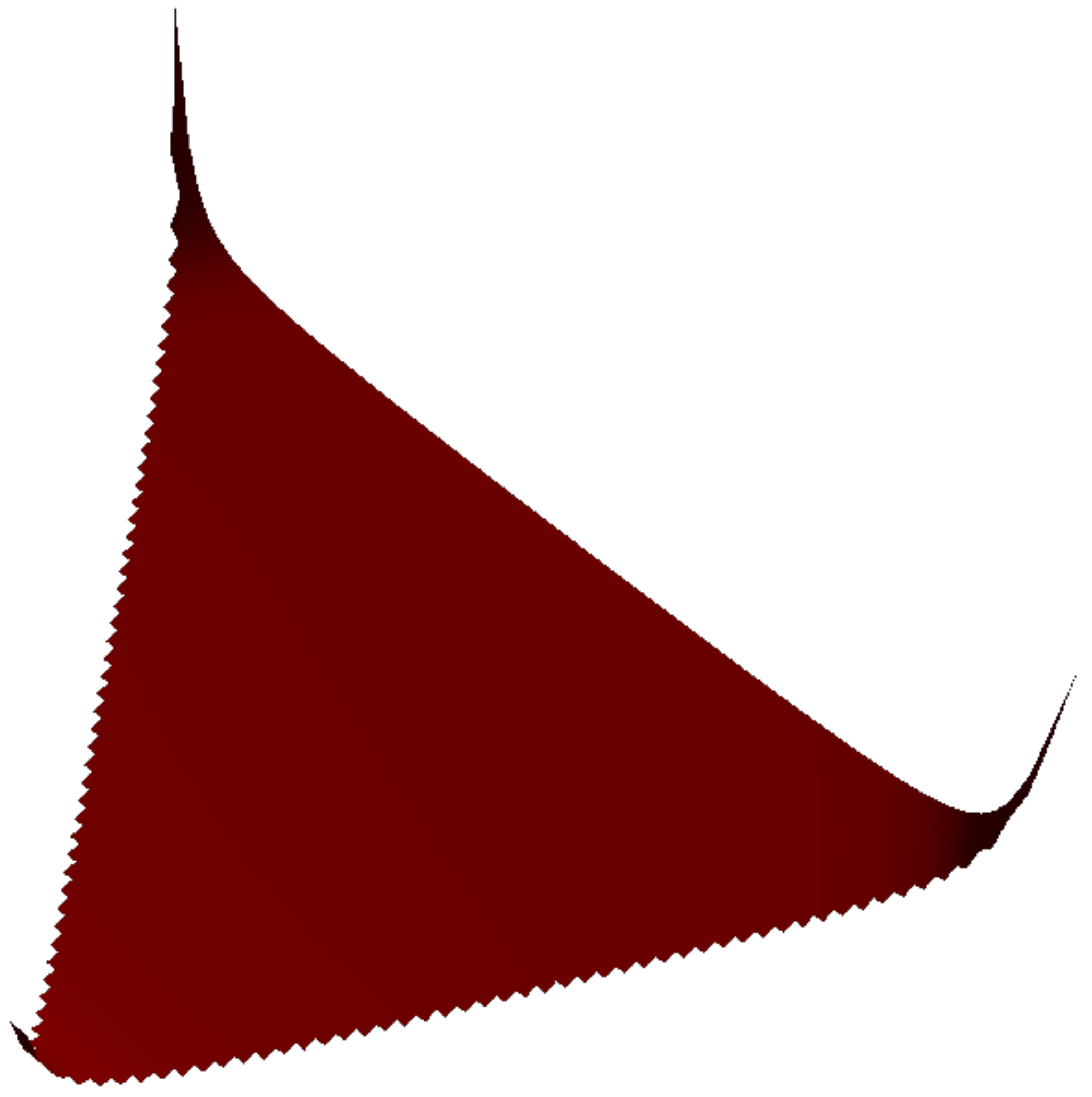}
    \end{center}
    \caption{\label{fig:shape-evolve}Evolution of the dimensionless
    bispectrum
    $\dimlessB(\alpha, \beta)$ at fixed $k_t$ in the axion-quartic
    model~\eqref{eq:axion-quartic}.}
\end{figure}

\para{Bispectrum shape}
In Fig.~\ref{fig:shape-evolve} we plot the evolution of the bispectrum
shape
as the wavenumbers move
from subhorizon to superhorizon scales.
The panels show the dimensionless
bispectrum $\dimlessB(k_1, k_2, k_3)$ as a function of the shape
parameters $\alpha$ and $\beta$ at fixed $k_t$,
scaled to have unit amplitude at the equilateral point~\cite{Fergusson:2008ra}.
The value of $k_t$ is chosen so that the time of horizon
exit for $k_t/3$ is $14.0$ e-folds later than the initial conditions.
Notice that these plots have a six-fold redundancy
corresponding to the $3!$ permutations of the momenta $\vect{k}_1$, $\vect{k}_2$,
$\vect{k}_3$ that leave
$\langle \zeta(\vect{k}_1) \zeta(\vect{k}_2) \zeta(\vect{k}_3) \rangle$
invariant.
Therefore only one-sixth region of each triangle contains independent information,
although we follow Ref.~\cite{Fergusson:2008ra} in plotting the whole triangle
to aid visualization.

The top panel shows the shape at early times,
where the short wavenumbers represented by points near the
centre of the triangle are on subhorizon scales
and the correlations should effectively be those of
quantum fluctuations in Minkowski space.
The squeezed limits
near $(\alpha,\beta) = (1,0)$, $(-1,0)$ and $(0.1)$ are nearly empty
because long--short correlations are absent.
Only short--short correlations are significant, giving the bispectrum
an `equilateral-like' appearance
peaking around the equilateral configuration $(\alpha,\beta) = (0,1/3)$.

Eventually, the central short modes approach their horizon exit point
and become aware of the cosmological background.
At this time long--short correlations begin to develop,
pushing up the amplitude near the squeezed limits;
see middle panel.
(The earliest stages of this
process are already visible in the top panel, where there
is a tiny upturn near the vertices of the triangle.)
Finally, when the short modes are well outside the horizon,
the bispectrum approaches the `local-like'
shape shown in the bottom panel.

\begin{figure}
    \begin{center}
        \includegraphics[scale=0.3]{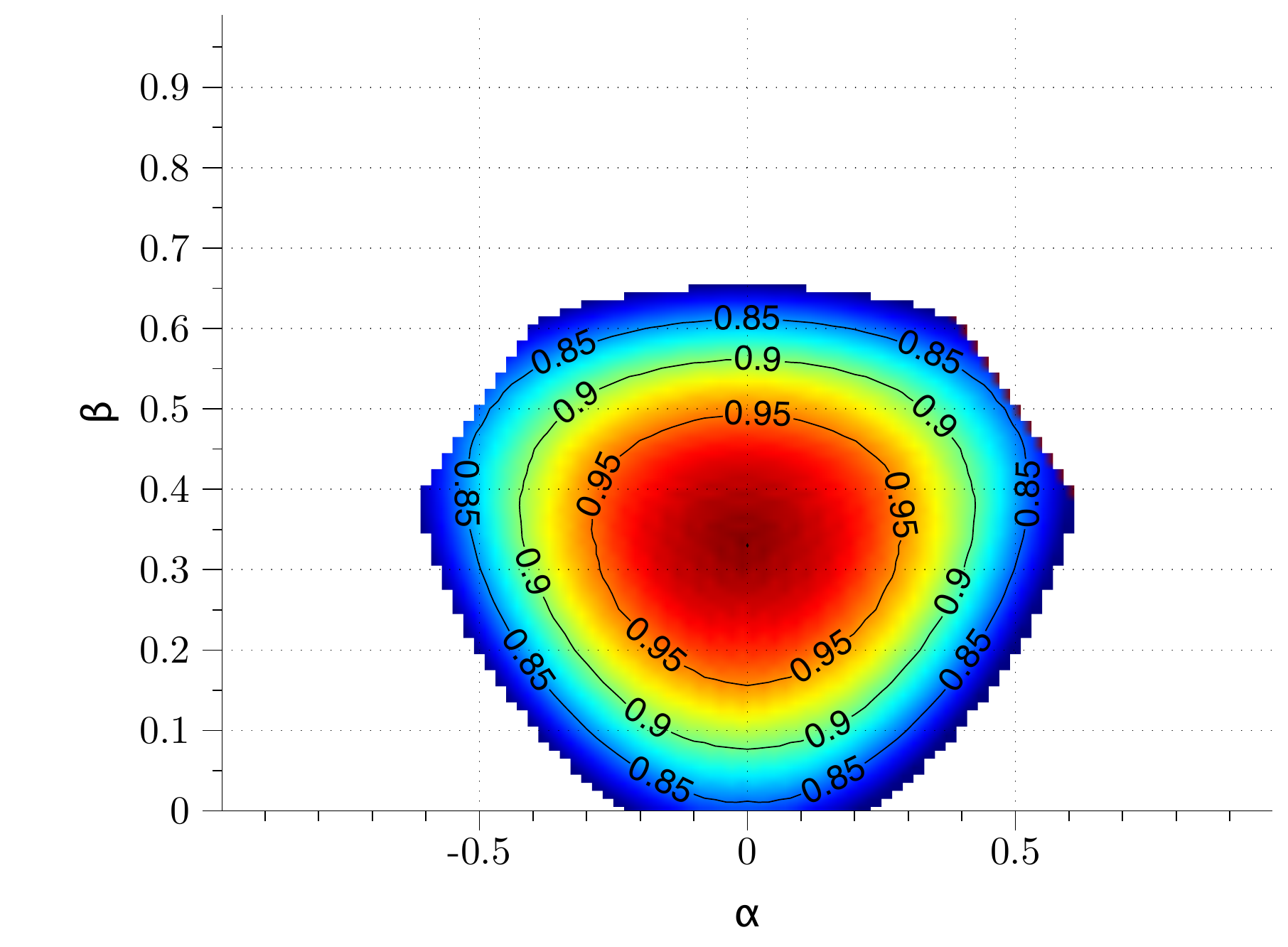}
        \qquad
        \includegraphics[scale=0.28]{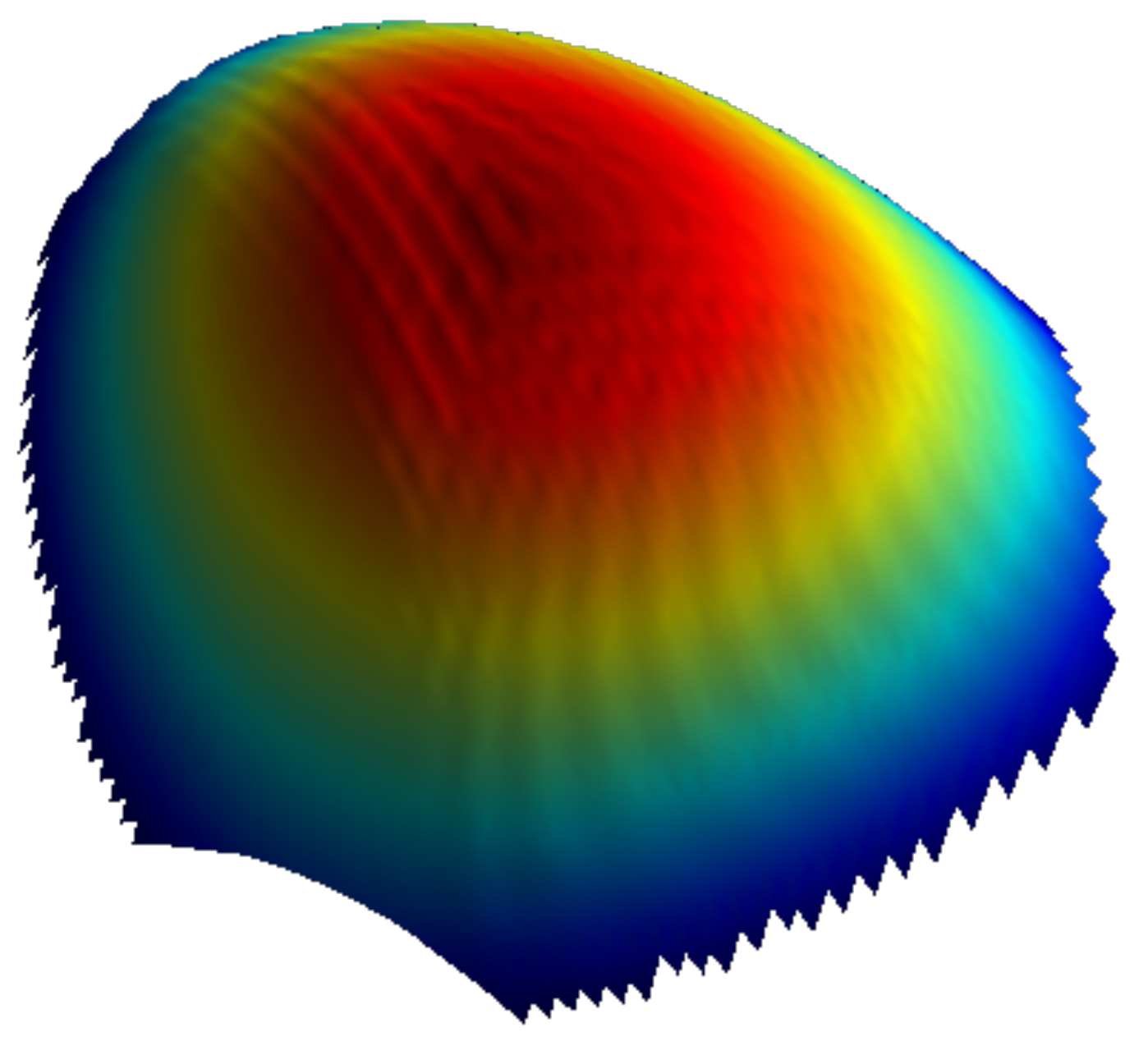}
    \end{center}
    \vspace{2mm}
    \begin{center}
        \includegraphics[scale=0.3]{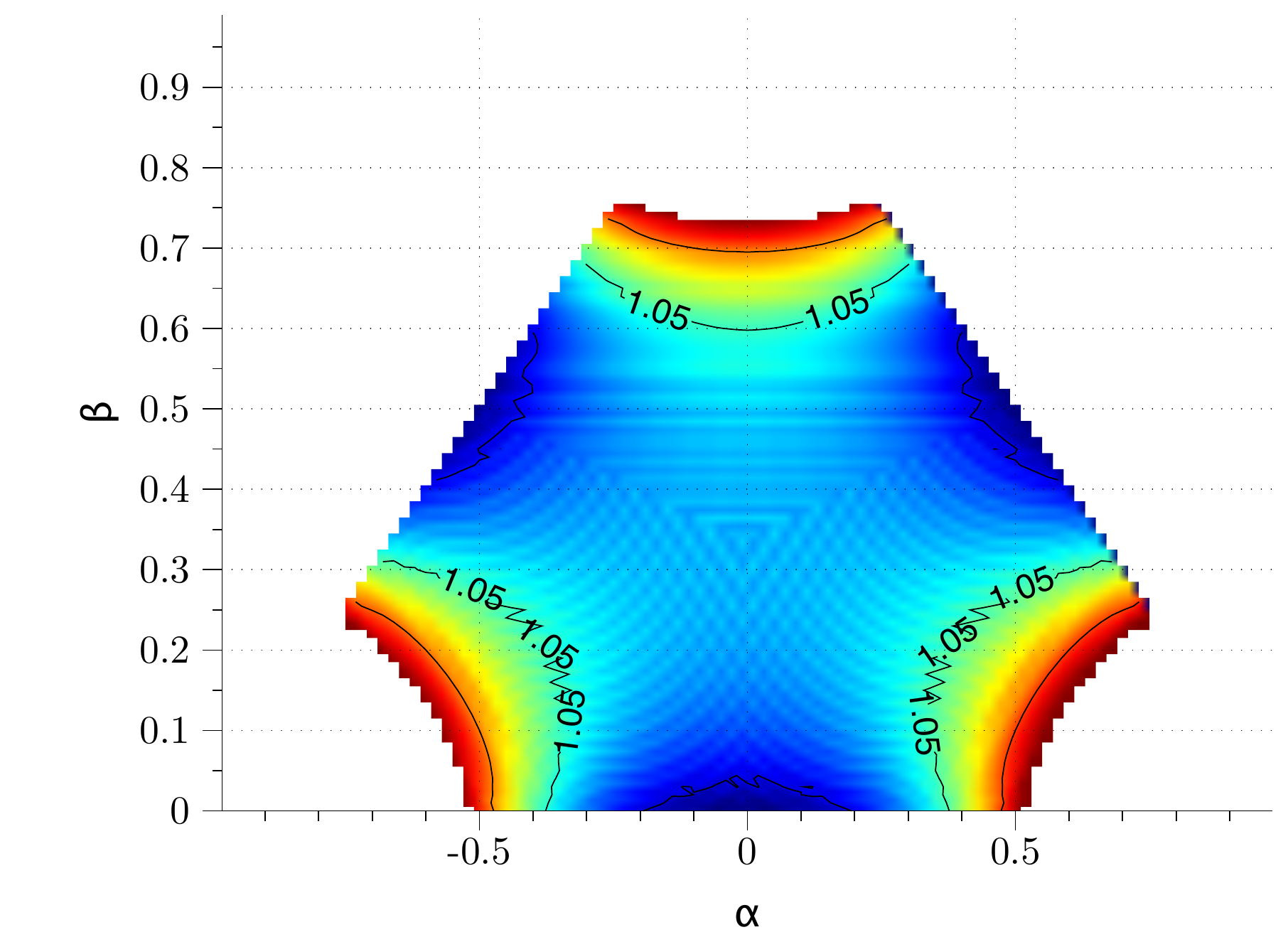}
        \qquad
        \includegraphics[scale=0.22]{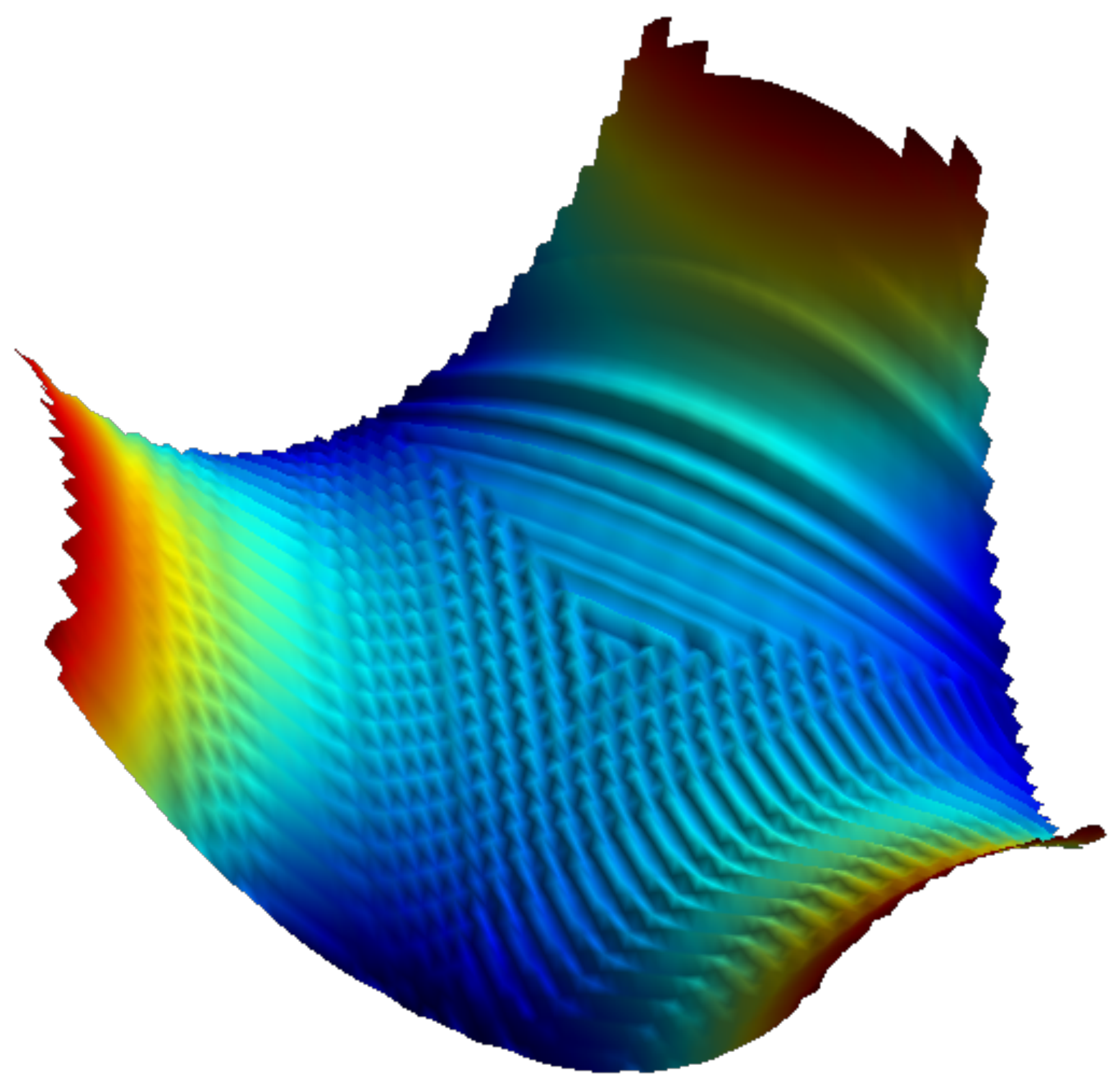}
    \end{center}
    \vspace{2mm}
    \begin{center}
        \includegraphics[scale=0.3]{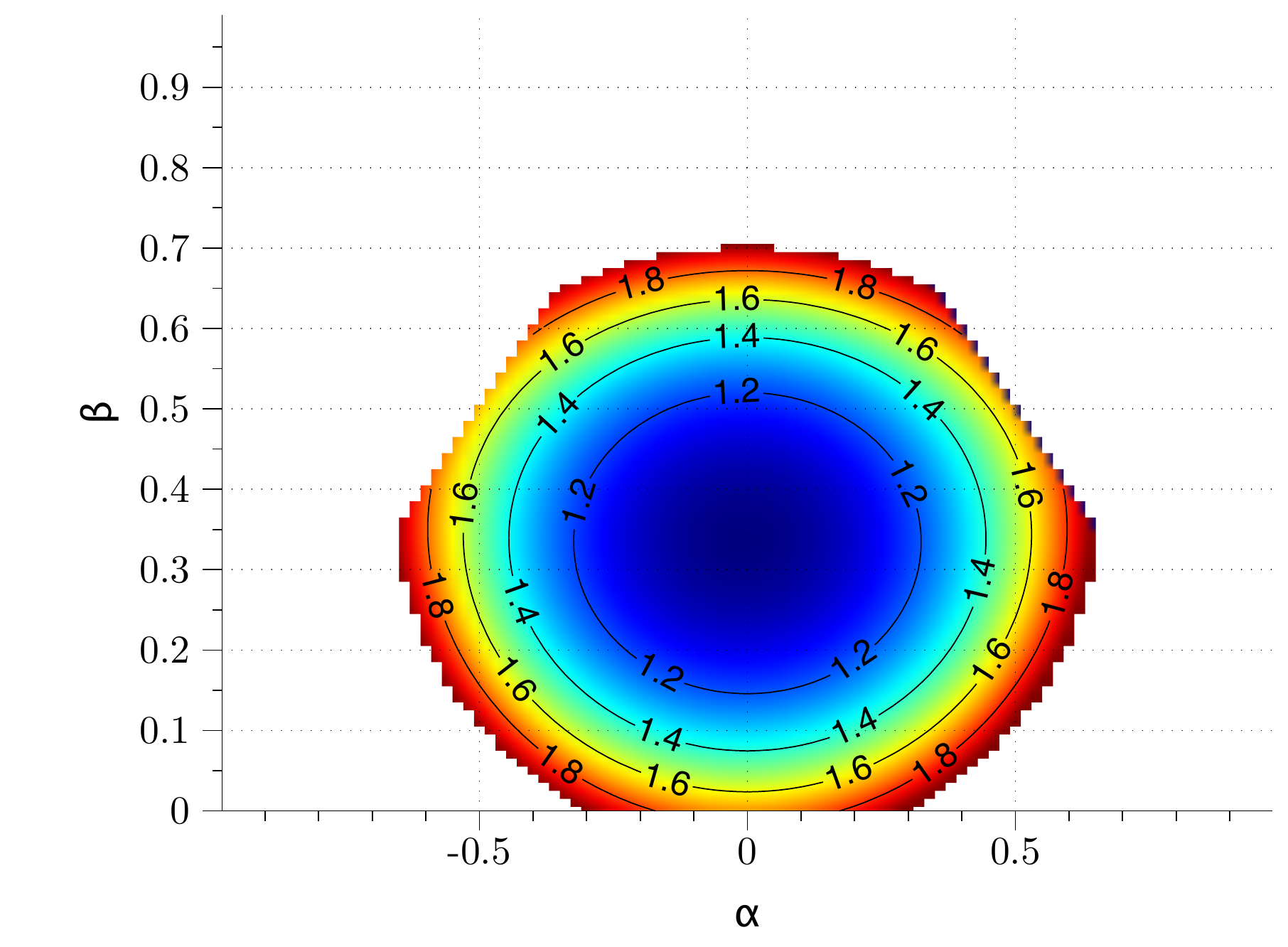}
        \qquad
        \includegraphics[scale=0.26]{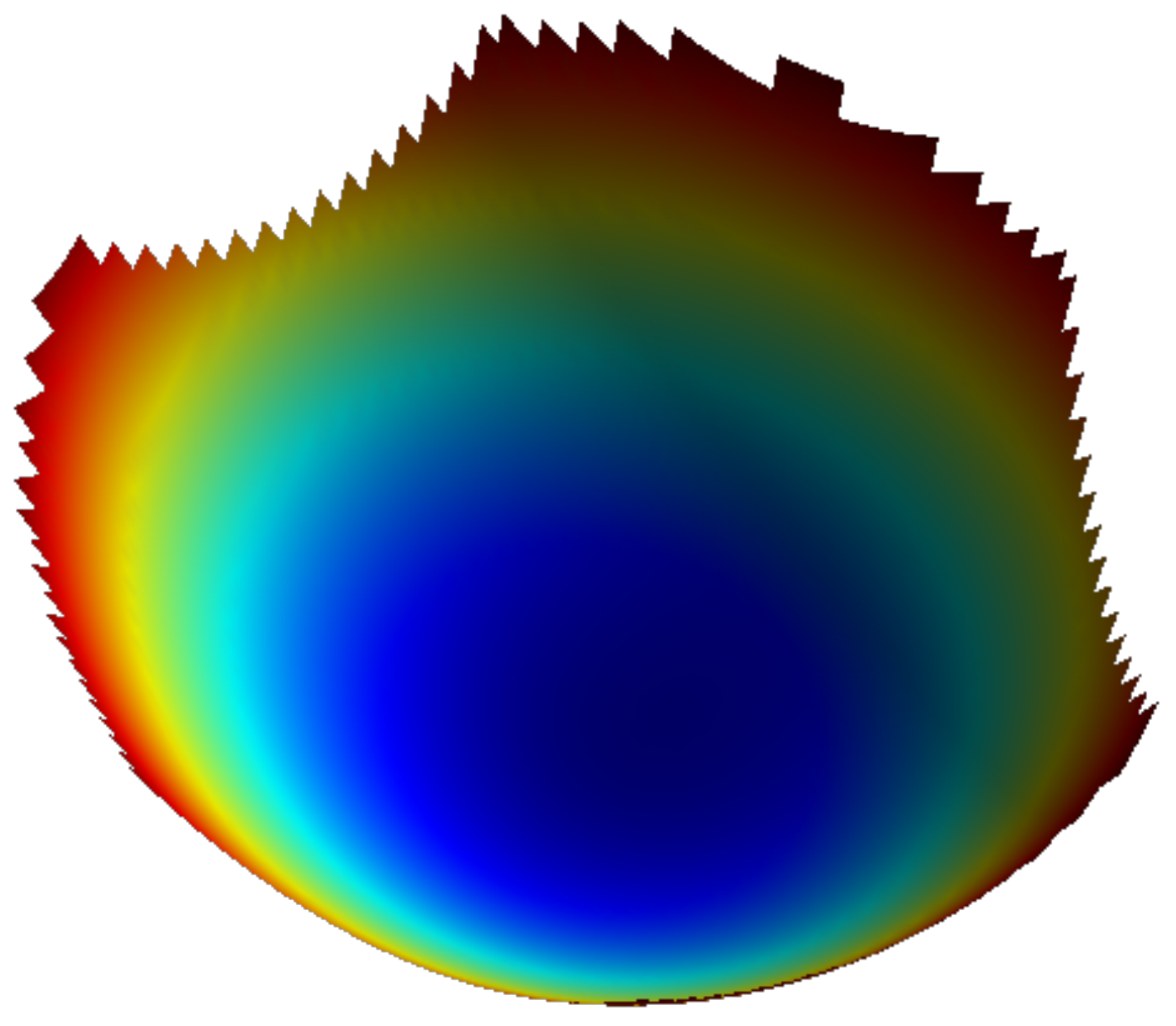}
    \end{center}
    \caption{\label{fig:central-shape-evolve}Zoom into the central
    regions of the dimensionless bispectrum
    $\dimlessB(\alpha, \beta)$ in the axion-quartic
    model~\eqref{eq:axion-quartic}.
    Regions near the vertices of the triangle are masked away,
    removing the strong spikes near squeezed configurations.}
\end{figure}

Our numerical method is capable of capturing very subtle effects
in the bispectrum shape.
In Fig.~\ref{fig:central-shape-evolve}
we highlight these effects by
masking away regions near the vertices,
leaving only the central region of the $(\alpha,\beta)$
triangle.
As in Fig.~\ref{fig:shape-evolve},
the top panel shows the shape at very early times.
Notice that contours of equal amplitude
are roughly triangular, with rounded corners.
The overall appearance is smooth
but there is a delicate `wrinkled' structure
inherited from the momentum dependence of the vacuum
fluctuations.

The central panel captures a transition occurring near
$N = 57.1$ e-folds from the initial conditions.
(The time of evaluation for this panel is not the
same as in Fig.~\ref{fig:shape-evolve}.)
At this time the central short modes are far outside the horizon
and the bispectrum amplitude is preparing to undergo rapid changes
as the $\chi$ field,
previously frozen by Hubble friction, rolls
towards its minimum.
Although the bispectrum shape is still dominated
by the local-like spikes at each vertex,
subtle bumps appear
near the folded configurations along each edge.
Meanwhile, in the central region,
a pattern of oscillations is clearly visible,
inherited from interference effects around horizon exit.

Finally, the bottom panel shows the configuration
at late times when the system has stabilized
near an adiabatic limit.
The equilateral configuration
$(\alpha,\beta) = (0,1/3)$ is now a minimum of the amplitude
rather than a maximum,
and the surrounding contours
are much more circular than in the top panel.
The central region is very smooth, with the oscillations
visible in the middle panel having been damped away.

\para{Time evolution}
Alternatively, we can study the time history of the correlation functions for
a single configuration. As we now explain, this is often a useful tool with which
to detect contamination of the numerical solution.

\begin{figure}
    \begin{center}
        \includegraphics[scale=0.5]{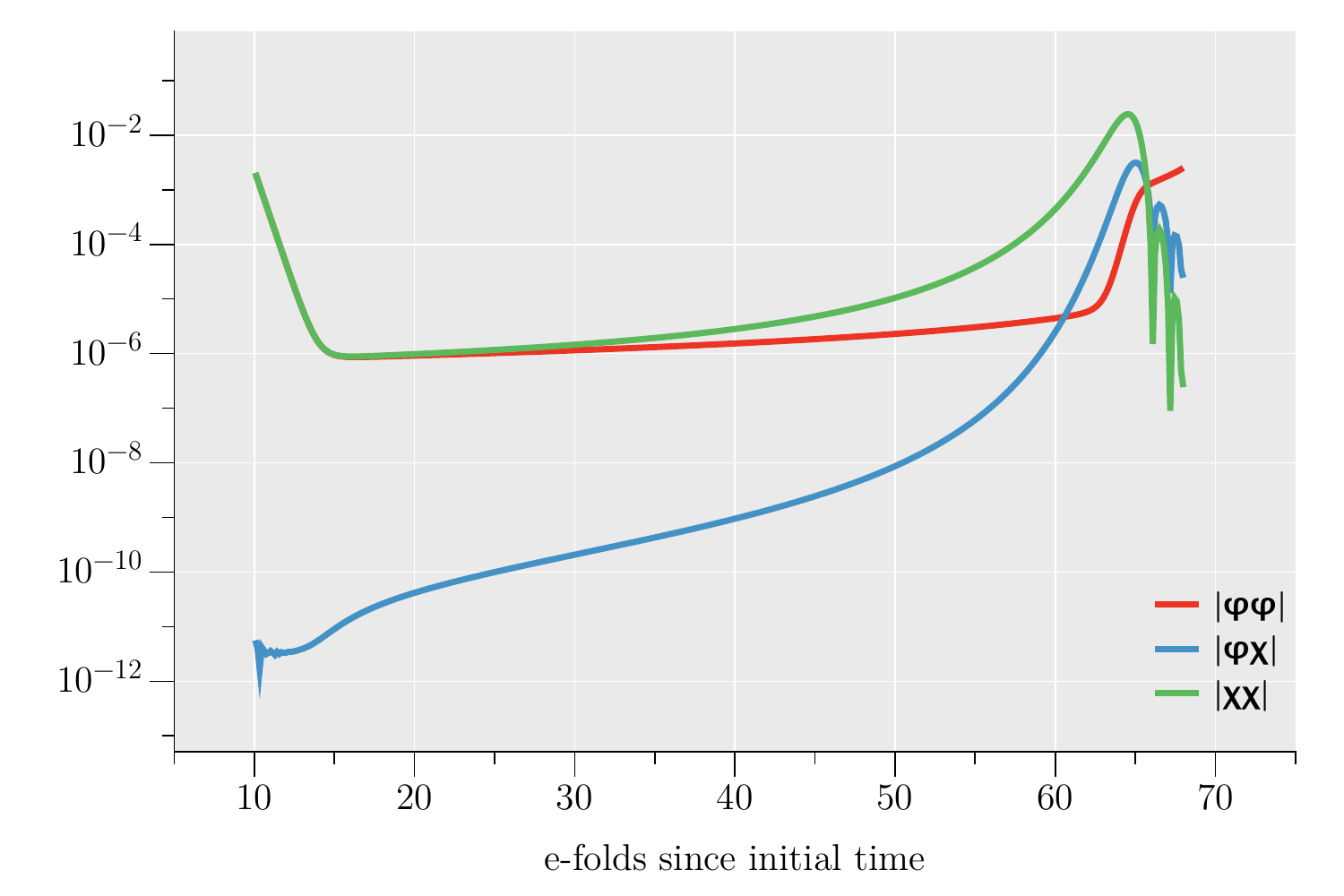}
        \includegraphics[scale=0.5]{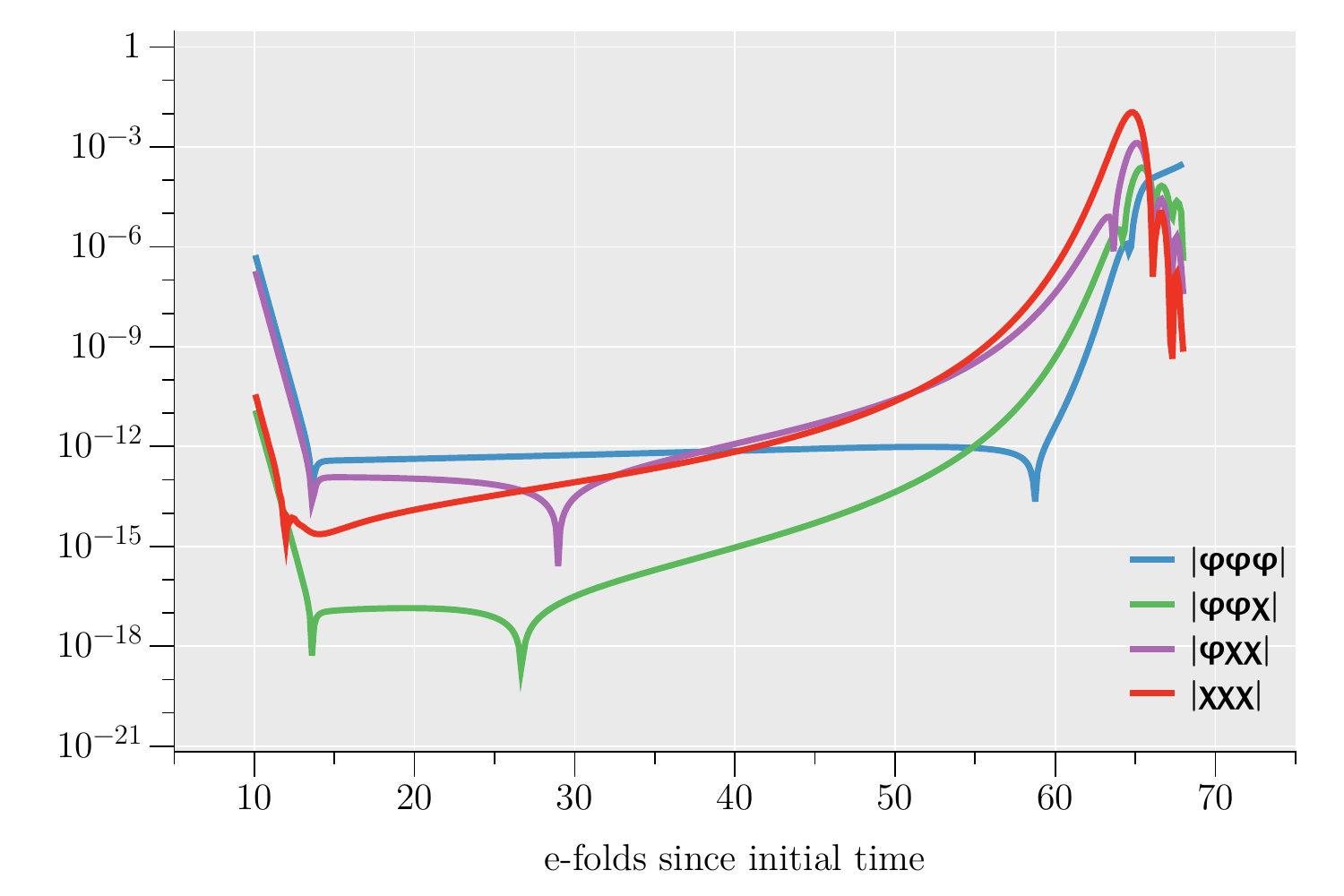}
    \end{center}
    \begin{center}
        \includegraphics[scale=0.5]{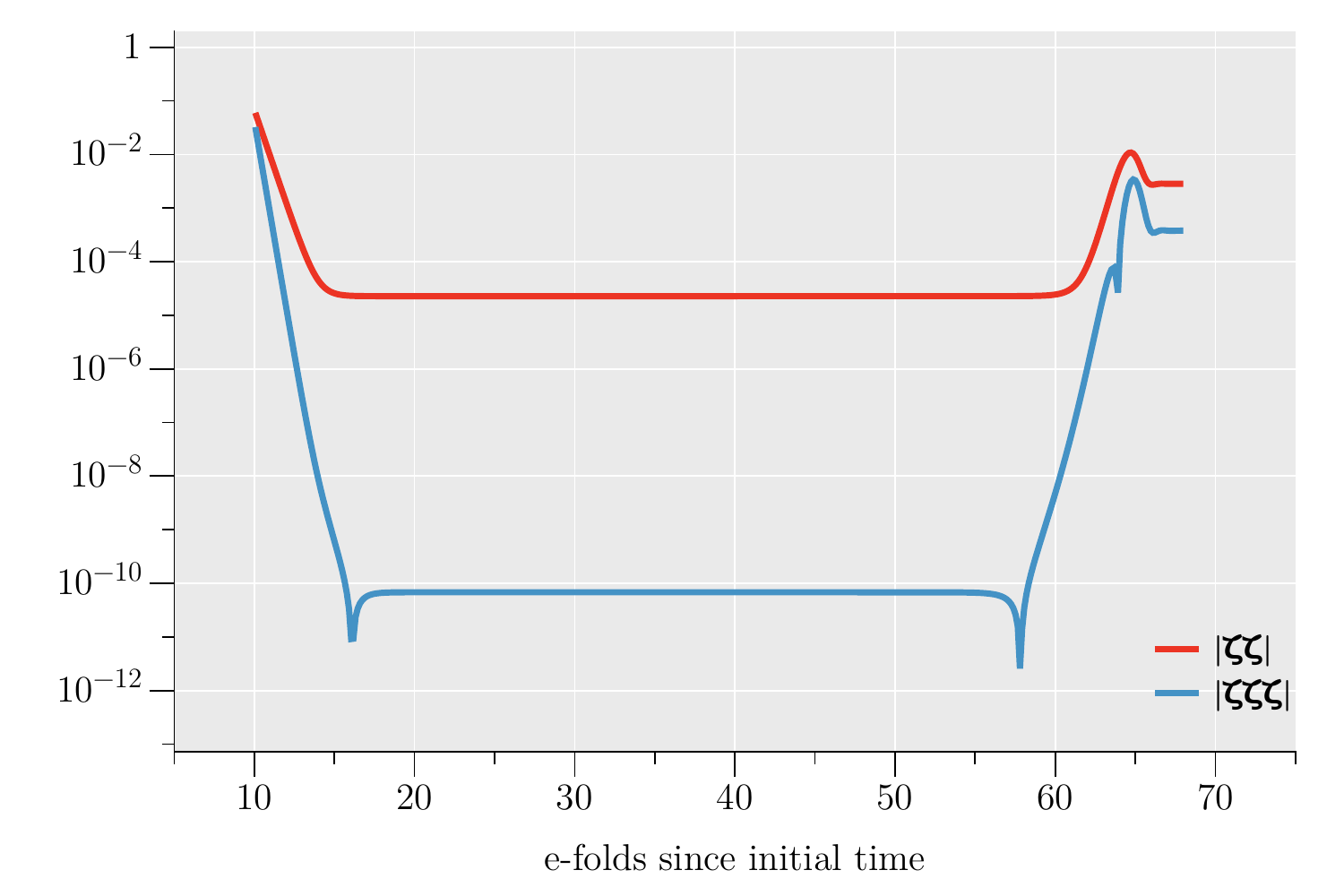}
        \includegraphics[scale=0.5]{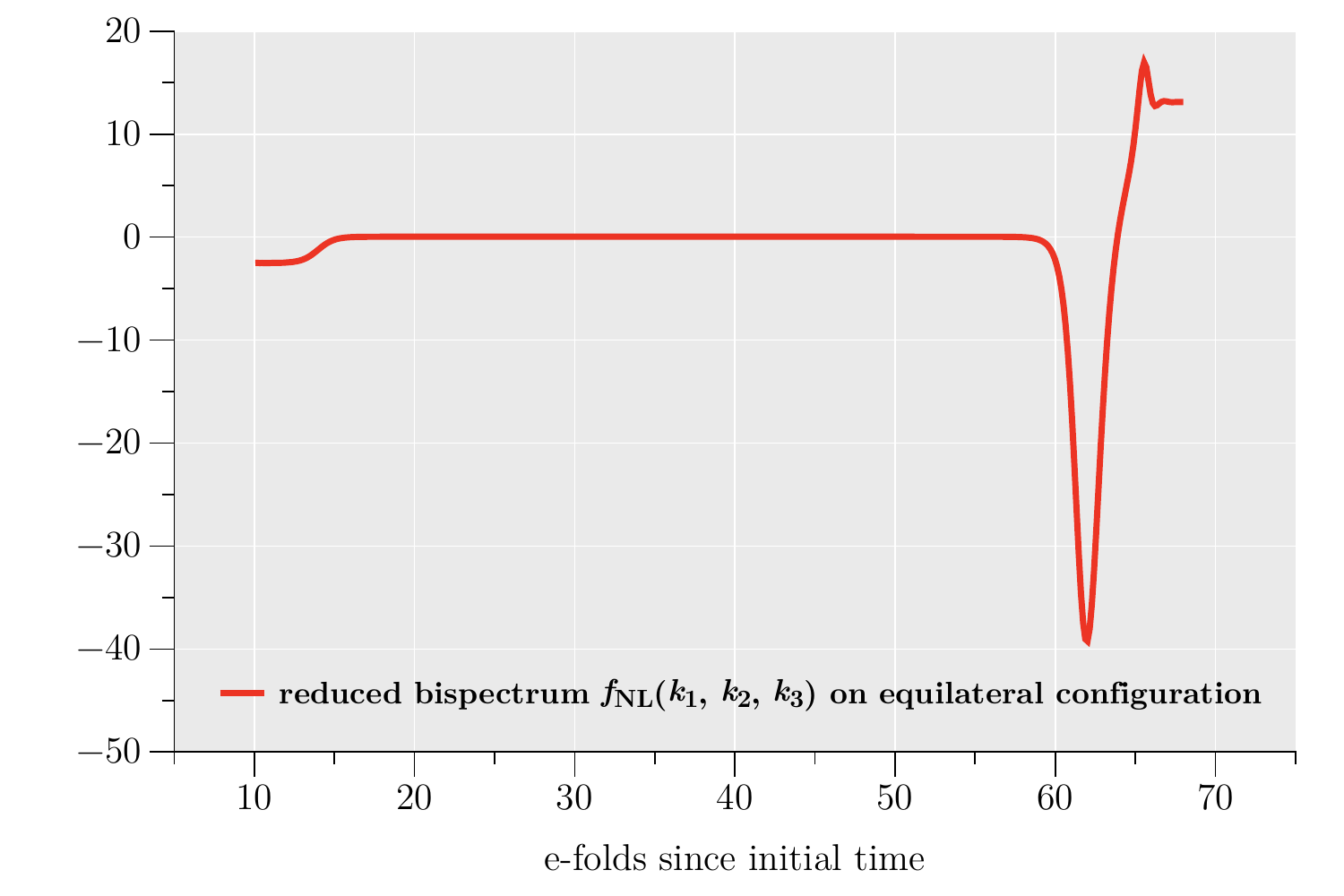}
    \end{center}
    \caption{\label{fig:equi-time-evolution}Time evolution of correlation functions
    for an equilateral configuration.
    Top left panel: field-space two-point functions.
    Top right panel: field-space three-point functions.
    Bottom left panel: $\zeta$ two- and three-point functions.
    Bottom right panel: reduced bispectrum.
    All plots except the bottom right panel show absolute values.}
\end{figure}

In Fig.~\ref{fig:equi-time-evolution} we plot the evolution of the
field-space and $\zeta$
three-point functions for an equilateral configuration
with $k_t$ adjusted so that $k_t/3$ exits the horizon $14.0$ e-folds after
the initial time;
this configuration corresponds to the equilateral point
in Figs.~\ref{fig:shape-evolve} and~\ref{fig:central-shape-evolve}.
We also plot the two-point functions for the
scale $k_t/3$ that appears on each side of the momentum triangle.

The critical property appearing in these plots is the smooth, exponential
decay of the correlation functions on subhorizon scales.
The presence of this feature is easy to understand by comparison with the
classical result~\eqref{eq:2ptdeSitter}
for the two-point function of a massless scalar field
in de Sitter space.
Specializing to the equal-time limit, the corresponding dimensionless power
spectrum is
\begin{equation}
    \dimlessP = \frac{H^2}{4\pi^2} ( 1 + k^2 \tau^2 )
    =
    \frac{H^2}{4\pi^2}
    \Bigg(
        1 + \frac{k^2}{a^2 H^2}
    \Bigg) ,
\end{equation}
where we have used the approximate relation $a = -(H\tau)^{-1}$.
If $H$ is almost constant then $(k/aH)^2 \propto \e{-2N}$, reproducing the
exponential decay visible in Fig.~\ref{fig:equi-time-evolution}.
For the three-point function the decay can be faster
if a higher power of $k/aH$ is dominant; cf. the bottom left panel of
Fig.~\ref{fig:equi-time-evolution}, where the blue line drops more steeply.
Our procedure for computing initial conditions
in~\S\ref{sec:initial-conditions}
amounted to keeping only the exponentially largest terms at early times.

In~\S\ref{sec:performance}
we will see that it is very expensive to integrate
during this exponentially-decaying subhorizon phase.
This is a problem
that afflicts all attempts to numerically solve the inflationary
perturbations.
In a traditional `Feynman calculus' code such as
{\FieldInf} or {\ModeCode}/{\MultiModeCode}
it appears through the obligation to track
exponentially rapid phase oscillations\
$\sim \e{\im k/aH}$
of the constituent wavefunctions
in~\eqref{eq:2ptdeSitter}.
Because our approach is based on evolution of correlation functions rather than
their constituent wavefunctions these explicit phase oscillations are absent,
but their influence remains through the nonzero imaginary parts of the
two-point function
that
couple off-diagonal terms in our matrix equations.
In either case, the net result is that the integrator must
dramatically reduce its step size
in order to accurately reproduce this smooth exponential profile.

This property is unfortunate from the perspective of pure numerical
performance, but it can be exploited to provide a sanity check on
the accuracy of the numerical solution.
Even small discrepancies can disrupt the cancellations that conspire to
produce smooth decay.
An example can be seen in the top-left panel of Fig.~\ref{fig:equi-time-evolution}
where the blue line representing the $\phi\chi$ cross-correlation function
shows small oscillations at early times.
We believe these oscillations stem from a small inaccuracy in the massless
approximation,
which remains relevant in these correlation functions because of their
small amplitude.
The initial condition could perhaps be made more accurate by the inclusion
of next-order terms.

In practice we have found these oscillations around components
with small amplitude
to be largely harmless; they damp out as phase oscillations decay
near horizon exit and do not disrupt the superhorizon epoch.
Of more concern is the appearance of noise or
uncontrolled oscillations
around an exponentially decaying profile
such as those appearing in the autocorrelation functions.
A feature of this type is likely to be symptomatic
of numerical errors
that will produce an inaccurate result at late times.
The normal response should be
to tighten the numerical
tolerances, increase the number of e-folds
of massless evolution (see~\S\ref{sec:massless-efolds}),
or both.

\subsection{Single-field model with feature}
\label{sec:single-field-feature}

The axion--quartic model yields a significant bispectrum
from smooth field-space evolution.
An alternative is to introduce a sharp `feature' into the
inflationary potential.
In this section we show that our method handles such
scenarios equally well.
As an example, consider the single-field
step model
\begin{equation}
    V(\phi) = \frac{1}{2} m^2 \phi^2
    \Big(
        1
        +
        c \tanh \frac{\phi-\phi_0}{d}
    \Big) .
    \label{eq:feature-potential}
\end{equation}
This model was studied numerically by Chen, Easther \& Lim~\cite{Chen:2006xjb,Chen:2008wn},
who computed both the two- and three-point functions.
The step occurs at $\phi = \phi_0$.
We take
$m = 10^{-5} \Mp$,
$\phi_0 = 14.84 \Mp$,
$c = 0.0018$
and
$d = 0.022 \Mp$.
The initial condition is
$\phi = 16.5 \Mp$, and its velocity is estimated using the slow-roll
approximation.

\begin{figure}
    \begin{center}
        \includegraphics[scale=0.5]{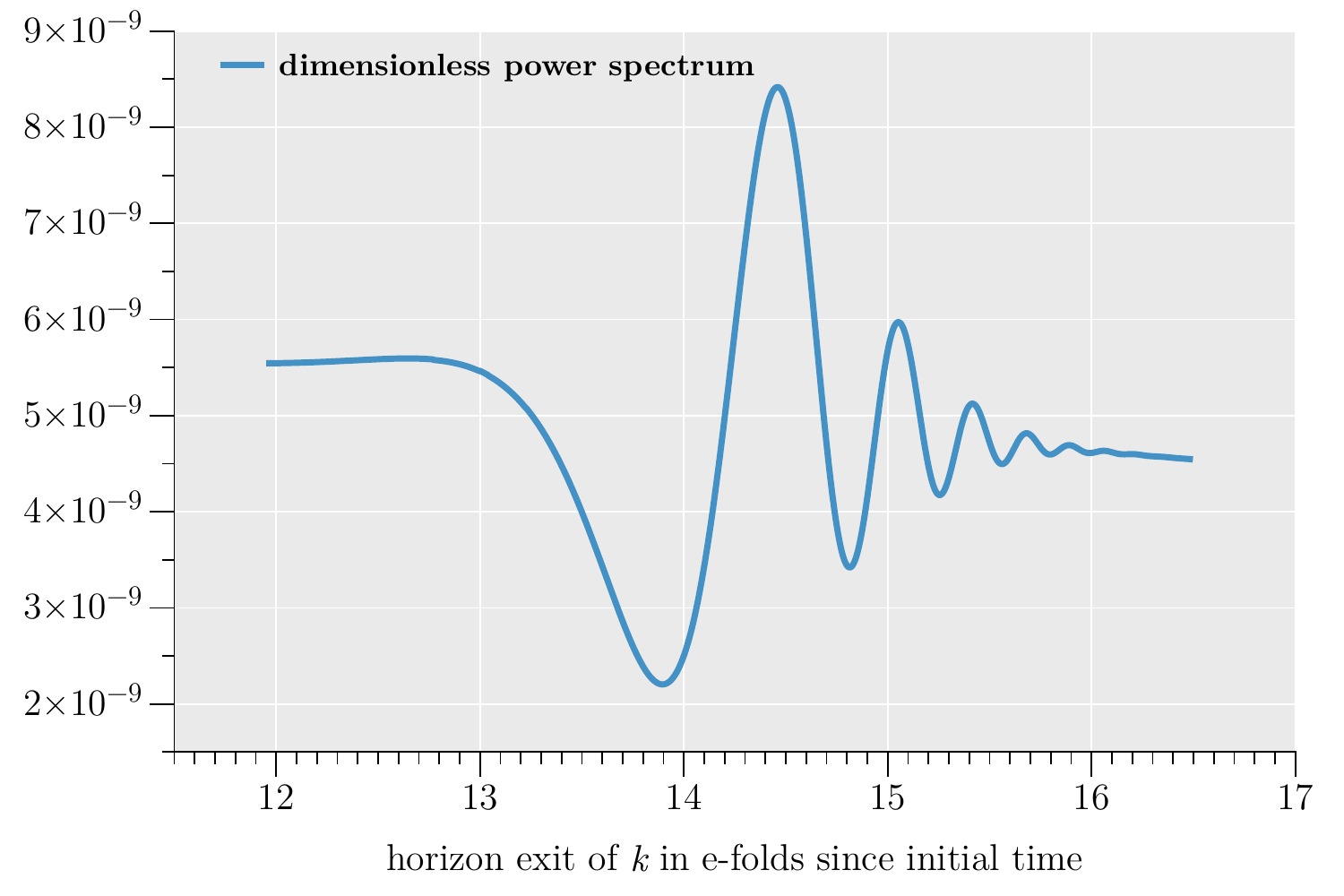}
        \includegraphics[scale=0.5]{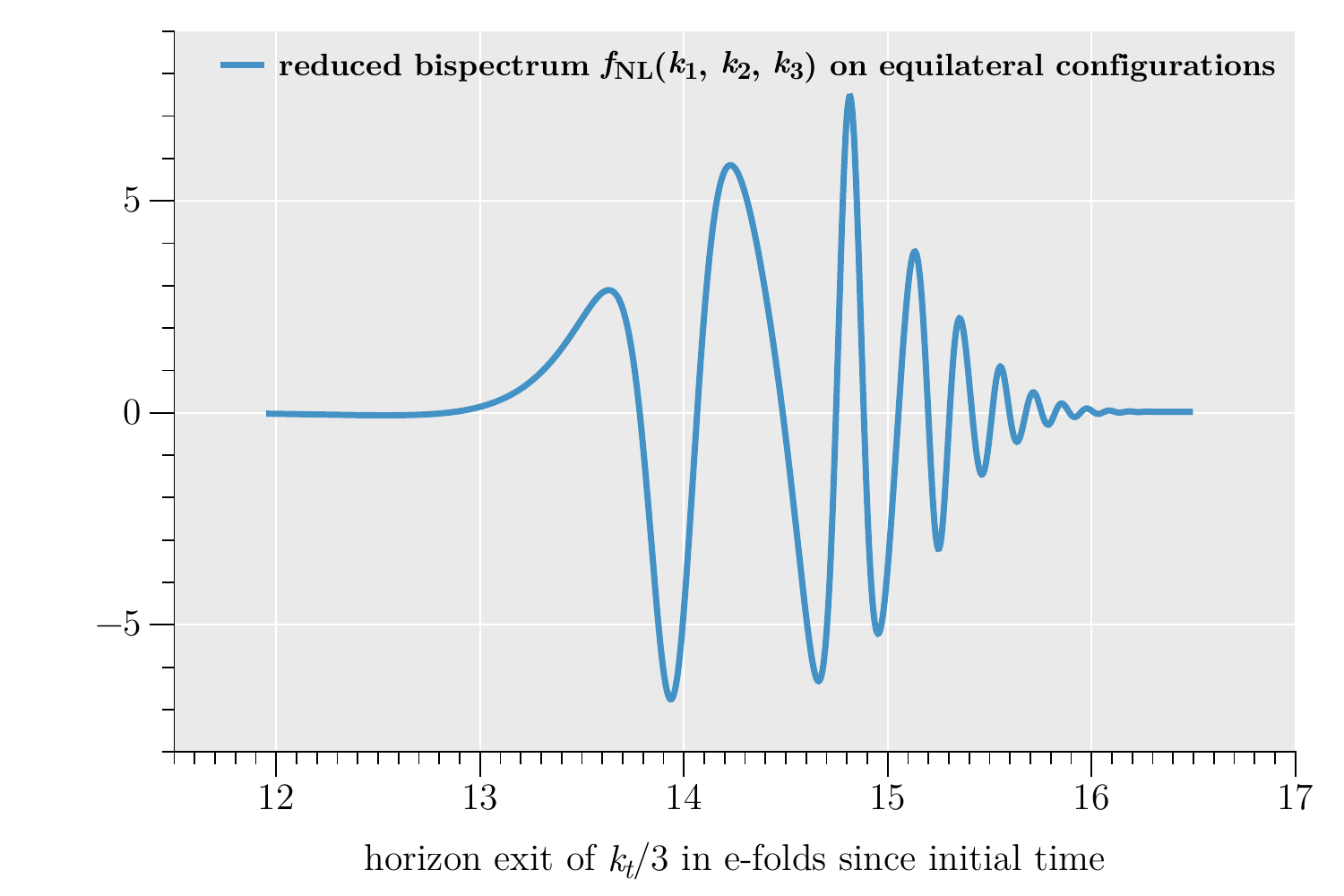}
    \end{center}
    \caption{\label{fig:feature-outputs}Left panel: dimensionless power spectrum
    in single-field feature model~\eqref{eq:feature-potential}.
    Right panel: reduced bispectrum $\fNL(k_1, k_2, k_3)$ on equilateral
    configurations as a function of scale $k_t$.}
\end{figure}

\begin{figure}
    \begin{center}
        \includegraphics[scale=0.3]{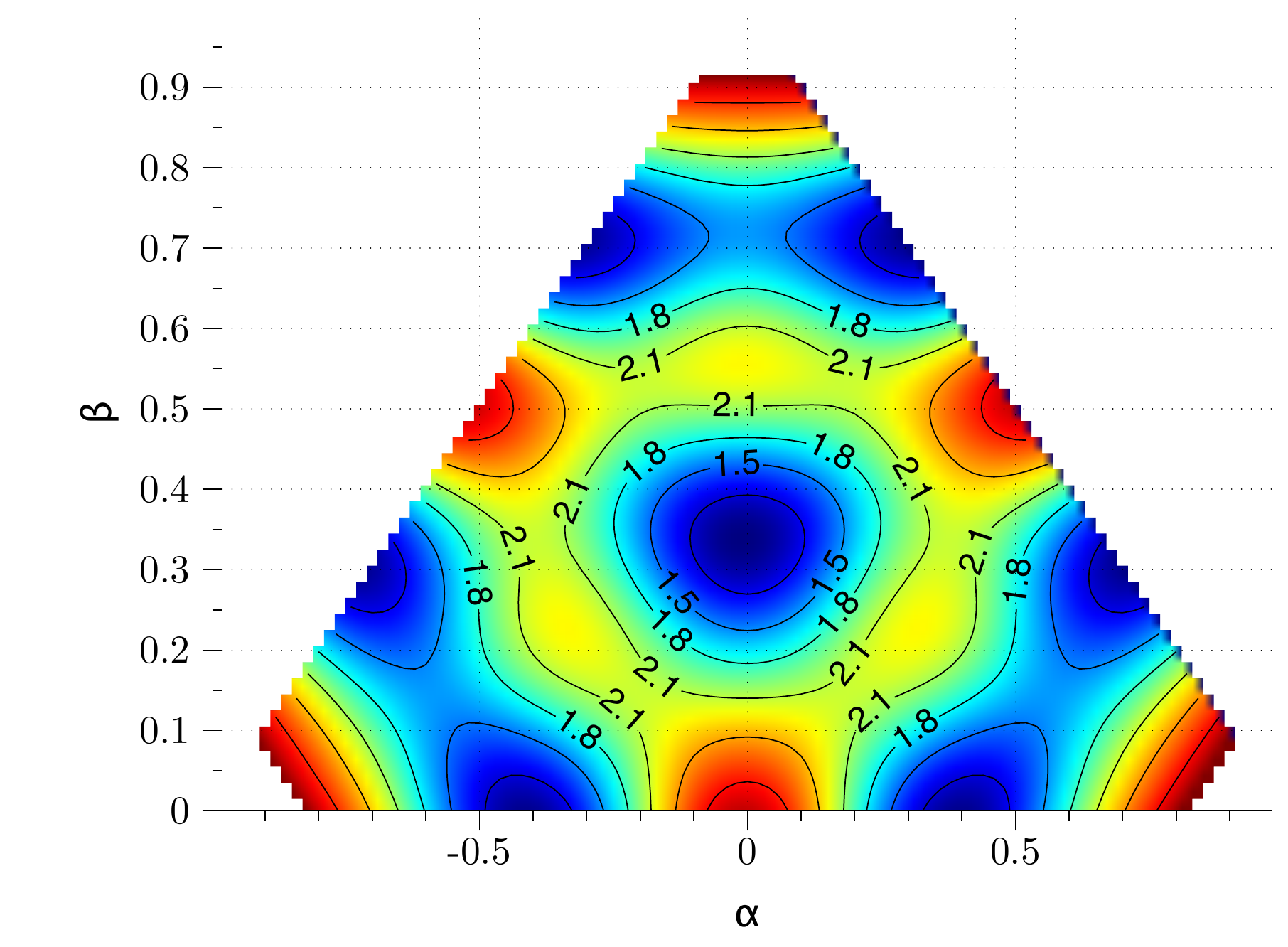}
        \qquad
        \includegraphics[scale=0.23]{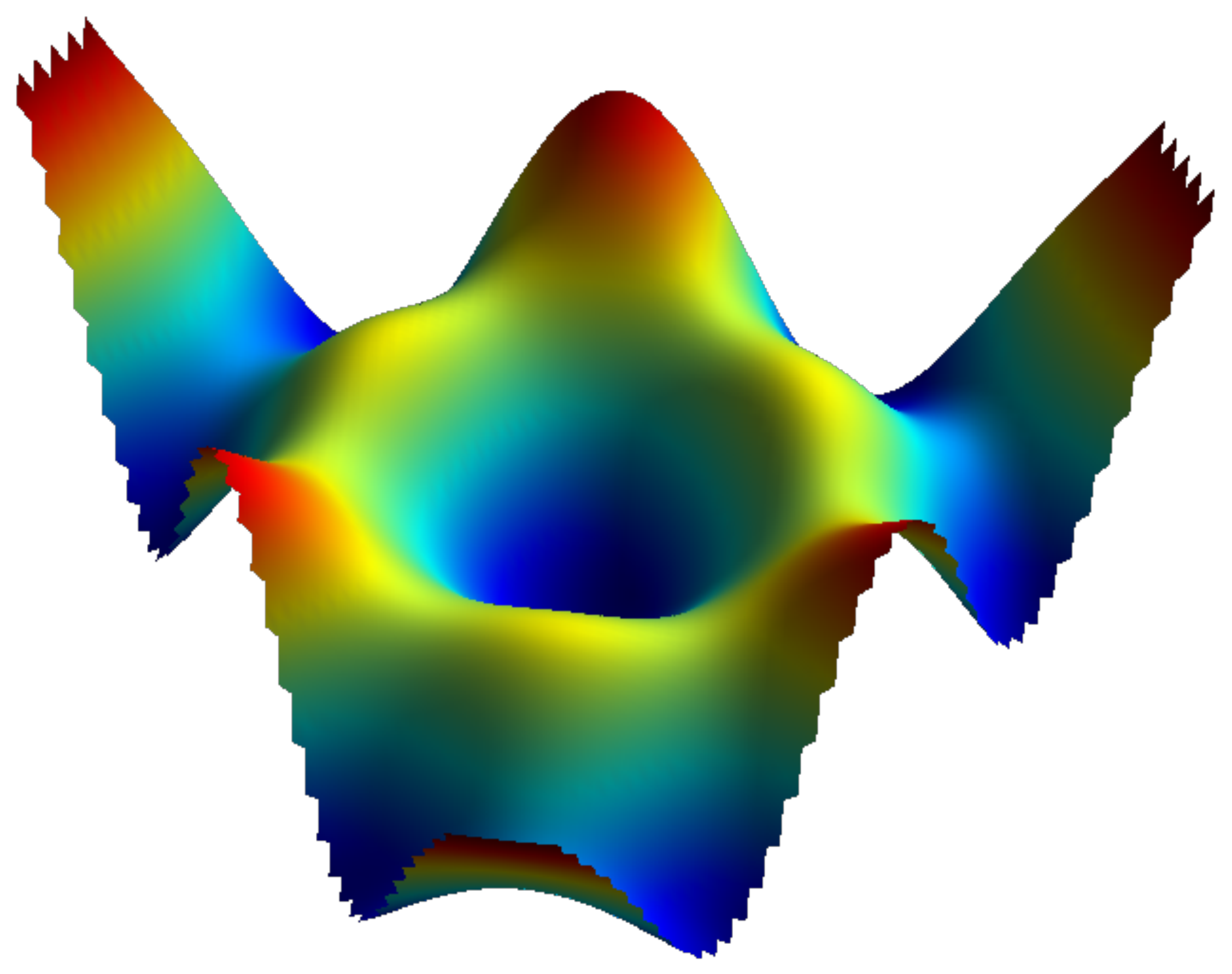}
    \end{center}
    \begin{center}
        \includegraphics[scale=0.3]{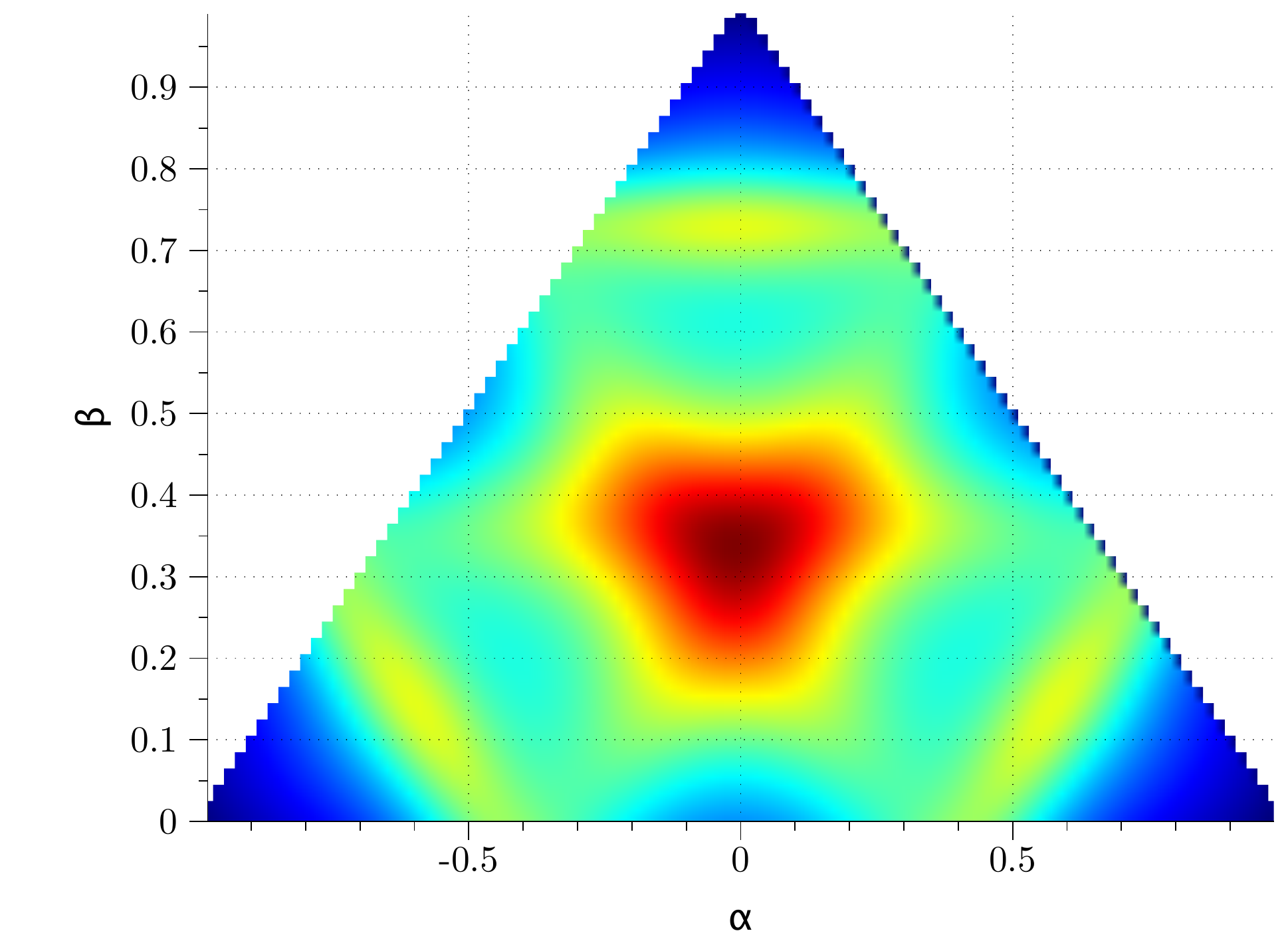}
        \qquad
        \includegraphics[scale=0.25]{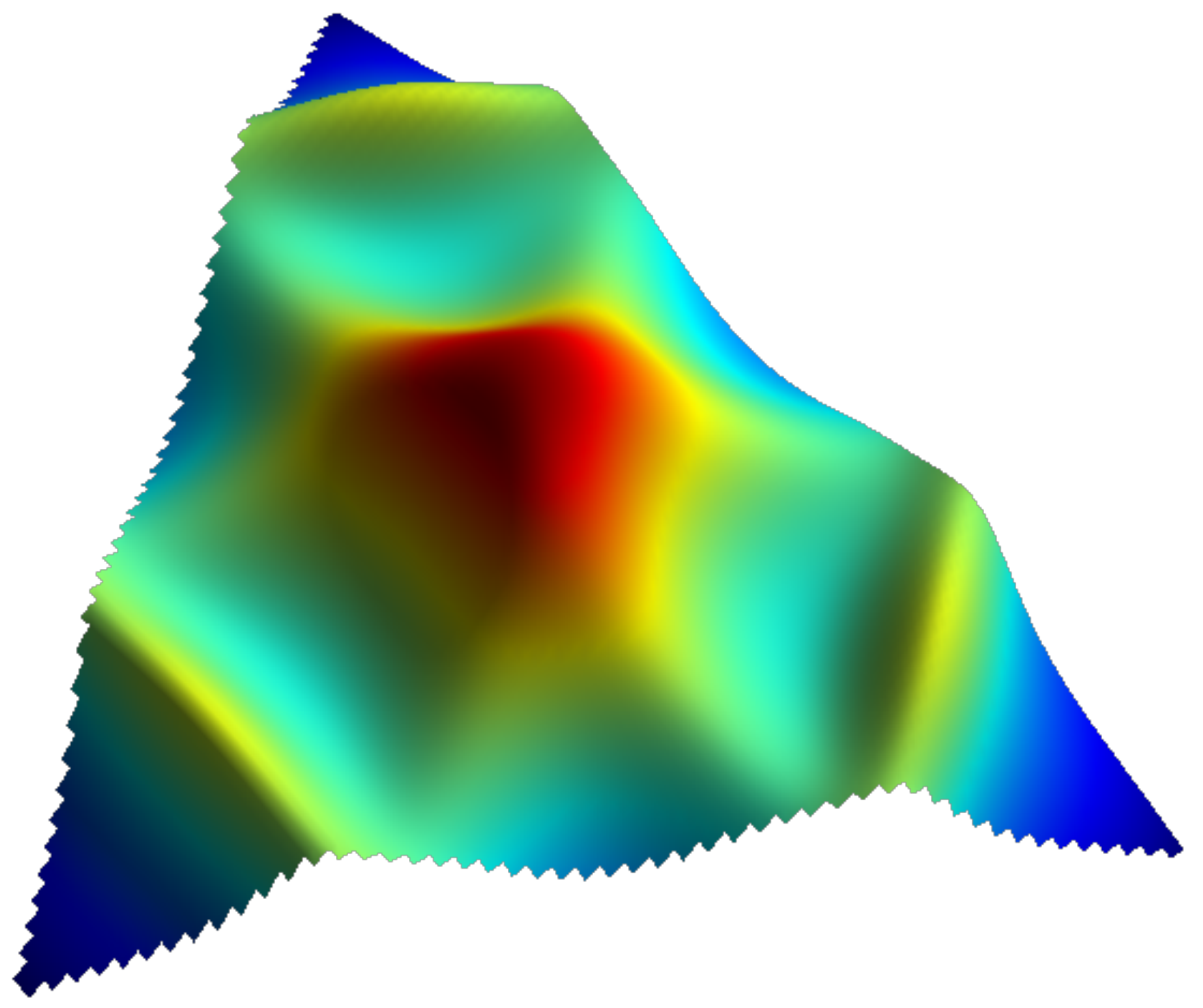}
    \end{center}
    \caption{\label{fig:feature-shape}Top row: dimensionless bispectrum
    $\dimlessB(\alpha,\beta)$ as a function of the shape parameters
    $\alpha$, $\beta$.
    The squeezed regions near the vertices of the triangle have been removed.
    Bottom row: reduced bispectrum, also as a function of shape.}
\end{figure}

The dimensionless power spectrum and reduced bispectrum on equilateral
configurations
are shown in Fig.~\ref{fig:feature-outputs}.
These are in good agreement with the results of Chen et al.~\cite{Chen:2008wn}.
In Fig.~\ref{fig:feature-shape} we plot the shape of the dimensionless bispectrum
$\dimlessB(k_1, k_2, k_3)$ and reduced bispectrum
$\fNL(k_1, k_2, k_3)$
at a fixed scale $k_t$,
chosen
so that $k_t/3$ exits at $N=14.8$ e-folds
from the initial time,
corresponding to the peak of the largest positive spike
of $\fNL$ in Fig.~\ref{fig:feature-outputs}.
The conclusion is that
not only the scale dependence but also
the shape dependence is very significant.
Consequently, making a prediction for late-time observables
that are sensitive to the bispectrum
is a complex question.
Accurate predictions are likely to require a detailed numerical study.

\subsection{Heavy modes: adiabatic and non-adiabatic effects}

The discussion in~\S\S\ref{sec:axion-quartic}--\ref{sec:single-field-feature}
established that our numerical method can successfully track
very delicate features in the shape- and scale-dependence
of the bispectrum.
We now study
examples that illustrate the effect of heavy modes
as described in~\S\ref{sec:why-automated}
and demonstrate that we can successfully predict their
$n$-point functions.

\subsubsection{Adiabatic-like models: gelaton and QSFI}

First consider the gelaton and QSFI scenarios.
These are `adiabatic-like' scenarios in which the
expectation value of the heavy field tracks the minimum
of its effective potential.
As we now explain,
such models
appear to be rare
if we attempt to engineer
them from a Cartesian field-space metric and a turning trajectory.
In fact,
we have not yet managed to construct
an explicit
model that yields a significant
bispectrum from either of these effects.

Consider a model with canonically-normalized
Cartesian fields $X$ and $Y$,
and translate to polar coordinates
$X = R \cos \theta$, $Y = R \sin \theta$.
The action for $R$ and $\theta$ becomes
\begin{equation}
    S = -\frac{1}{2} \int \d^4 x \; \sqrt{-g}
    \Big[
        (\partial R)^2
        +
        R^2 (\partial \theta)^2
        +
        2 V( R \cos \theta, R \sin \theta )
    \Big]
    \label{eq:polar-action}
\end{equation}
where $V(X, Y)$
is the original potential.
To study heavy-field effects we should choose the radial
direction $R$ to be heavy and the angular direction
$\theta$ to be light.
The effective metric experienced by $\theta$ is non-Euclidean.
This idea was suggested by Chen \& Wang~\cite{Chen:2009we};
see also Assassi et al.~\cite{Assassi:2013gxa}.

We can assume that $R$ is stabilized at some radius $R_0$.
If the motion is purely rotational with
angular velocity $\omega = \dot{\theta}/H = \d \theta / \d N$
then the slow-roll
parameter $\epsilon$ satisfies
\begin{equation}
    \epsilon \approx \frac{R_0^2}{2\Mp^2} \omega^2 .
\end{equation}
The adiabatic power spectrum $\dimlessP$
will be
\begin{equation}
    \dimlessP \sim \frac{H^2}{\epsilon \Mp^2}
    \sim \frac{H^2}{R_0^2} \frac{1}{\omega^2} .
\end{equation}
To achieve a suitable normalization
requires $R_0$
to be much larger than $H$,
roughly $R_0 \sim H/\omega \dimlessP^{1/2}$.
Successful inflation requires $\epsilon < 1$
and therefore $H \lesssim \Mp \dimlessP^{1/2}$.

\para{Gelaton scenario}
In a gelaton scenario the radial mass
should be
some scale $M$ at least modestly larger than $H$.
We set $V''(R_0) = M^2 \gtrsim H^2$,
and take all higher derivatives to be negligible.
For a circular trajectory
we expect a significant gelaton-like
renormalization of the sound speed
when $\omega \gg M/H$~\cite{Achucarro:2010da}.
Such a large angular velocity implies that
a full $2\pi$-rotation in field space
occurs
in substantially less than an e-fold.

As a consequence of this constraint
we have been unable to find a parameter combination
that gives a significant gelaton effect
over a meaningful range of scales.
The conclusion appears to be that
prolonged gelaton-like behaviour requires
a more non-canonical metric
than can be produced by angular rotation in the $X$, $Y$ model.%
    \footnote{As an alternative
    one could consider a spiral-like model,
    which might accommodate a large $\omega$
    by allowing multiple windings around the origin.
    Unfortunately it is still not easy to construct an explicit model.
    As we will discuss below, centrifugal forces associated with
    rotation will cause the $R$ expectation value to be displaced
    away from its bare minimum $R_0$ by an amount $\Delta$.
    Assuming this displacement to be controlled by the quadratic
    term,
    choosing $\omega \sim M/H$ gives $\Delta \sim R_0$.
    Therefore no more than a few spirals can be
    packed into the region $0 \leq R \leq R_0$.}

\para{QSFI scenario}
Now consider a QSFI-like scenario.
In this case the radial mass and radial cubic coupling should both be
close to $H$,
and the angular velocity $\omega$ should be a little smaller
than unity in order to give efficient transfer of the
radial fluctuations into the adiabatic mode~\cite{Chen:2009zp}.
Therefore there is no longer a problem associated with excessively
rapid rotation; for example, we could pick $\omega \approx \pi / 30 \approx 0.1$.
This would allow $N \approx 30$ e-folds for a rotation of $\pi$ around the
origin.

The centrifugal force associated with angular motion will cause the $R$
expectation value to be displaced from its bare
minimum $R_0$ by an amount $\Delta$.
The effective potential experienced by radial fluctuations $\delta R$
around this displaced expectation value is
\begin{equation}
\begin{split}
    \Veff(\delta R)
    &
    \approx
    \frac{1}{2}
    \Big[
        V''(R_0)
        + V'''(R_0) \Delta
        + \frac{1}{2} V''''(R_0) \Delta^2
    \Big] \delta R^2
    +
    \frac{1}{3!}
    \Big[
        V'''(R_0)
        + V''''(R_0) \Delta
    \Big] \delta R^3
    + \cdots ,
    \\
    &
    =
    \frac{1}{2} \mu^2 \delta R^2 + \frac{1}{3!} g \delta R^3
\end{split}
\end{equation}
where a prime $'$ denotes a derivative
with respect to $R$,
and we have introduced the effective mass-squared $\mu^2$
and cubic coupling $g$.
In each term we have kept derivatives up to $V''''(R_0)$.

The presence of a large scale $R_0 \gg H$ makes it difficult
to keep the dressed mass $\mu$ smaller than $H$.
To illustrate the problem, assume that the bare mass-squared $V''(R_0)$
is of order $H^2$ or smaller. Adopting a larger bare value can only make
tuning problems worse.

First assume that $V''''(R_0)$ is negligible.
The cubic derivative $V'''(R_0)$ will control the displacement $\Delta$
when $|V'''(R_0)| \gtrsim \dimlessP^{1/2} H / \omega$, which will always
be true if we choose $V'''(R_0) \sim H$
in order to achieve a QSFI phenomenology.
The displacement is then
\begin{equation}
    \Delta \approx \left( \frac{2H^2}{V'''(R_0) R_0} \right)^{1/2}
    \omega R_0 .
\end{equation}
This makes a contribution to the mass-squared of order
\begin{equation}
    \delta \mu^2
    \supseteq
    V'''(R_0) \Delta
    \sim
    \omega^{1/2} \dimlessP^{-1/4} H^{3/2} \sqrt{V'''(R_0)}
    \gg H^2 .
\end{equation}
Unless it is cancelled,
this large contribution to the mass
causes radial fluctuations to decay rapidly.
It
suppresses the transfer of any isocurvature
bispectrum into the adiabatic mode~\cite{Chen:2012ge}.%
    \footnote{As explained in~\S\ref{sec:loop-twopf}, this conversion
    \emph{is} captured by a tree-level calculation because it
    occurs via quadratic mixing rather than decay of two or more
    particles.
    Therefore we expect our tree-level codes to fully support
    the QSFI phenomenology.}

\begin{figure}
    \begin{center}
        \includegraphics[scale=0.25]{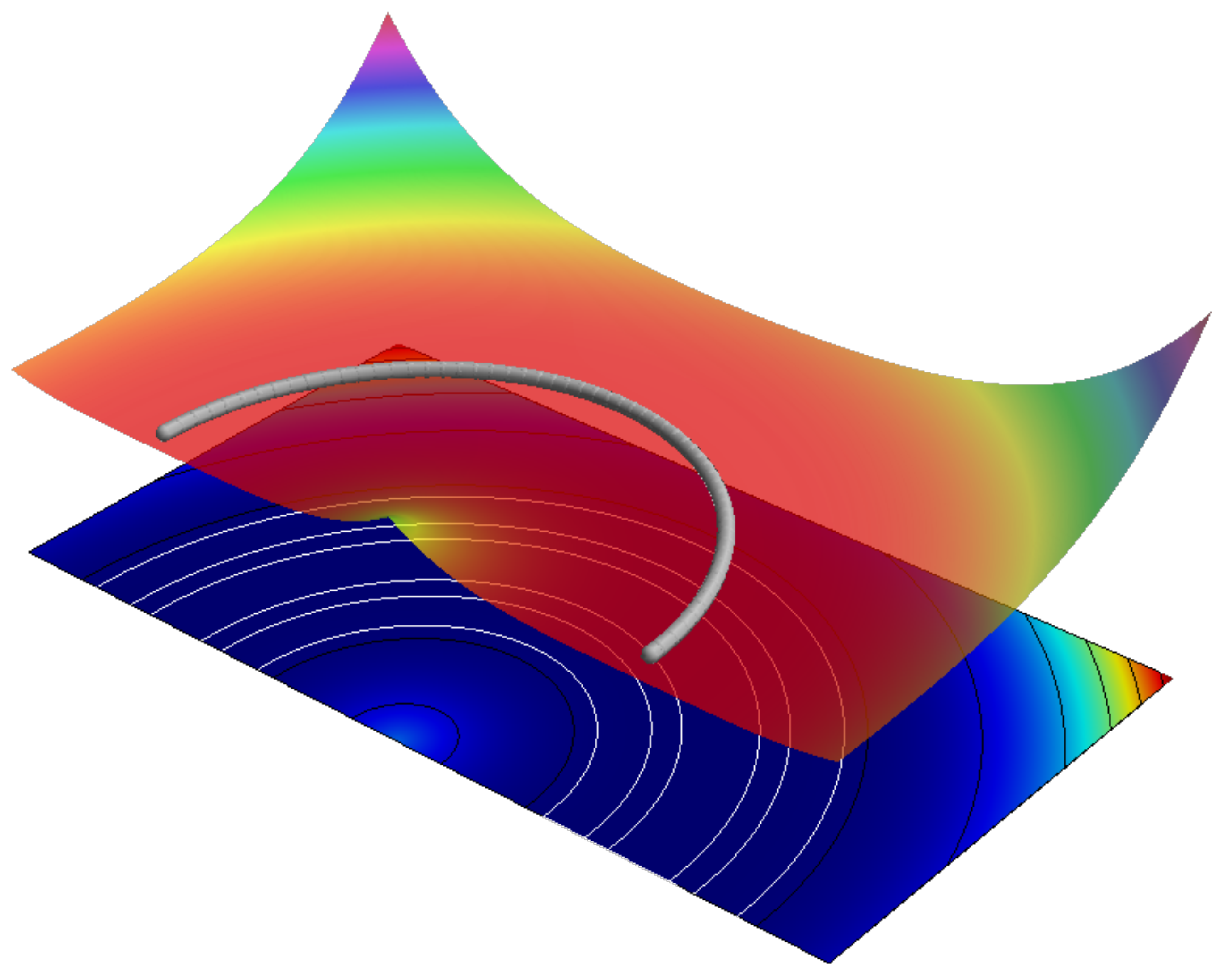}
        \includegraphics[scale=0.5]{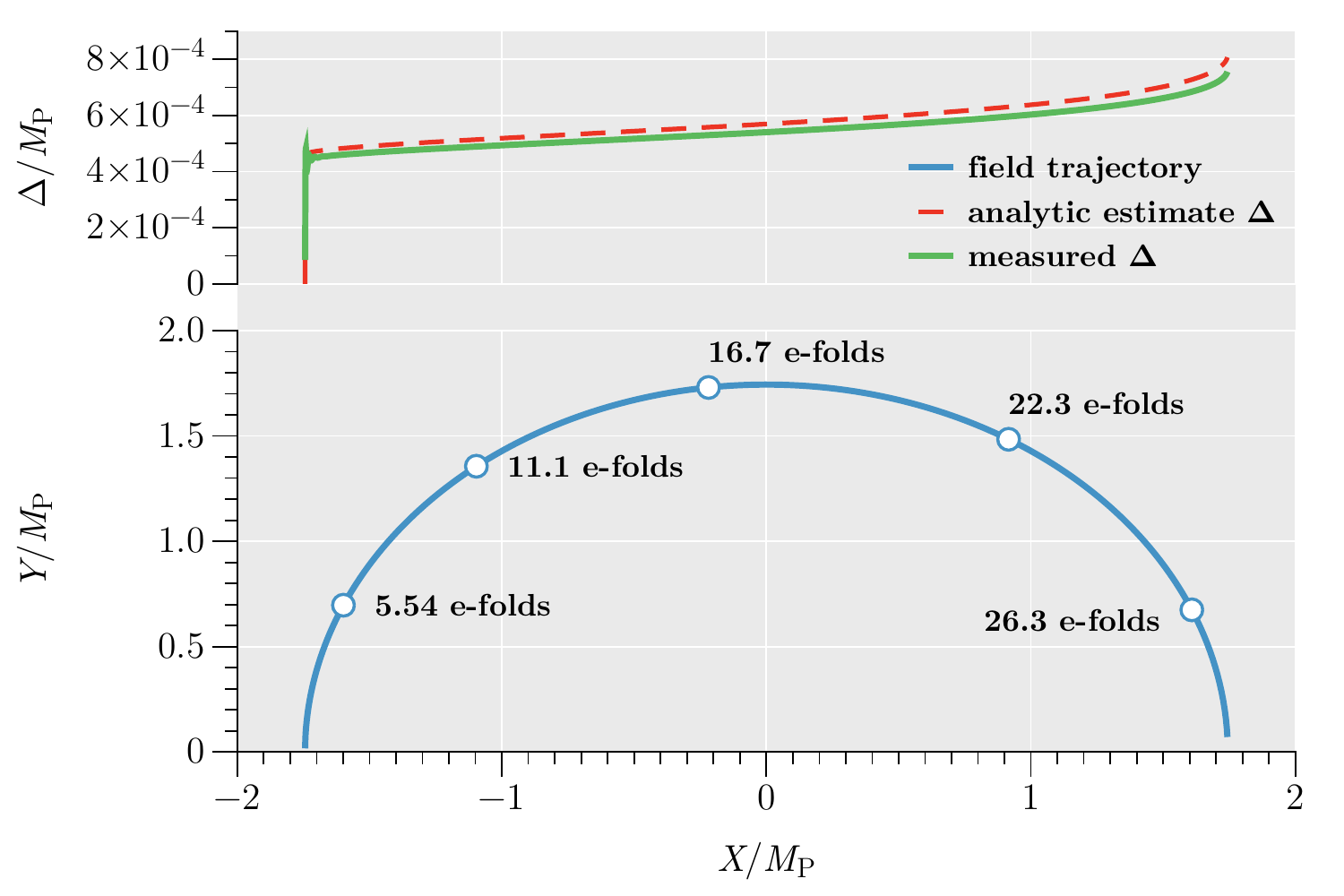}
    \end{center}
    \caption{\label{fig:qsfi-trajectory}Potential and trajectory
    for the model~\eqref{eq:qsfi-potential}.
    Left: Potential and trajectory. The fields roll from the far to the near side,
    and the lower plane shows equipotential contours.
    Right: zoomed plot of the trajectory, showing the displacement $\Delta$.}
\end{figure}

Similar estimates can be made for more complex
parameter combinations,
but in the cases we have
checked there is always a contribution to $\delta \mu^2$
whose scale is set by $H^2$.
The possibility of constructing a model in which the radial
mass is $< H^2$ then rests on the precise
$\Or(1)$ factors that occur.
A representative example is
\begin{equation}
    V = V_0 \left(
        1
        + \frac{29\pi}{120} \theta
        + \frac{1}{2} \frac{\eta_R}{\Mp^2} (R-R_0)^2
        + \frac{1}{3!} \frac{g_R}{\Mp^3} (R-R_0)^3
        + \frac{1}{4!} \frac{\lambda_R}{\Mp^4} (R-R_0)^4
    \right)
    ,
    \label{eq:qsfi-potential}
\end{equation}
with the parameters
$V_0 = 10^{-10} \Mp^4$,
$\eta_R = 1/\sqrt{3}$,
$g_R = \Mp^2 V_0^{-1/2}$
and
$\lambda_R = 0.5 \Mp^3 \omega^{-1/2} V_0^{-3/4}$.
We plot the potential and inflationary trajectory in
Figure~\ref{fig:qsfi-trajectory}.
\begin{figure}
    \begin{center}
        \includegraphics[scale=0.5]{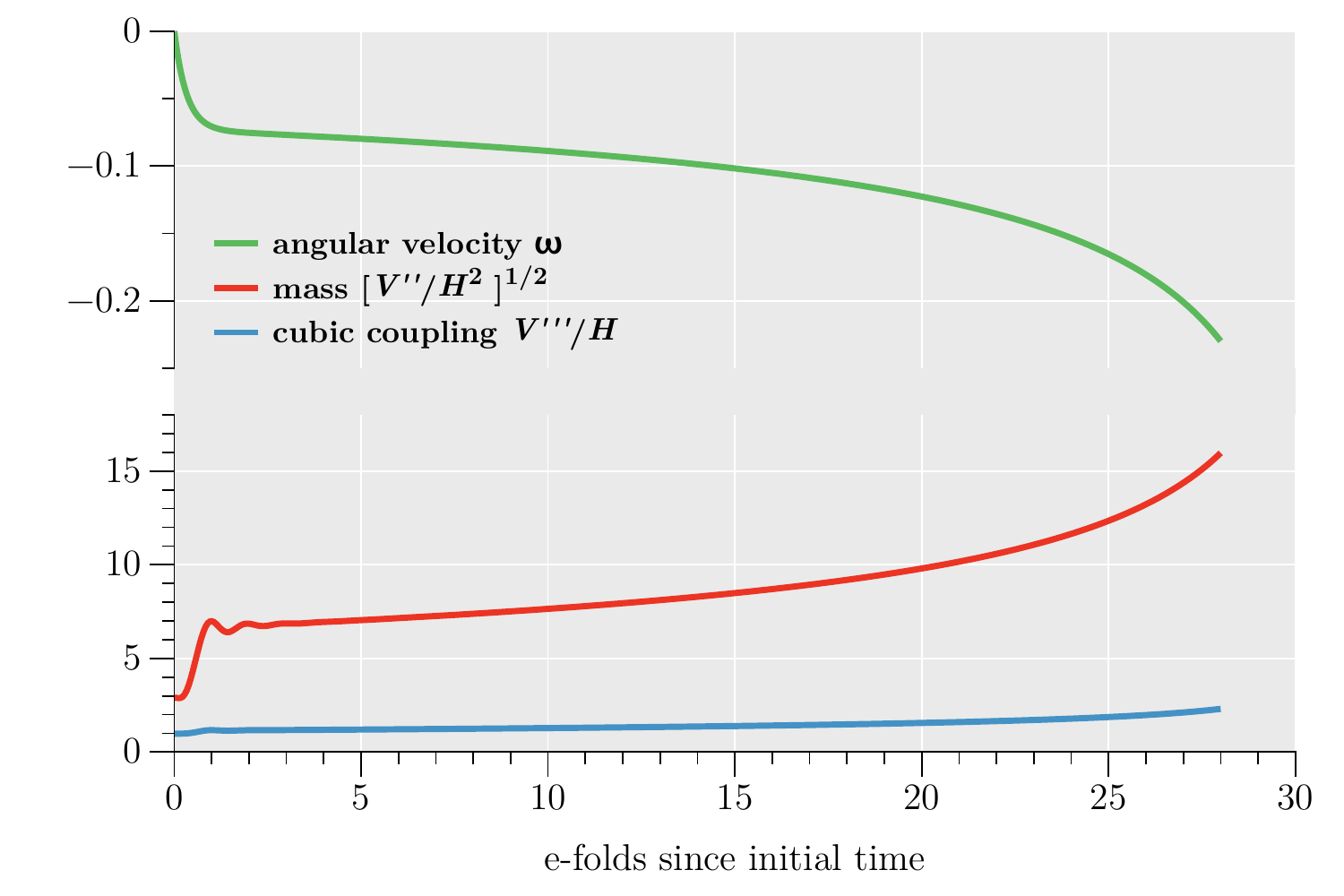}
        \includegraphics[scale=0.5]{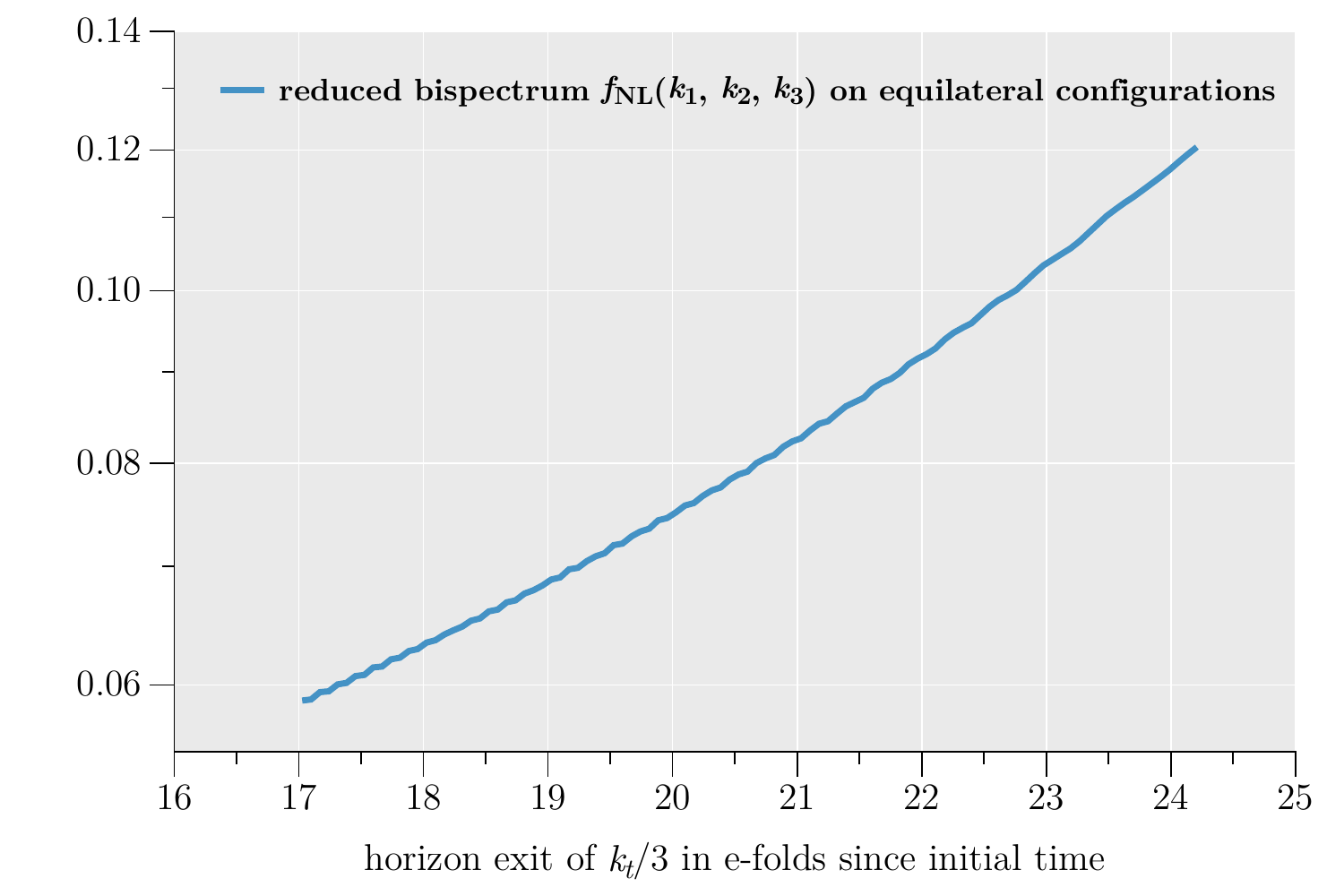}
    \end{center}
    \caption{\label{fig:qsfi-results}Left: evolution of
    angular velocity $\omega = \dot{\theta}/H$,
    together with the mass $\mu/H$ and coupling $g/H$.
    Right panel: reduced bispectrum evaluated on equilateral configurations
    as a function of scale $k_t$.}
\end{figure}

In Fig.~\ref{fig:qsfi-results} we show the time evolution
of the angular velocity $\omega = \dot{\theta}/H$
together with the dressed mass $\mu/H$ and cubic coupling $g/H$.
The mass lies roughly in the range $5 H$ to $15 H$
and the cubic coupling lies roughly between $H$ and $2H$.
In the right panel we show the reduced bispectrum
produced by this model on equilateral configurations.
The amplitude is of order $10^{-1}$.
It would be interesting to search for a parameter combination
(even if very finely tuned)
that yields a more significant bispectrum,
but we leave this issue for future work.

Although we have not successfully reproduced a QSFI effect with
significant amplitude,
this model demonstrates the ability of our numerical
tools to handle
scenarios with Hubble-scale masses or sizeable cubic couplings.

\subsubsection{Non-adiabatic model: particle production}
\label{sec:non-adiabatic-model}

It is much easier to construct models that realize
nonadiabatic effects.
In this section
we study a model introduced by
Gao, Langlois \& Mizuno~\cite{Gao:2012uq},
which is designed so that the inflationary trajectory contains
a turn through an angle $\Delta \theta$.
If the turn is sufficiently sharp then it
will force the heavy field away from its minimum,
and the subsequent relaxation will be nonadiabatic.

\begin{figure}
    \begin{center}
        \includegraphics[scale=0.25]{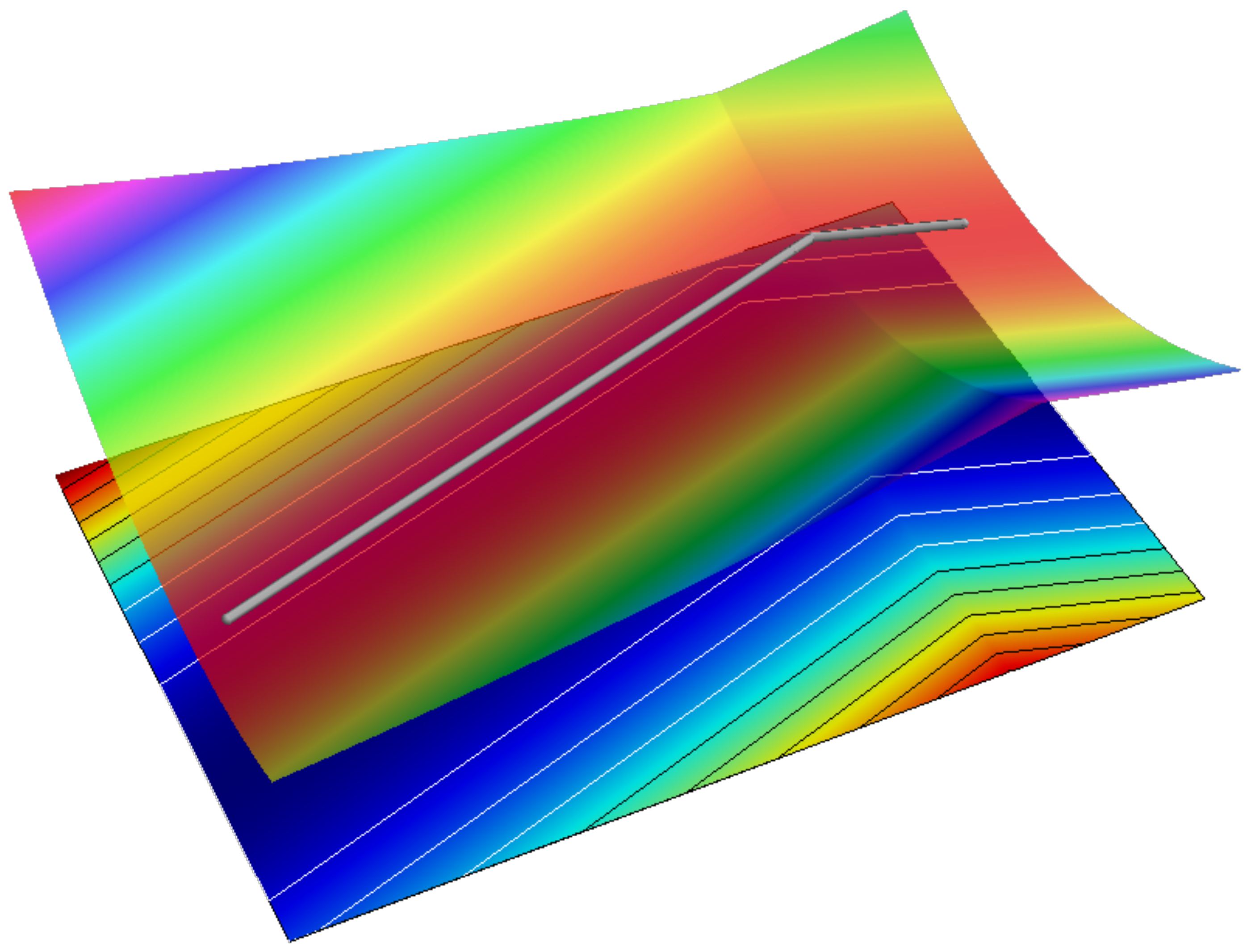}
        \includegraphics[scale=0.5]{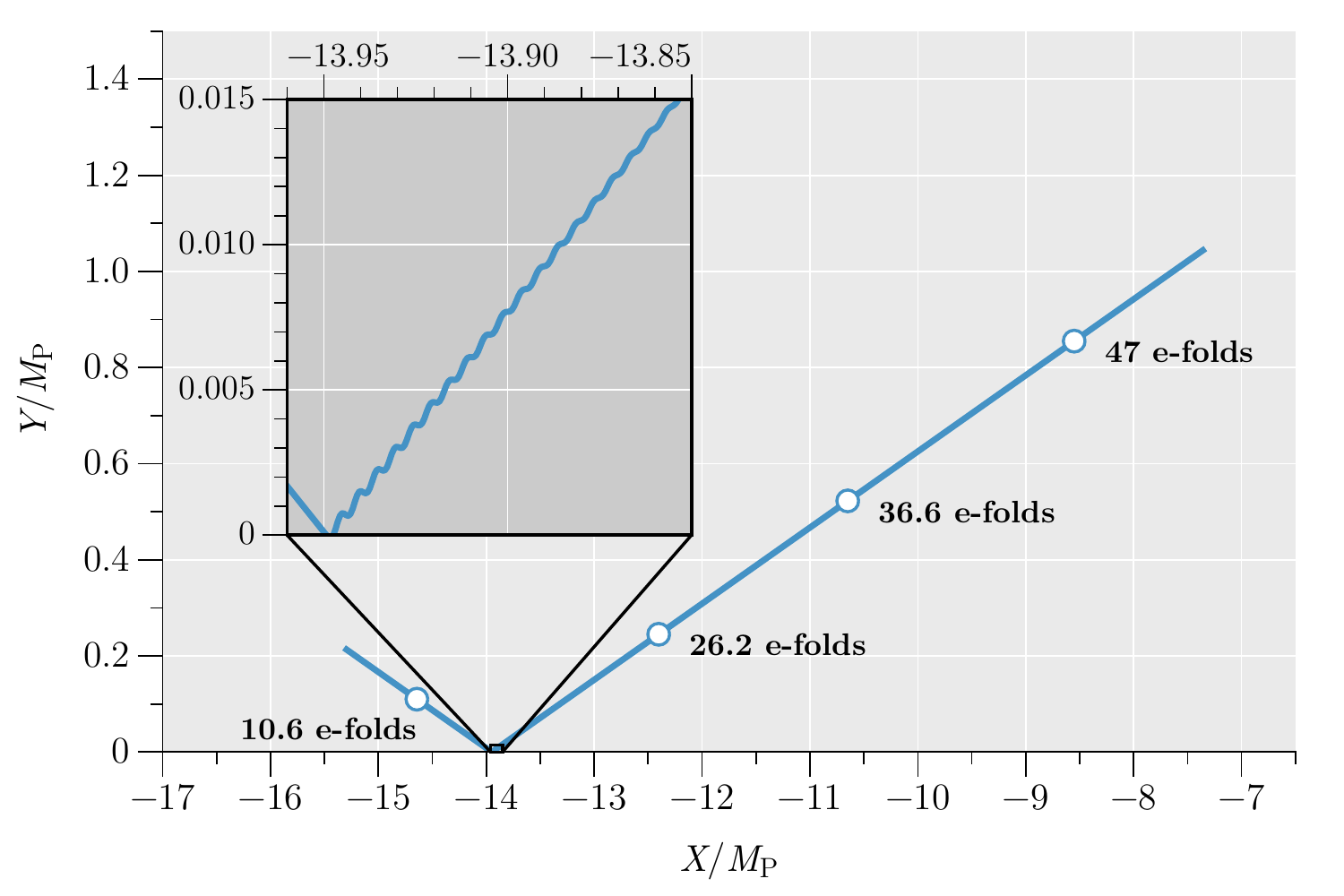}
    \end{center}
    \caption{\label{fig:nonadiabatic-trajectory}Potential and trajectory
    for the nonadiabatic model~\eqref{eq:nonadiabatic-potential}.
    Left: Potential and trajectory. The fields roll from the far to the near side,
    and the lower plane shows equipotential contours.
    Right: zoomed plot of the trajectory.}
\end{figure}

The potential used by Gao et al. was
\begin{equation}
    V(X, Y) = \frac{1}{2} m_X^2 X^2
    +
    \frac{1}{2} M^2 \cos^2 \frac{\Delta \theta}{2}
    \Big[
        Y
        -
        (X - X_0) \tan \Xi(X)
    \Big]^2 ,
    \label{eq:nonadiabatic-potential}
\end{equation}
where $\Xi(X)$ is defined by
\begin{equation}
    \Xi(X) = \frac{\Delta \theta}{\pi}
    \arctan s(X - X_0) .
\end{equation}
The parameter $s$ controls the sharpness
of the turn, which occurs near $(X, Y) = (X_0, 0)$.
Larger values of $s$ generate a sharper turn.
We take $M = 10^{-4} \Mp$,
$m_X = 10^{-7} \Mp$, and
$\Delta \theta = \pi/10$.
The turn is positioned
at $X_0 = (231 - 100\sqrt{6}) \Mp \approx -14 \Mp$
and
the sharpness parameter is $s = 1000 \sqrt{3} / \Mp \approx 1700/\Mp$.
We choose initial conditions
$X = (229 - 100 \sqrt{6}) \Mp \approx -16 \Mp$
and
$Y = 2 \Mp \tan (\pi/20)$.
The potential and trajectory are shown in
the left-hand panel of Fig.~\ref{fig:nonadiabatic-trajectory}.
The right-hand panel shows the trajectory in more detail, including
a zoomed region focused on the turn.
When viewed as a whole the trajectory has the appearance of two
straight lines joining abruptly
at $N \approx 16$ e-folds from the initial time.
With sufficient magnification, however, the small oscillations
characterizing nonadiabatic relaxation
are clearly visible.
The changing mass associated with these oscillations leads
to stimulated particle production~\cite{Konieczka:2014zja}.

\begin{figure}
    \begin{center}
        \includegraphics[scale=0.5]{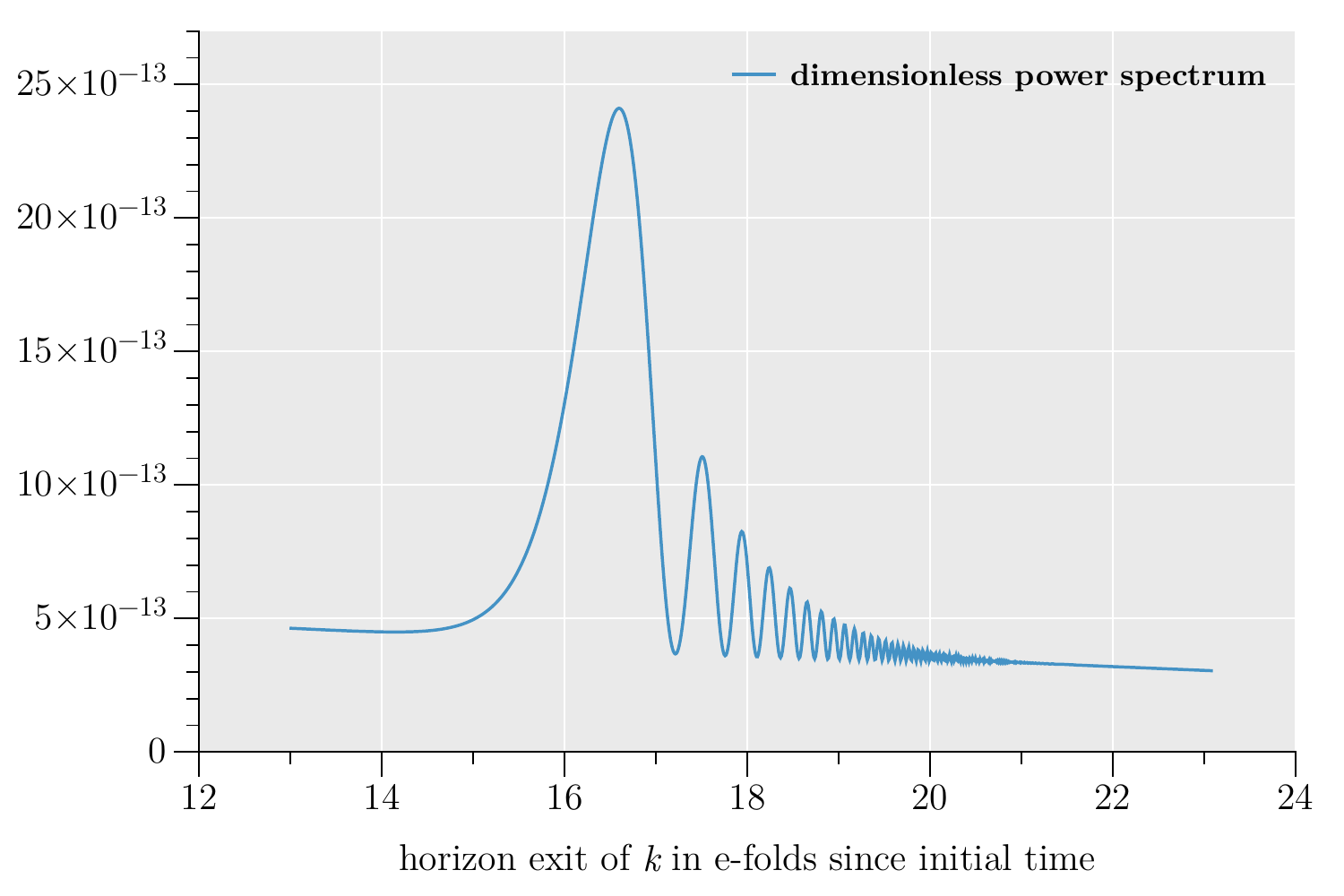}
        \includegraphics[scale=0.5]{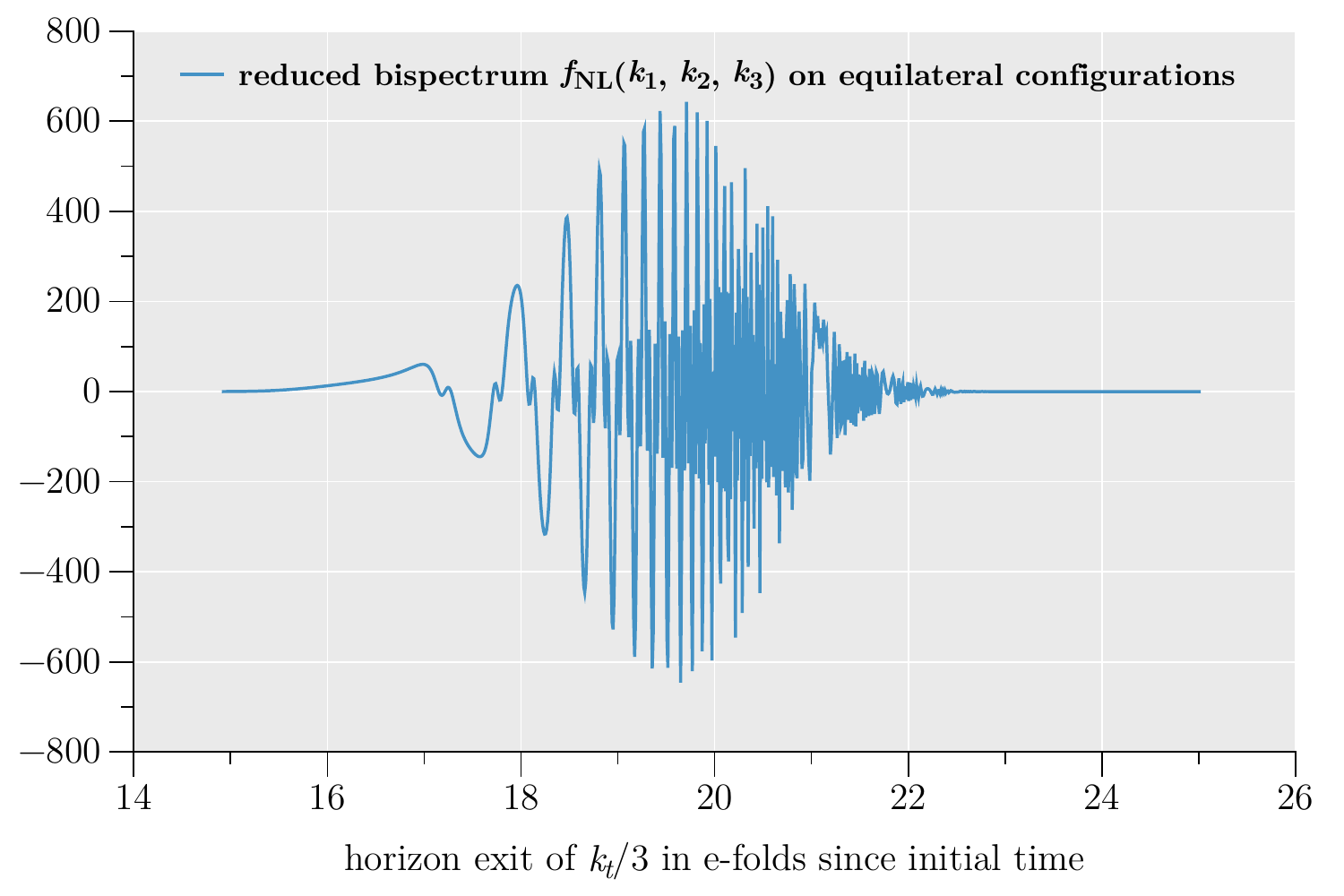}
    \end{center}
    \caption{\label{fig:nonadiabatic-outputs}
    Left panel: dimensionless power spectrum produced in the nonadiabatic
    model~\eqref{eq:nonadiabatic-potential}.
    Right panel: reduced bispectrum
    evaluated on equilateral configurations
    as a function of scale $k_t$.}
\end{figure}

We plot the dimensionless power spectrum
and reduced bispectrum on equilateral configurations in
Fig.~\ref{fig:nonadiabatic-outputs}.
The power spectrum shows a significant enhancement of power
associated with the turn,
with the enhancement peaking for scales that exit the
horizon just after $N=16$ e-folds from the initial time.
Smaller scales show decaying oscillations,
eventually returning to the continuum level
when the non-adiabatic evolution has died away.
In the power spectrum these oscillations become negligible
for scales that exit the horizon $N \gtrsim 21$
e-folds from the initial time.

The reduced bispectrum
also shows significant enhancement, but with a more complex structure.
There is some growth around the turn at $16$ e-folds, but the most dramatic
effects occur later during the oscillating phase.
There are rapid, large-amplitude oscillations
achieving up to $|\fNL(k_1, k_2, k_3| \sim 600$.
As the heavy field returns to adiabatic evolution
these oscillations decrease in amplitude,
and the bispectrum returns to its background level.
The transition occurs on slightly smaller scales than for the
power spectrum, for scales $k_t$ that exit the horizon
$N \gtrsim 23$ e-folds from the initial time.

\para{Comparison with analytic estimates}
Gao, Langlois \& Mizuno developed an analytic framework
in which to estimate the power spectrum produced by~\eqref{eq:nonadiabatic-potential},
but did not discuss the bispectrum~\cite{Gao:2012uq,Gao:2013ota}.
Ach\'{u}carro and collaborators
introduced a similar framework
and used it to study contributions to both the power spectrum and bispectrum,
but their analysis assumed that any features were only
moderately sharp~\cite{Achucarro:2013cva,Achucarro:2014msa}.
Chen et al. studied single-field models with sharp features~\cite{Chen:2008wn}.
Adshead et al. later developed an alternative
single-field framework based on
a Green's function solution to the mode equations~\cite{Adshead:2013zfa}.
More recently, Flauger et al. estimated the bispectrum from
particle production in two field model which exhibits
an approximate discrete shift symmetry~\cite{Flauger:2016idt}.

None of these formalisms are exactly applicable to~\eqref{eq:nonadiabatic-potential}
with our parameter choices.
However, the general structure visible in Fig.~\ref{fig:nonadiabatic-outputs}---a sudden
increase in the two-point function, followed by decaying oscillations,
and rapid oscillations embedded within a `pulse' of large amplitude
in the three-point function---are qualitatively in very good agreement with
Refs.~\cite{Achucarro:2013cva,Achucarro:2014msa,Adshead:2013zfa}.

\section{Performance and scaling behaviour}
\label{sec:performance}

Before concluding, we pause to discuss the performance of our method---%
and especially how the
integration time scales with
different choices for key parameters such as the number
of e-folds of massless subhorizon evolution.
To illustrate these properties
we use the model of double quadratic inflation,
\begin{equation}
    V = \frac{1}{2} m_\phi^2 \phi^2 + \frac{1}{2} m_\chi^2 \chi^2 .
\end{equation}
This potential was used by Rigopoulos, Shellard \& van Tent
to provide a simple test-case for numerical methods~\cite{Rigopoulos:2005xx,Rigopoulos:2005us};
see also Refs.~\cite{Mulryne:2009kh,Mulryne:2010rp, Huston:2011fr,Frazer:2013zoa,Peterson:2010np,Price:2014xpa}.
Because the potential is sum-separable
it falls into a small class where analytical
results for the factorization coefficients
$N_a$ and $N_{ab}$ can be computed using the slow-roll
approximation;
these calculations were performed by
Vernizzi \& Wands~\cite{Vernizzi:2006ve}.
In our computations we
take $m_\phi = 9 \times 10^{-5} \Mp$
and
$m_\chi = 1 \times 10^{-5} \Mp$.
The initial conditions are
$\phi = 10 \Mp$
and $\chi = 12.9 \Mp$,
and
their derivatives are set using the slow-roll approximation.

\subsection{Number of massless subhorizon e-folds}
\label{sec:massless-efolds}

It was explained in~\S\ref{sec:initial-conditions}
that our strategy of computing
initial conditions using the massless approximation
forces us to position the initial time
sufficiently early
that $(k/a)^2 \gg m^2$,
where $m^2$ is the largest eigenvalue of the mass matrix.
Typically $m^2$ will itself be time dependent.
We define the \emph{massless time} for the mode
$k$
to occur
when
$(k/a)^2 = m^2$,
or at horizon crossing $(k/a)^2 = H^2$, whichever is earlier.%
    \footnote{This terminology
    is a convenient shorthand, but one should not be misled.
    When $(k/a)^2 = m^2$
    the kinetic energy strictly
    balances the potential energy from the mass.
    To be in the massless regime
    we need $(k/a)^2 \gg m^2$.}

The initial time
for a given wavenumber configuration
should be placed at least a few e-folds earlier than the
earliest massless time with which it is associated.
(By an extension of terminology we say that this earliest massless
time is the massless time for the configuration as a whole.)
The earlier we take the initial time, the more accurate the
calculation of initial conditions should be.
Typically this means that numerical accuracy improves
as we increase the number of e-folds spent in the massless phase.
On the other hand, integration in this phase is expensive
and we must often make a choice that balances runtime against
accuracy.

The {\CppTransport} platform allows a user to
automatically position the initial time for
each configuration a fixed number of e-folds $\Npre$
prior to its massless time.
The same choice can be made with {\PyTransport}
using a set of supplied scripts.
In this subsection we study how runtime and accuracy vary
as we adjust $\Npre$.

\begin{figure}
    \begin{center}
        \includegraphics[scale=0.5]{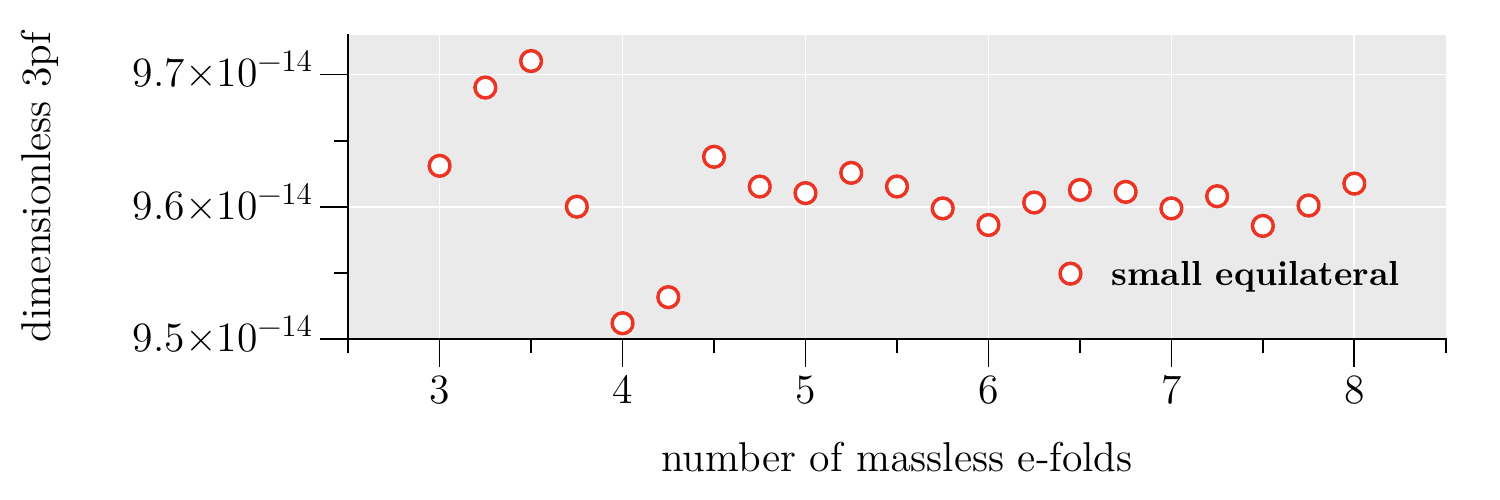}
        \includegraphics[scale=0.5]{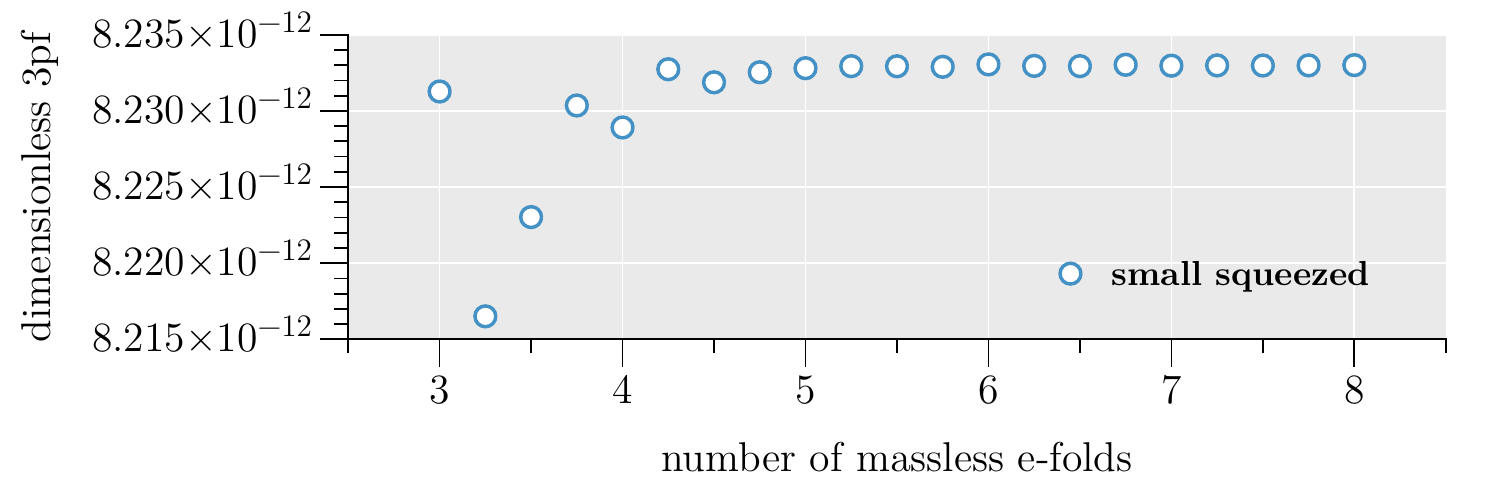}
    \end{center}
    \begin{center}
        \includegraphics[scale=0.5]{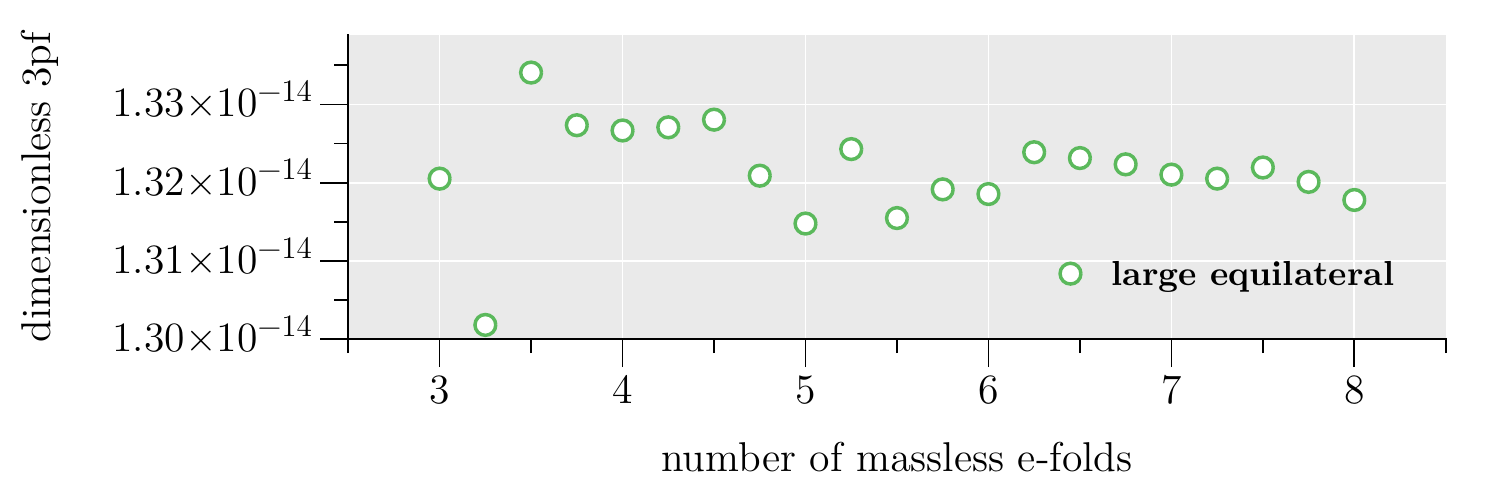}
        \includegraphics[scale=0.5]{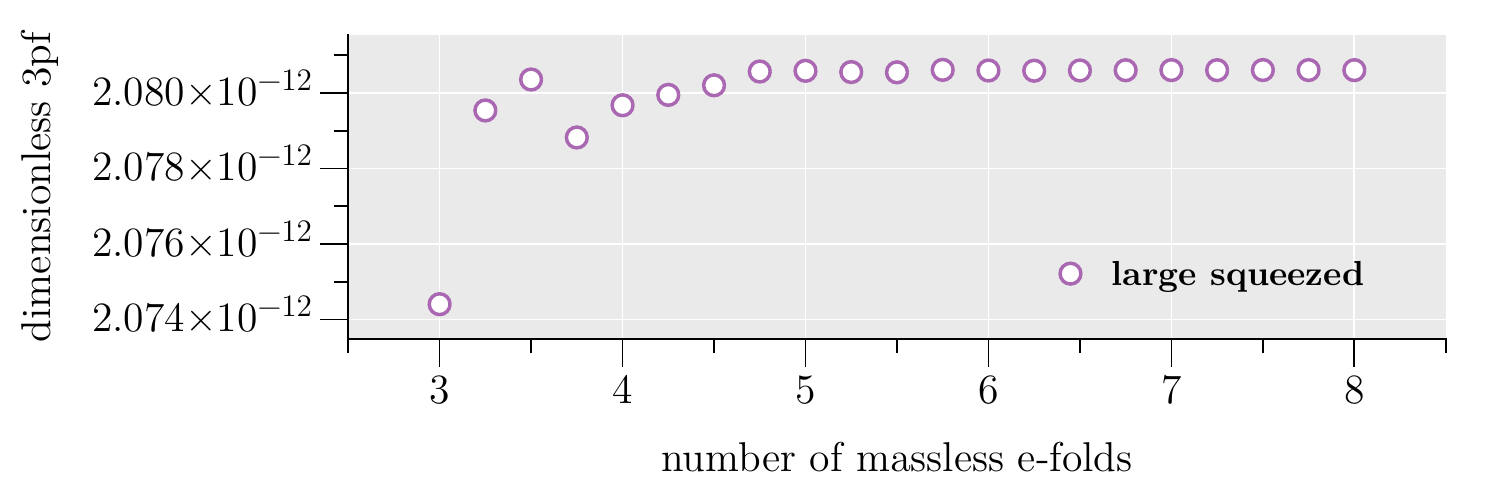}
    \end{center}
    \caption{\label{fig:convergence}Convergence properties of the numerical
    solution with increasing number of massless subhorizon e-folds.}
\end{figure}
\begin{figure}
    \begin{center}
        \includegraphics[scale=0.5]{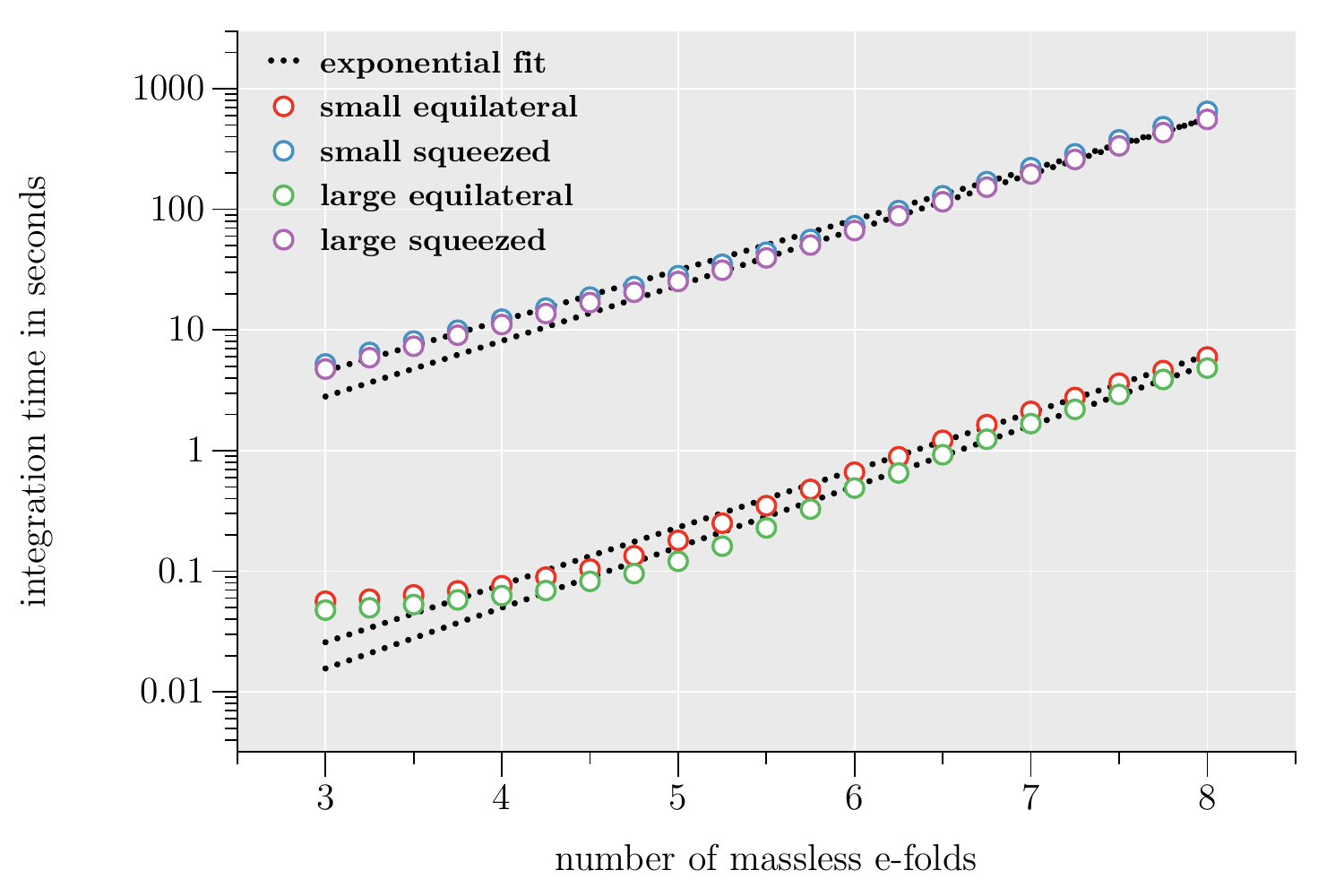}
    \end{center}
    \caption{\label{fig:timing}Scaling of integration time
    with increasing number of massless (or subhorizon) e-folds.}
\end{figure}

\para{Convergence}
In Fig.~\ref{fig:convergence}
we plot the final value of the dimensionless
$\zeta$ bispectrum
$\dimlessB(k_1, k_2, k_3)$
for four wavenumber configurations:
\begin{enumerate}
    \item \semibold{Top left}: a `small' equilateral configuration for which $k_t/3$
    has horizon-exit time $\approx 19.0$ e-folds after the initial conditions.

    \item \semibold{Top right}: a `small' squeezed configuration with the same $k_t$ as
    above, but $\alpha = 0$ and $\beta = 0.99$.

    \item \semibold{Bottom left}: a `large' equilateral configuration for which $k_t/3$
    has horizon-exit time $\approx 24.5$ e-folds after the initial conditions.

    \item \semibold{Bottom right}: a `large' squeezed configuration with the same
    value for $k_t$, but $\alpha = 0$ and $\beta = 0.99$.
\end{enumerate}
It is clear that the results are very stable, especially for the squeezed
configurations.
The equilateral configurations shows some scatter
but the numerical value settles down for
$\Npre \gtrsim 4.5$.

\para{Scaling of integration time with $\Npre$}
In Fig.~\ref{fig:timing}
we plot the corresponding integration times $T$
required by the {\CppTransport} platform,
together with dotted lines showing fits to
the scaling law
$T \propto \e{\alpha \Npre}$.%
    \footnote{The test configuration was an
    {\packagefont iMac13,2}: 3.4 GHz Intel Core i7-3770 (Ivy Bridge), macOS 10.11.5, clang-703.0.29, Boost 1.59.}
The squeezed configurations fall nearly exactly on such a law
with $\alpha \approx 1$,
and the same is true
for the equilateral configuration
at sufficiently large $\Npre$.
The same scaling is seen with {\PyTransport}.

The conclusion is that increasing the number of massless
e-folds is exponentially expensive.
The reasons for this behaviour were explained in~\S\ref{sec:axion-quartic}
above,
and can ultimately be traced to increasingly rapid phase oscillations
on subhorizon scales.
Fortunately,
Fig.~\ref{fig:convergence}
shows that convergence is fairly rapid,
and beyond a certain point there are diminishing returns.
Therefore very large values of $\Npre$ are seldom necessary;
choices in the range $3$ to $5$ give reasonable
results for typical models.
In more complex cases where a dynamical feature occurs near horizon exit
it is necessary to use larger $\Npre$, in some cases as large as $7$ or $8$.

\subsection{Shape-dependence}
\label{sec:timing-shape}

We have seen that the number of e-folds of massless evolution
is a key factor in determining the integration time.
A second factor is the shape of the wavenumber configuration, as measured
by $\alpha$ and $\beta$.
In particular, strongly squeezed configurations
contain one wavenumber that is much smaller than the other two.
Without loss of generality we can take this to be $k_3$, so
we are focusing on configurations for which
$k_3 \ll k_2 \sim k_3 \sim k_t / 2$.

\begin{figure}
    \begin{center}
        \includegraphics[scale=0.5]{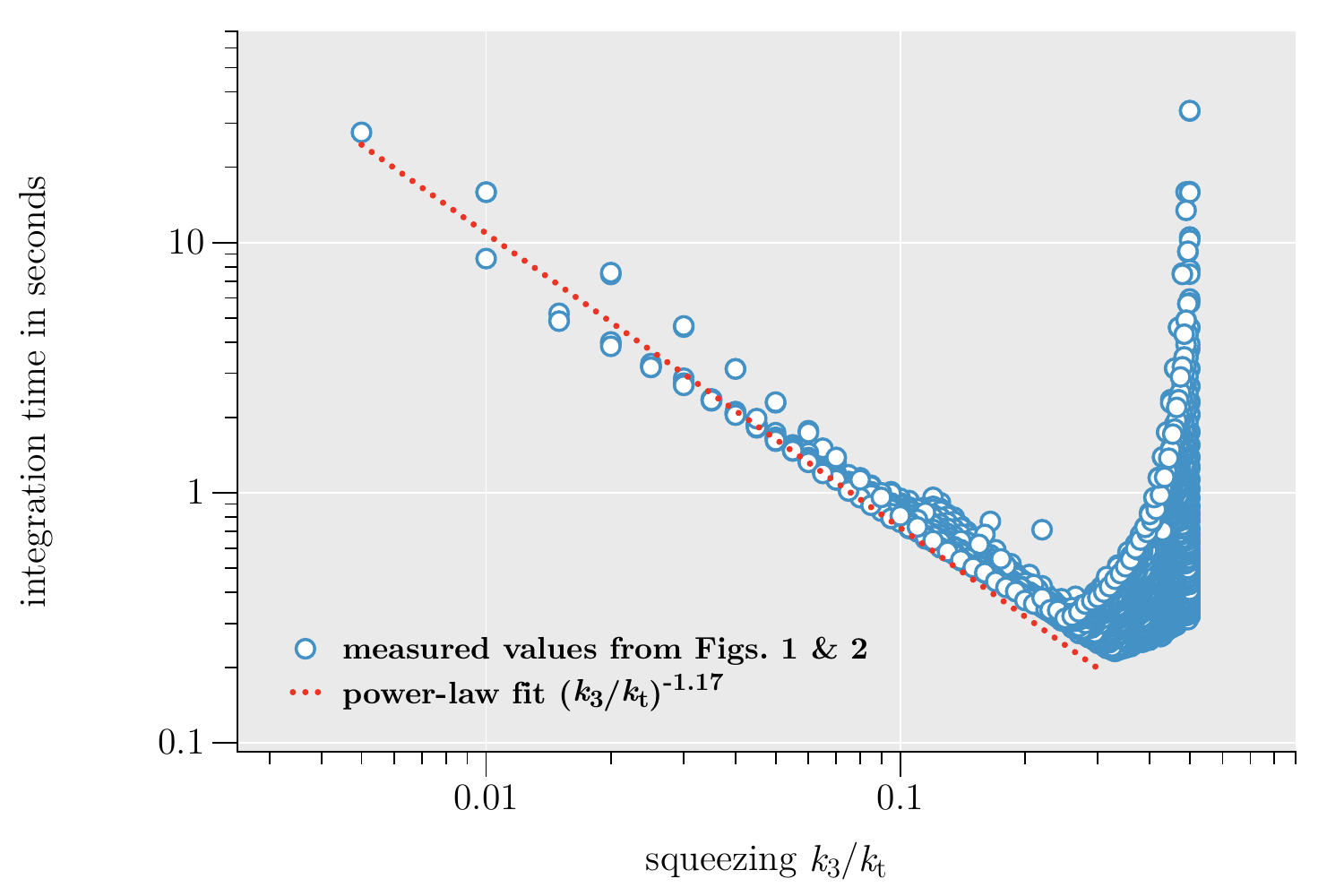}
    \end{center}
    \caption{\label{fig:timing-squeezed}Dependence of integration time on the
    squeezing parameter $k_3/k_t$.}
\end{figure}

The massless time for this configuration will be determined by the massless
time for $k_3$, which may be many e-folds earlier than the massless time
for $k_1 \sim k_2$.
This means that there will be a long phase of exponential decay associated
with terms of the form $k_1/aH$ and $k_2/aH$,
and on the basis of what has been said in~\S\ref{sec:massless-efolds}
this phase will be expensive
to integrate.
Therefore we should expect the integration time to scale strongly with the
`squeezedness' of the configuration.
Since we are taking $k_3$ to be the squeezed wavenumber this limit
occurs when $\beta \uparrow 1$,
but to measure the scaling it proves to be more convenient to work in terms
of $k_3/k_t = (1-\beta)/2$.
In Fig.~\ref{fig:timing-squeezed}
we plot the integration time for each configuration
used to construct the bispectrum shape plots in Figs.~\ref{fig:shape-evolve}
and~\ref{fig:central-shape-evolve}.
Recall that these configurations have fixed $k_t$
but varying $\alpha$ and $\beta$.

In Fig.~\ref{fig:timing-squeezed}
the different $\alpha$ configurations are visible as a cluster of points
near $k_3/k_t \sim \frac{1}{3}$.
For $k_3/k_t < 1$ the allowed triangular configurations become rarer,
and eventually for very small $k_3/k_t$ there are
only isosceles configurations with $\alpha = 0$.
In the squeezed region
the timing data can be roughly fit by a power law $(k_3/k_t)^{-1.17}$.

The precise details of the scaling vary with the model under discussion,
but the dependence on $k_3/k_t$ is normally a power-law.
In some models there can be one or more breaks
between different power laws, depending
(among other possibilities) on the dynamical
behaviour of the background.
Broadly speaking we find the power-law index lies between $-1$ and $-2$.
For example, for the complex scaling model discussed in~\S{4}
of Byrnes et al.~\cite{Byrnes:2015dub},
we find that the time required in the deeply squeezed limit
scales like $(k_3/k_t)^{-1.64}$.
We conclude that strongly squeezed configurations are
expensive to compute. This is unfortunate given that some observables
are principally sensitive to these configurations, such as the scale-dependent
galaxy bias or the position-dependent power spectrum.

\section{Conclusions}
\label{sec:conclusions}

The major result of this paper is a complete formalism for numerical
calculation of the tree-level correlation functions produced during an epoch
of early-universe inflation.
This formalism was described
in~\S\S\ref{sec:numerical-computation}--\ref{sec:gauge-transform}.
It does not require the slow-roll approximation
except to obtain the
estimates of initial
conditions given in~\S\ref{sec:initial-conditions}.
As explained in~\S\ref{sec:tree-level},
the tree-level approximation means that our formalism
should produce accurate estimates unless
multiparticle
production channels make a significant
contribution to the curvature perturbation $\zeta$,
for example by nontrivial scattering processes
or decays.
In certain scenarios, such as warm inflation or trapped
inflation,
the tree approximation may also fail to capture
processes by which energy is drained from
the zero-mode into finite-wavenumber excitations.
These restrictions should be carefully considered
before applying our tools---or
any others based on a tree-level approximation---to study
some particular inflationary model.

We are supplying two concrete implementations of this
formalism,
described in~\S\ref{sec:automated-codes}
and
available for download under open source licences.
These are not just bare implementations of the evolution equations;
instead, they both
support \emph{automated} analysis of models
from a high-level Lagrangian description,
and can be
used to produce immediate high-resolution numerical results for
models
whose bispectra were previously intractable.

In~\S\S\ref{sec:numerical-computation}--\ref{sec:gauge-transform}
we focused on the correlation functions generated by a system of
canonically-normalized scalar fields (and their contribution
to the curvature perturbation $\zeta$), and
the two bispectrum-level implementations
described in~\S\ref{sec:automated-codes} are
currently restricted to scenarios of this type.
However, this restriction
is not necessary as a matter of principle.
Extensions to more complex models, such as those with a nontrivial
kinetic sector
$G_{\alpha\beta}(\phi) \partial_a \phi^\alpha \partial^a \phi^\beta$
or interactions of Galileon-type,
are limited only by algebraic complexity.
All that would be required
are replacements for the tensors
${u^{\ext{a}}}_{\ext{b}}$
and
${u^{\ext{a}}}_{\ext{b}\ext{c}}$,
and appropriate initial conditions that account for the
new interactions and kinetic structure.
An older Mathematica-based implementation
capable of handling models with nontrivial field-space metric is
already available, although
its output is limited to the two-point function~\cite{Dias:2015rca}.
All three implementations,
together with links to further resources,
can be found at the website
\href{https://transportmethod.com}{transportmethod.com}.

In~\S\ref{sec:examples} we exhibited results for a
selection of concrete models
exemplifying the wide range of mass spectra and coupling constants
that can be accommodated.
Collectively, these demonstrate that our formalism successfully tracks highly nuanced
features of the bispectrum amplitude and shape.
Just as important, because it includes all relevant
effects (especially from
gravitational-strength couplings),
it can be used to predict
the complete scale- and shape-dependent bispectrum generated in models
where the reduced bispectrum $\fNL(k_1, k_2, k_3)$ is order unity.
These models are a target for the next generation of galaxy surveys,
including Euclid, DESI and LSST~\cite{Alvarez:2014vva}.
To obtain robust predictions with these low amplitudes it is not sufficient
to rely on approximations that discard physical effects occurring
on subhorizon scales or around horizon exit,
or that do not accurately account for the effect of hierarchies among the
wavenumbers appearing in each correlation function.
Our formalism, and especially the reusable implementations we provide,
supply a means for these models to be accurately analysed for the first time.

Meanwhile, to achieve a suitable level of preparation for a Euclid-, DESI-
or LSST-like survey it will be insufficient
merely to improve the accuracy of primordial calculations.
Reliable forecasts for models where
$|\fNL(k_1, k_2, k_3)| \lesssim 1$
must at least account for gravitational evolution after horizon exit
and the characteristics of the survey.
These details are now well-understood; what is required is an integrated
toolchain that links them all together.
At the simpler level required by CMB experiments, an analysis
such as that given in Ref.~\cite{Byrnes:2015dub}---%
which relied on numerical bispectra produced using the {\CppTransport}
platform---%
demonstrates how accurate, high-resolution
calculations of primordial correlation functions
can be integrated into a numerical toolchain
producing accurate, reliable results customized
to a specific experiment.
(In Ref.~\cite{Byrnes:2015dub} the customization was for a Planck-like
CMB survey,
but the point we are making is much more general.)
As we stockpile datasets of ever-increasing accuracy
there is a corresponding burden on theorists to
generate predictions with matching refinement.
No one tool, or single approach, will be sufficient---but we
hope that that the software tools we are making available
constitute a step towards this goal.

\begin{acknowledgments}

\para{\PyTransport}
Development of {\PyTransport} has been supported by the Royal Society through
the support given to DJM by a Royal Society University Research Fellowship and
by Science and Technology Facilities Council grant ST/J001546/1.

\para{\CppTransport}
Development of {\CppTransport} has been supported
significantly
by an ERC grant:
\begin{itemize}
    \item \emph{Precision tests of the inflationary
    scenario},
    funded by
    the European Research Council under the European Union's
    Seventh Framework Programme (FP/2007--2013) and ERC Grant Agreement No. 308082.
    (MD, DS)
\end{itemize}
In addition,
some development of {\CppTransport} has been supported by
other funding sources.
Portions of
the work described in this document have been supported by:
\begin{itemize}
    \item The UK
    Science and Technology Facilities Council via grants
    ST/I000976/1 and ST/L000652/1,
    which funded the science programme at the University of Sussex Astronomy
    Centre from April 2011--March 2014
    and April 2014--March 2017, respectively.
    (MD, JF, DS)
    \item The Leverhulme Trust via a Philip Leverhulme Prize. (DS)
    \item The National Science Foundation Grant No. PHYS-1066293
    and the hospitality of the Aspen Center for Physics. (MD, JF, DS)
    \item The hospitality of the Higgs Centre for Theoretical
    Physics at the University of Edinburgh,
    and the Centre for Astronomy \& Particle Physics
    at the University of Nottingham. (DS)
\end{itemize}

In addition, MD acknowledges funding from the German Science Foundation (DFG) within the Collaborative Research Centre 676 \emph{Particles, Strings and the Early Universe} and by the ERC Consolidator Grant STRINGFLATION under the HORIZON 2020 contract no. 647995, and JF acknowledges funding from the ERC Consolidator Grant STRINGFLATION under the HORIZON 2020 contract no. 647995 and a grant from the Simons Foundation.

\para{Data availability statement}
No new data were collected for the preparation of this paper.
All plots and figures are generated by computer codes.
These codes are freely available for download as described
in~\S\ref{sec:automated-codes}.

\end{acknowledgments}

\end{fmffile}

\bibliographystyle{JHEP}
\bibliography{paper}

\end{document}